\documentclass[useAMS]{mn2e}
\usepackage{graphicx,amssymb,amsmath}
\usepackage{hyperref,xcolor}
\usepackage{subfigure}
\usepackage{pdflscape}
\usepackage{mathrsfs}

\title[Unveiling  Vela ]
  { Unveiling  Vela - Time Variability of Na I D lines in the Direction of the Vela Supernova Remnant \thanks{Based on
observations obtained with  The Vainu Bappu Telescope of the Indian Institute of Astrophysics .}}
\author[N. Kameswara Rao, S Muneer, David L. Lambert \& B.A. Varghese ]
  { N.Kameswara Rao$^1$$^,$$^2$, S.Muneer$^1$, David L. Lambert$^2$ \& B.A. Varghese $^1$\\
        \thanks{E-mail: nkrao@iiap.res.in (NKR);
muneers@iiap.res.in (SM); dll@astro.as.utexas.edu (DLL); baba@iiap.res.in (BAV)},
      \\
  $^1$Indian Institute of Astrophysics, Bangalore 560034, India\\
       $^2$The W.J. McDonald Observatory and Department of Astronomy, The University of Texas, Austin, TX 78712-1083, USA\\ }

\begin{document}

\pagerange{\pageref{firstpage}--\pageref{lastpage}} \pubyear{2014}

\maketitle

\label{firstpage}

\begin{abstract}
   High-resolution  spectral profiles of  Na\,{\sc i} D lines from the interstellar medium
  towards 64 stars in the direction of  the Vela supernova
  remnant are presented. This survey conducted mostly between 2011-12
   complements an earlier
  survey of the same stars by Cha \& Sembach done  in the 1993-96 period.   The interval  of 15 to 18 years provides a base line to search
   for changes in the interstellar profiles.
    Dramatic disappearance of strong
   absorption components at low radial velocity is seen
   towards three stars  -- HD 63578, HD 68217, HD 76161  -- over 15-18
    years; HD 68217 and HD 76161 are associated with the Vela SNR but
 HD 63578 is likely associated with the
   wind bubble of $\gamma^2$ Velorum. The vanishing of these cold neutral
   clouds in  the short time of 15 to 18 years needs some explanation. Other changes are seen in high-velocity Na D components.

\end{abstract}


\section{Introduction} 
The supernova  responsible for the Vela supernova remnant (SNR)  exploded about 11000 years ago (Reichley,  Downs \& Morris 
1970).\footnote{ Paleoindians in the southern Peruvian Andes who had high altitude settlements  spanning the past 12.4  thousand years
  (Rademaker et al. 2014) might have witnessed this event ! }   Within the Vela SNR lies a pulsar at a distance of 287$\pm$19 pc, as measured by its VLBI parallax
  (Dodson et al. 2003).  This distance is consistent with that determined from a search for high-velocity components in high-resolution Ca\,{\sc ii} and Na\,{\sc i} interstellar absorption
  line profiles towards OB stars with {\it Hipparcos} and spectroscopic parallaxes. These high-velocity components discovered by Wallerstein \& Silk (1971) which are attributed to interactions between the local
  interstellar medium and the SNR  were assigned the distance of 290$\pm$30 pc (Sushch, Hnatyk \& Neronov  2011; Cha, Sembach \& Danks 1999). 

\begin{figure*}
\includegraphics[width=14cm,height=14cm]{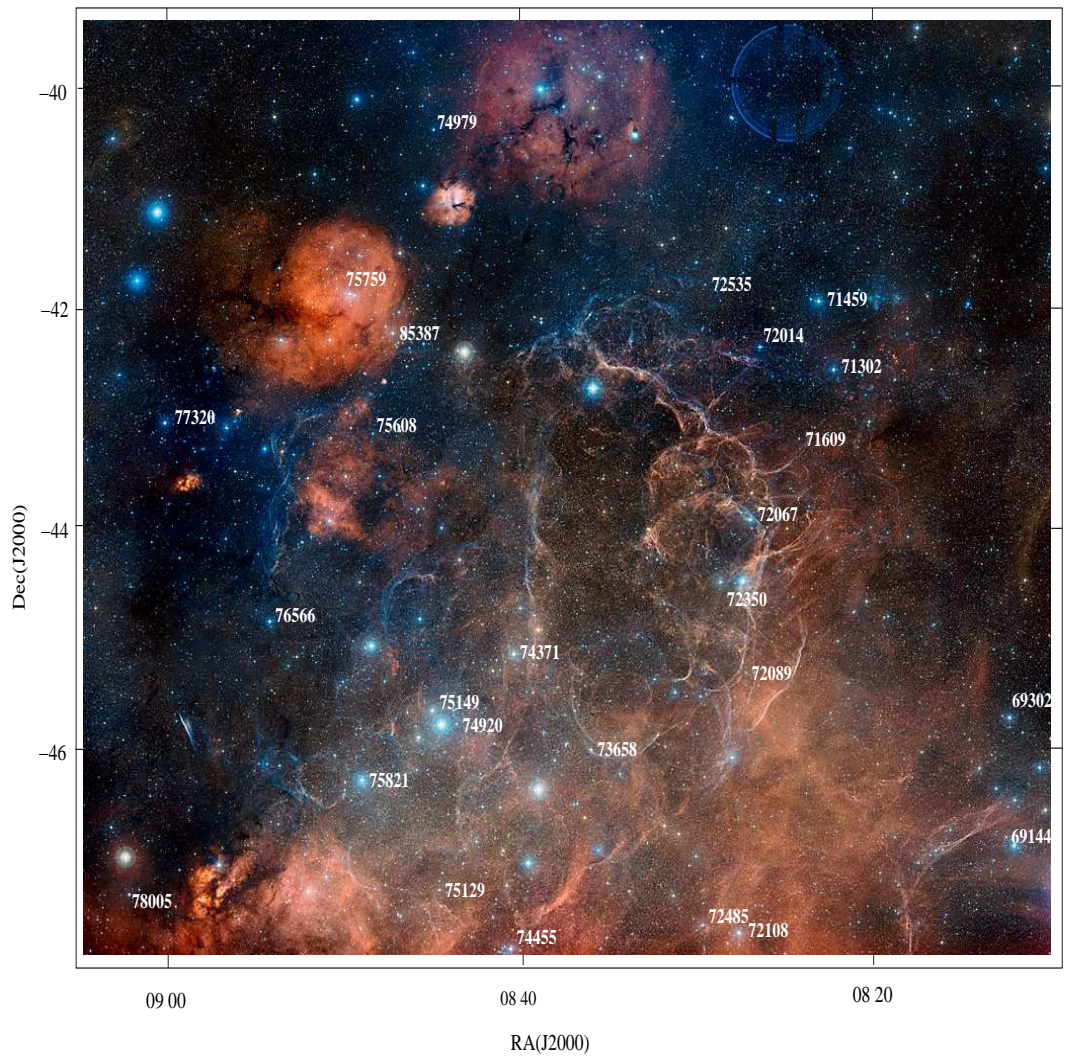}
\caption{Location of the stars observed towards the Vela SNR. This optical image covering a field of 9.3$\degr$ x 8.5$\degr$ is from Davide de Martin (WWW.Skyfactory.org)- by permission.  Stars are identified by their HD number.}
\end{figure*}

 Various studies have expanded observations of the  interstellar
  components seen in stars in the direction of the Vela remnant with some studies emphasizing searches for variability of high and low velocity  components with efforts concentrated on the Ca\,{\sc ii} K and Na\,{\sc i} D
  lines.  This paper  reports on a campaign mainly in 2011-2012 to obtain high-resolution profiles of the Na D lines towards many OB stars in the direction of the Vela SNR and to compare
  profiles with similar data obtained in 1993-1996 by Cha \& Sembach (2000) of Ca\,{\sc ii} K and Na\,{\sc i} D lines.  As appropriate, we comment on the relationship
  between our Na\,{\sc i} D line profiles and other published observations  towards stars behind the Vela SNR.

            Na\,{\sc i} D lines are important in several respects.
They  represent the behaviour of neutral gas. It is now clear that the Vela SNR's
 expansion proceeded in a cloudy interstellar medium. Earlier studies of the
 absorption lines towards  various stars suggested that the high velocity
 ($> 100$  km s$^{-1}$) components,  mainly present in Ca\,{\sc ii} lines,
 represent the shocked gas that had interacted
  with the SNR (Wallerstein, Silk \& Jenkins 1980; Jenkins, Silk \& Wallerstein 1984). However, as
 pointed out by Sushch, Hnatyk \& Neronov  (2011), although the distance to the SNR is 290
 $\pm$ 20 pc,
 the high velocity absorption components are present  only in stars more 
 distant than 500 pc and stars with distances smaller than 350 pc do not
 show evidence of 100  km s$^{-1}$ absorption components.  Stars at or 
 immediately behind the SNR show components of intermediate velocity, which
  are well represented by Na\,{\sc i} lines along with the Ca\,{\sc ii} K lines.
 Thus, the Na\,{\sc i} components might be from the neutral ISM that is 
  interacting with the SNR and surroundings. 
  
 The Vela SNR has a diameter of about  7.3 degrees on the sky (Aschenbach 1993;
Aschenbach, Egger \& Trumper 1995).  Figure 1  shows the remnant  and the locations of  most of the stars observed in our
 Na D survey. 
  The optical image covering a field of 9.3$\degr$ x 8.5$\degr$ from Davide
 de Martin (WWW.Skyfactory.org) is compiled from AAO, UK Schmidt and Digital sky
 survey images.
 Cha \& Sembach's (2000)  multi-star survey was based on a target list of 68 OB stars observed at high spectral resolution
  in 1993, 1994 and 1996.  A  selection of the target list was observed more than once  and  slightly
more than half of the  68 stars observed for Ca\,{\sc ii} were observed also at Na\,{\sc i}.  Of the 13 stars observed twice, seven proved to have
variable absorption line profiles.  Our survey principally from 2011 and 2012 provides only Na\,{\sc i} profiles for the majority of Cha \& Sembach's 68 stars and expands their baseline
of three years to 15 to 18 years.   Whenever possible,  other high-resolution profiles in the literature are included in the discussion of variability.

\begin{table*}
\centering
\begin{minipage}{160mm}
\caption{\Large Observations \& Other Parameters  }
\begin{footnotesize}
\begin{tabular}{lcrcrrrrrccrr}
\hline
 Star   &\multicolumn{2}{c}{$\alpha$,$\delta$(2000)}&& &  \multicolumn{3}{c}{}&&\multicolumn{2 }{c}{ Distance} \\
\cline{2-3} \cline{6-8} \cline{10-11} \\
  &    &      & & epoch& V   & (B-V) &E(B-V)& &d(Hip)$^{a}$ &d(Sp) & (S/N)$^{b}$  \\
     &   &      & &  & &     & & &pc  &pc  & \\
\hline
HD 63308&07$^{h}$46$^{m}$33$^{s}$.4 &-40$^{o}$03$^{'}$34$^{"}$.2&  &2011.1.7 & 6.57&-0.13 & 0.11&&606$\pm$108& 490&112 \\
       &           &           &  &2011.3.16&     &      &     &&           &    & 63\\
       &           &           &  &2012.11.14&    &      &     &&           &    & 122\\
HD 63578&07$^{ }$ 47$^{ }$ 31$^{ }$.5 &-46$^{ }$ 36$^{ }$ 30$^{ }$.5&  &2011.1.8 & 5.23&-0.14 & 0.11&&481$\pm$45 & 480&149 \\
       &           &           &  &2011.2.19&     &      &     &&           &    & 80\\
       &           &           &  &2012.1.16&     &      &     &&           &    &116\\
       &           &           &  &2012.11.14&    &      &     &&           &    &175\\
HD 65814&07$^{ }$ 58$^{ }$ 50$^{ }$.4 &-40$^{ }$ 20$^{ }$ 34$^{ }$.6&  &2011.3.16& 8.77& 0.28 & 0.52&& &2590:&34\\
HD 68217&08 09 35.9 &-44 07 22.0&  &2007.1.3 & 5.21&-0.19 & 0.05&&383$\pm$26 &340 & \\
       &           &           &  &2011.1.7 &     &      &     &&           &    &132\\
       &           &           &  &2011.2.19&     &      &     &&           &    &107\\
       &           &           &  &2012.1.16&     &      &     &&            &    &155\\
       &           &           &  &2012.11.14&    &      &     &&            &    &163\\
HD 68243&08 09 29.3 &-47 20 43.0&  &2011.4.2 & 4.27&-0.24 & 0.02&&            &240 &153 \\
       &           &           &  &2012.2.14&     &      &     &            &    &200\\
HD 68324&08 09 43.2 &-47 56 13.9&  &2011.4.2 & 5.23&-0.21 & 0.05&&339$\pm$21  &520 &98 \\
       &           &           &  &2012.2.14&     &      &     &&            &    &116 \\
HD 69144&08 13 36.2 &-46 59 29.9&  &2012.4.2 & 5.13&-0.16 &0.07 &&417$\pm$32  &300 &112 \\
HD 69302&08 14 23.9 &-45 50 04.3&  &2011.3.16& 5.84&-0.19 &0.05 &&362$\pm$29  &450 &70 \\
HD 70309&08 19 05.6 &-48 11 52.3&  &2011.3.16& 6.45&-0.14 &0.06 &&288$\pm$24  &390 &52 \\
HD 70930&08 22 31.7 &-49 29 25.4&  &2011.2.19& 4.82&-0.15 &0.10 &&526$\pm$76  &550 &103\\
HD 71302&08 24 57.2 &-42 46 11.4&  &2011.3.4 & 5.95&-0.15 &0.05 &&549$\pm$114 &320 &120\\
HD 71459&08 25 51.9 &-42 09 11.1&  &2011.3.4 & 5.47&-0.14 &0.06 &&220$\pm$9   &250 &113\\
       &           &           &  &2012.2.24&     &      &     &&            &    &119\\
HD 71609&08 26 38.2 &-43 24 30.9&  &2011.3.17& 7.87& 0.10 &0.28 &&1149$\pm$440&2090&77 \\
HD 72014$^{c}$&08 28 52.1 &-42 35 14.9&  &2011.3.16& 6.25&-0.07 &0.23 &&1176$\pm$307&670 &65\\
HD 72067&08 29 07.6 &-44 09 37.5&  &2011.12.24&5.83&-0.16 &0.08 &&439$\pm$78  &360 &103\\
HD 72088&08 29 12.6 &-44 53 05.6&  &2011.3.18& 9.07& 0.03 &0.23 &&            &1630&28 \\
HD 72089&08 29 07.0 &-45 33 26.9&  &2011.3.20& 8.02&-0.09 &0.06 &&            &1600&46 \\
HD 72108&08 29 04.8 &-47 55 44.1&  &2011.2.19& 5.33&-0.15 &0.10 &&653$\pm$118 &380 &67\\
       &           &           &  &2012.4.16&     &      &     &&            &    &129 \\
HD 72127A&08 29 27.5&-44 43 29.4&  &2004.2.14& 5.20&-0.18 &0.10 &&            &480 & \\
        &          &           &  &2007.1.7 &     &      &     &&            &    & \\
HD 72179&08 29 37.7 &-44 05 58.0&  &2011.3.20 & 8.14&-0.12 &0.04 &&            &640&48 \\
HD 72232&08 29 45.6 &-46 19 54.1&  &2011.3.4 & 5.99&-0.15 &0.00 &&177$\pm$7   &290 &93\\
HD 72350&08 30 39.2 &-44 44 14.4&  &2011.3.17& 6.30&-0.02 &0.16 &&606$\pm$132 &390 &116\\
HD 72485&08 31 10.6 &-47 51 59.8&  &2011.3.17& 6.38&-0.14 &0.09 &&365$\pm$34  &350 &101\\
HD 72535&08 31 47.6 &-42 01 59.7&  &2011.3.20& 8.20& 0.07 &0.31 &&            &770 &52 \\
HD 72555&08 31 39.6 &-47 14 27.7&  &2011.3.17& 6.76&-0.14 &0.09 &&439$\pm$60  &470 &96 \\
HD 72648&08 32 19.0 &-43 55 53.4&  &2011.3.18& 7.61& 0.12 &0.34 &&909$\pm$268 &1340&62 \\
HD 72800&08 32 53.8 &-47 36 19.8&  &2011.3.17& 6.64& 0.12 &0.12 &&2222$\pm$956&2260&75 \\
HD 73326&08 36 02.2 &-46 30 05.9&  &2011.3.18& 7.27&-0.03 &0.21 &&            &800 &51 \\
       &           &           &  &2011.3.23&     &      &     &&            &    &80 \\
HD 73478&08 36 42.7 &-47 59 54.2&  &2011.3.19& 7.37&-0.10 &0.10 &&581$\pm$133 &820:&103 \\
HD 73658&08 37 40.0 &-46 16 57.8&  &2011.3.19& 6.86& 0.04 &0.28 &&1515$\pm$440&1580&120 \\
HD 74194&08 40 47.8 &-45 03 30.2&  &2008.4.21& 7.57& 0.21 &0.50 &&            &2800&16 \\
HD 74234&08 40 53.4 &-48 13 31.8&  &2011.3.19& 6.95&-0.17 &0.07 &&719$\pm$141 &610 &98 \\
HD 74251&08 41 01.6 &-48 04 02.9&  &2011.3.19& 7.76&-0.11 &0.09 &&1111$\pm$397&840:&43 \\
HD 74273&08 41 05.3 &-48 55 21.6&  &2011.3.31& 5.92&-0.20 &0.05 &&606$\pm$68  &490 & \\
       &           &           &  &2012.2.24&     &      &     &&            &    &95 \\
HD 74319&08 41 34.9 &-44 59 30.9&  &2011.3.17& 6.69&-0.10 &0.06 &&538$\pm$87  &320 &38 \\
HD 74371&08 41 56.9 &-45 24 38.6&  &2011.3.16& 5.24& 0.21 &0.28 &&2325$\pm$787&1920&45\\
       &           &           &  &2012.3.16&     &      &     &&            &    &129 \\
HD 74436&08 42 07.6 &-48 14 40.8&  &2011.3.31& 8.24&-0.08 &0.12 &&662$\pm$293 &820:&51 \\
HD 74455$^{d}$&08 42 16.2 &-48 05 56.7&  &2012.3.30& 5.48&-0.17 &0.08 &&1389$\pm$437&390 &102\\
HD 74531&08 42 34.8 &-48 09 48.9&  &2011.3.31& 7.25&-0.16 &0.08 &&1351$\pm$524&1260&44 \\
HD 74650&08 43 16.6 &-47 48 23.8&  &2011.3.24& 7.35&-0.05 &0.01 &&            &220:&75 \\
\hline
\end{tabular}
\end{footnotesize}
\label{default}
\end{minipage}
\end{table*}

\begin{table*}
\flushleft{\Large Table 1 (continued)  }
\centering
\begin{minipage}{160mm}
\begin{footnotesize}
\begin{tabular}{lcrcrrrrrccrr}
\hline
 Star   &\multicolumn{2}{c}{$\alpha$,$\delta$(2000)}&& &  \multicolumn{3}{c}{}&&\multicolumn{2 }{c}{ Distance} \\
\cline{2-3} \cline{6-8} \cline{10-11} \\
  &    &      & & epoch& V   & (B-V) &E(B-V)& &d(Hip)$^{a}$ &d(Sp) & (S/N)$^{b}$  \\
     &        &     & &  & &     & & &pc  &pc  & \\
\hline

HD 74711&08$^{h}$43$^{m}$47$^{s}$.5 &-46$^{o}$47$^{'}$56$^{"}$.4&  &2011.3.5 & 7.11& 0.07 &0.33 && &1140&57 \\
HD 74753&08 43 40.3 &-49 49 22.1&  &2012.3.16& 5.16&-0.22 &0.08 &&568$\pm$58  &700 &118 \\
HD 74773&08 44 09.7 &-47 06 58.0&  &2008.4.16& 7.24&-0.12 &0.06 &&1351$\pm$524&480 &31\\
HD 74920&08 45 10.3 &-46 02 19.2&  &2011.3.24& 7.53& 0.03 &0.34 &&            &1500&61 \\
HD 74979&08 45 47.4 &-40 36 56.1&  &2011.3.24& 7.24&-0.05 &0.00 &&588$\pm$129 &1460&77 \\
HD 75009&08 45 47.5 &-44 14 52.8&  &2008.4.17& 6.70&-0.09 &0.02 &&362$\pm$41  &230 &67 \\
HD 75129&08 46 19.4 &-47 32 59.6&  &2011.3.23& 6.87& 0.26 &0.35 &&1282$\pm$412&1980:&32\\
HD 75149&08 46 30.5 &-45 54 45.0&  &2011.3.5 & 5.45& 0.27 &0.40 &&2702$\pm$948&1600&59\\
       &           &           &  &2012.3.16&     &      &     &&            &    &136\\
HD 75241&08 47 05.4 &-45 04 29.1&  &2008.4.17& 6.58& -0.13&0.03 &&356$\pm$37  &550 &69 \\
HD 75309&08 47 28.0 &-46 27 04.0&  &2011.3.30& 7.86& 0.01 &0.25 &&            &2610&71 \\
HD 75387&08 48 08.8 &-42 27 48.4&  &2011.1.7 & 6.42&-0.20 &0.04 &&483$\pm$61  &600 &63 \\
       &           &           &  &2011.4.1 &     &      &     &&            &    &52 \\
HD 75534&08 48 44.8 &-47 45 48.1&  &2011.3.30& 7.82& 0.36 &0.55 &&2083$\pm$1093&2610&65 \\
HD 75608&08 49 21.3 &-43 22 14.4&  &2011.3.30& 7.45&-0.09 &0.06 &&347$\pm$48  &420 &61 \\
HD 75759&08 50 21.0 &-42 05 23.3&  &2011.4.1 & 6.00&-0.11 &0.20 &&769$\pm$128  &860 &144\\
HD 75821&08 50 33.5 &-46 31 45.1&  &2011.3.31& 5.11&-0.23 &0.07 &&1000$\pm$213 &910 &107 \\
       &           &           &  &2012.1.16&     &      &     &&             &    &149 \\
HD 76004&08 51 50.0 &-44 09 03.4&  &2011.4.1 & 6.35&-0.17 &0.03 &&515$\pm$74   &390 &73 \\
HD 76161&08 52 38.6 &-48 21 32.8&  &2011.3.31& 5.90&-0.16 &0.04 &&318$\pm$57   &310 &32 \\
       &           &           &  &2011.12.25&    &      &     &&             &    &127 \\
HD 76534&08 55 08.7 &-43 27 59.9&  &2011.4.1 & 8.02& 0.13 &0.37 &&980$\pm$485  &650 &59 \\
HD 76566&08 55 19.2 &-45 02 30.0&  &2011.4.2 & 6.26&-0.16 &0.04 &&370$\pm$40   &530 &70 \\
HD 76838&08 57 07.6 &-43 15 22.3&  &2011.3.23& 7.31& 0.00 &0.20 &&336$\pm$59   &480 &87 \\
HD 77320&09 00 22.3 &-43 10 26.4&  &2011.3.5 & 6.02&-0.14 &0.09 &&320$\pm$24   &290 &53\\
       &           &           &  &2012.3.15&     &      &     &&             &    &145\\
HD 78005&09 04 05.8 &-47 26 29.2&  &2011.3.4 & 6.44&-0.15 &0.08 &&358$\pm$38   &550 &56 \\
HD 79275&09 11 33.4 &-46 35 02.1&  &2011.3.4 & 5.79&-0.22 &0.02 &&305$\pm$23   &460 &61\\
       &           &           &  &2012.2.24&     &      &     &&             &    &83 \\
\hline
\end{tabular}
\\
$^{a}$The distances are  Hipparcos revised parallaxes (van Leeuwen 2007) from SIMBAD. 
  The spectroscopic distances
   are from Cha \& Sembach (2000). V, B-V, and E(B-V) values are also from
 Cha \& Sembach (2000) except when they are uncertain(or differ too much from Simbad).
 In such cases Simbad values are
 used.  \\
 $^{b}$Signal to noise in the continuum near Na\,{\sc i} D lines of the telluric corrected spectrum . \\
$^{c}$ Vmag of Simbad is 6.577, B-V = -0.17 \\
$^{d}$Hipparcos and Spectroscopic distances differ too much. Unrevised Hipparcos parallax
 e.g. , in Cha \& Sembach (2000) gives a distance of 520$\pm$150 pc. \\
\end{footnotesize}
\label{default}
\end{minipage}
\end{table*}

  Our observations  were obtained with
 the fiber-fed
 cross-dispersed echelle spectrograph of the Vainu Bappu
2.3m reflector at the Vainu Bappu Observatory, Kavalur
(Rao  et al. 2005).
 The spectral resolving power, $R = \lambda/d\lambda $,
 employed was 72000. The spectrum covers 4000 to 10000\AA\ with gaps beyond about
5600\AA\ where the echelle orders were incompletely captured on the E2V
2048 x 4096 CCD.   The Na D lines are recorded but not  the Ca\,{\sc ii} H \& K lines.
 The wavelength calibration was done using 
  a Th-Ar  hollow cathode lamp soon after the stellar
 exposures.
   Our spectral resolving power,  as determined from the width of weak atmospheric (H$_2$O) lines,  is very close to that  ($R \simeq 75000$)
   used by
 Cha \& Sembach (2000) for their spectra and, thus,   matching of
  our line profiles  with theirs becomes easy and appropriate.
    Usually, two exposures of 30 to 45 minutes have been  combined for each night.
  A nearby
  hot star was observed to remove the telluric lines.
  Most of our observations were obtained between   early 2011 and late-2012 (see Table 1).
   Limited observations were also were obtained during
  2007 March-April and in 2008.

    We used IRAF routines for spectral reductions (flat field corrections, 
  wavelength calibration and telluric line corrections). 
  All heliocentric velocities are converted to the local standard of rest
  (LSR) adopted by   Cha \& Sembach (2000).  The Na D profiles are 
  fitted with gaussian components to determine the central radial velocity
  $V_{\rm LSR} $ and equivalent width of the components. The components 
  listed by Cha \& Sembach were taken as starting parameters for the fits and
  further adjustments were made such that both $D_{2}$ and $D_{1}$ profiles are 
  satisfied with the same number of components with the same $V_{\rm LSR} $
  and nearly same half-widths. The fits to the observed profiles are made such that
  no residuals are seen over the noise level of the surrounding continuum. 
   The signal to noise (S/N) in the continuum near Na\,{\sc i} D lines is given
  in  Table 1 for each telluric-corrected spectrum. If more than one
 observation is available then the average of equivalent width and  the  standard
  deviation for the components is given in tables.  


\begin{figure*}
\vspace{0.3cm}
\rotatebox{90}{\hspace{1.2cm}Normalised Intensity}
\includegraphics[width=5.7cm,height=5cm]{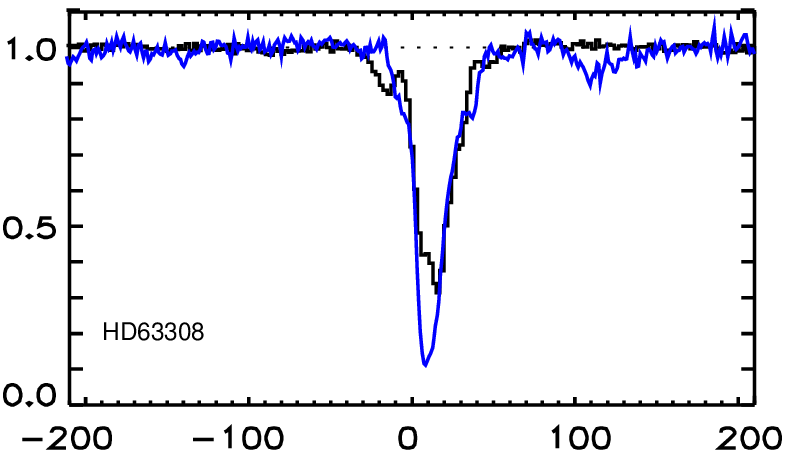}
\includegraphics[width=5.7cm,height=5cm]{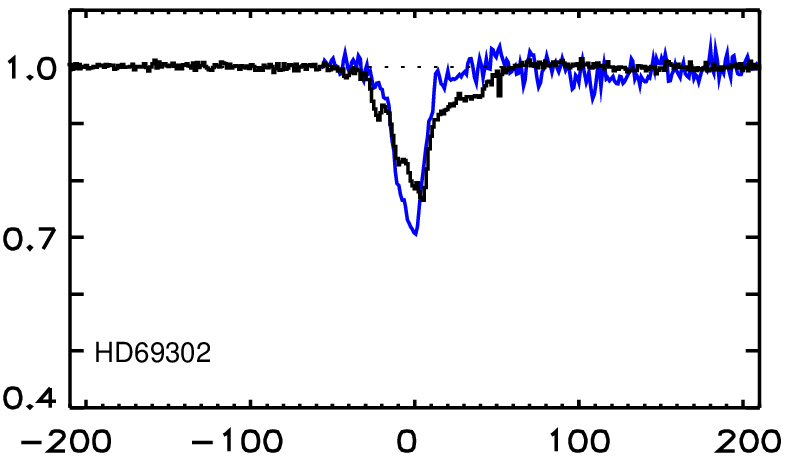}
\includegraphics[width=5.7cm,height=5cm]{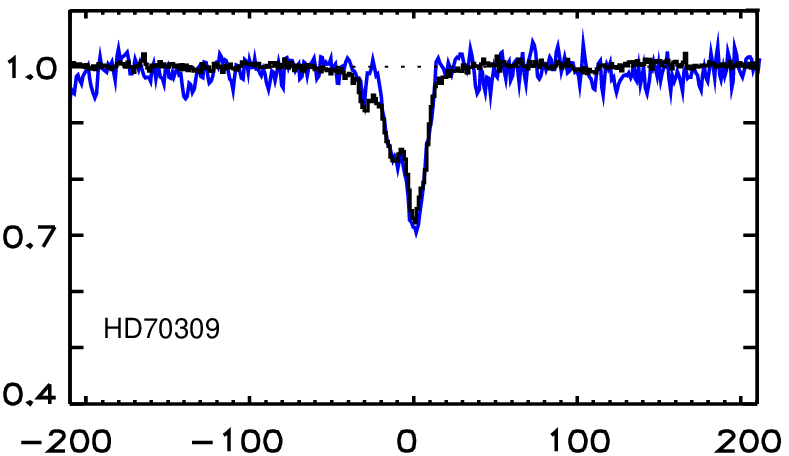}
\hspace{.5cm}
\vspace{.3cm}
\rotatebox{90}{\hspace{1.2cm}Normalised Intensity}
\includegraphics[width=5.7cm,height=5cm]{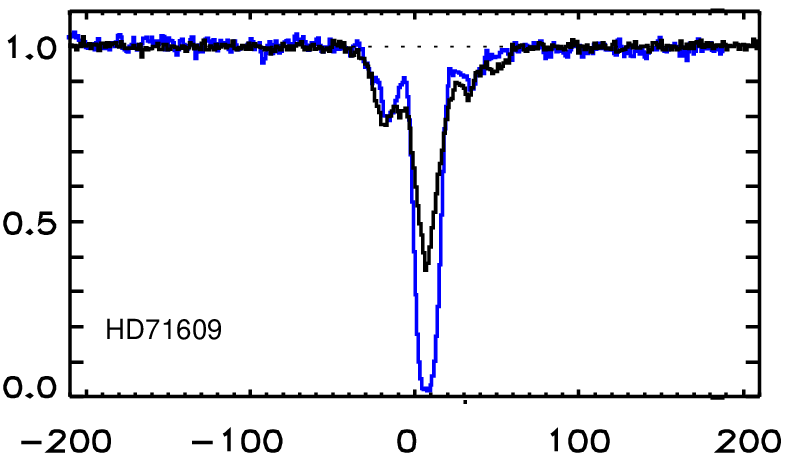}
\includegraphics[width=5.7cm,height=5cm]{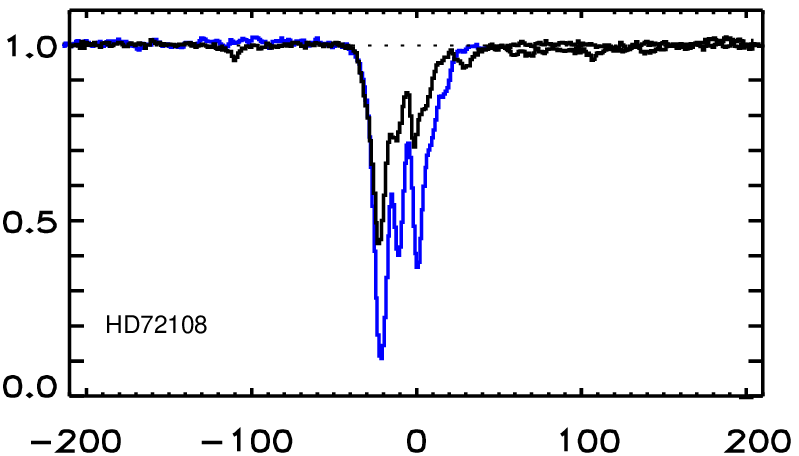}
\includegraphics[width=5.7cm,height=5cm]{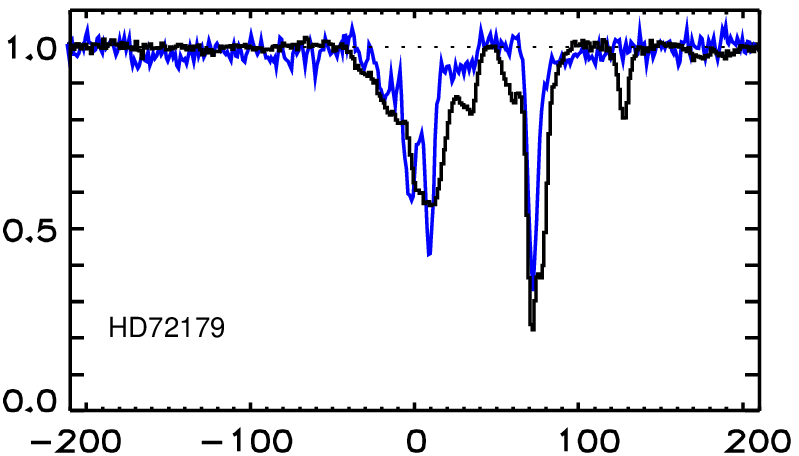}
\hspace{.5cm}
\vspace{0.1cm}
\rotatebox{90}{\hspace{1.2cm}Normalised Intensity}
\includegraphics[width=5.7cm,height=5cm]{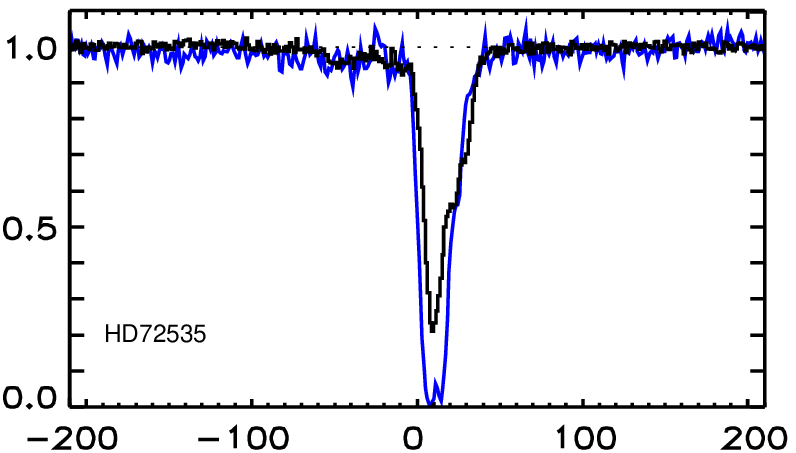}
\includegraphics[width=5.7cm,height=5cm]{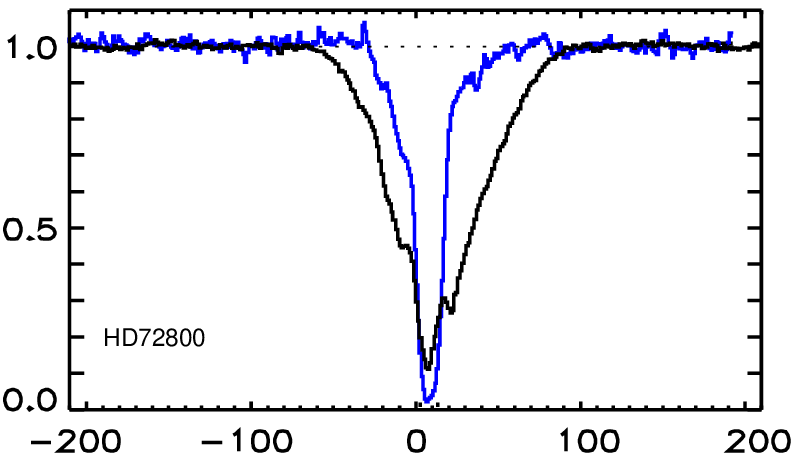}
\includegraphics[width=5.7cm,height=5cm]{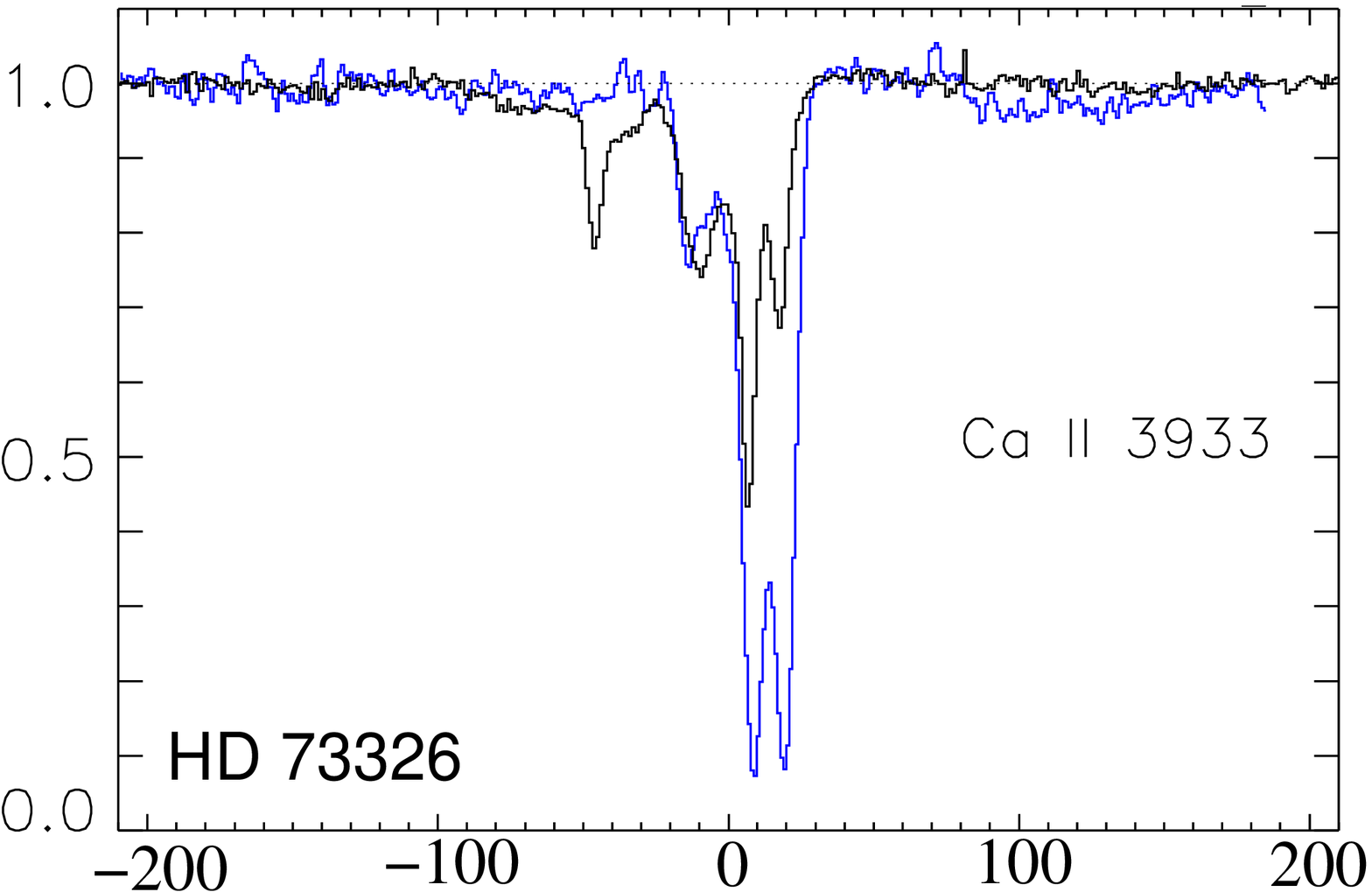}
\hspace{.5cm}
\vspace{-.1cm}
\rotatebox{90}{\hspace{1.2cm}Normalised Intensity}
\includegraphics[width=5.7cm,height=5cm]{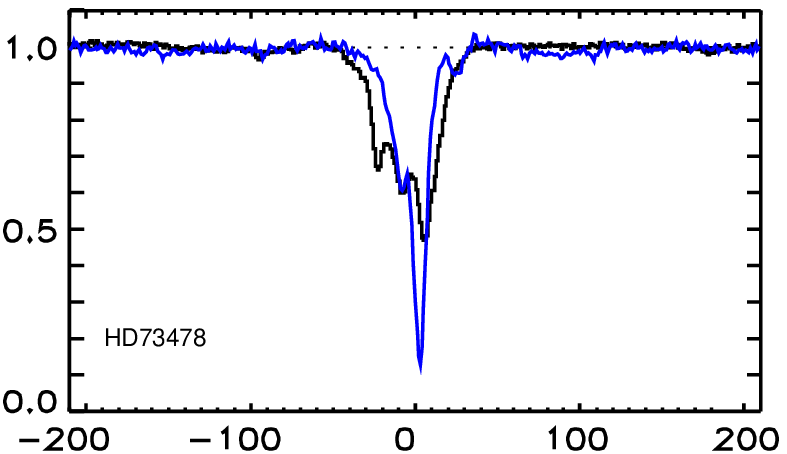}
\includegraphics[width=5.7cm,height=5cm]{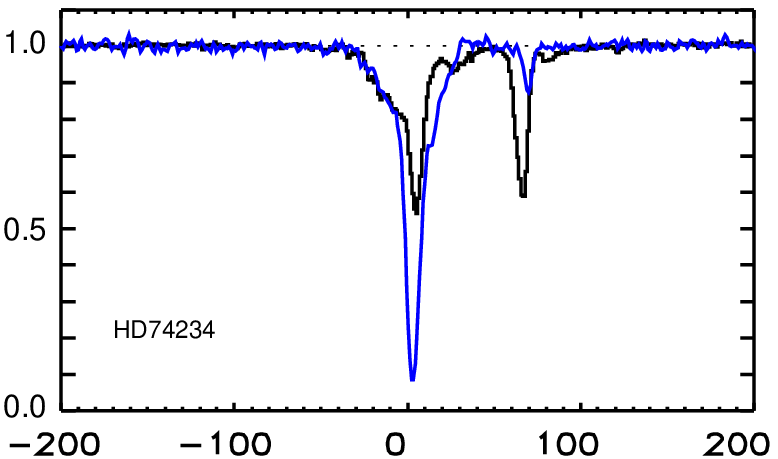}
\includegraphics[width=5.7cm,height=5cm]{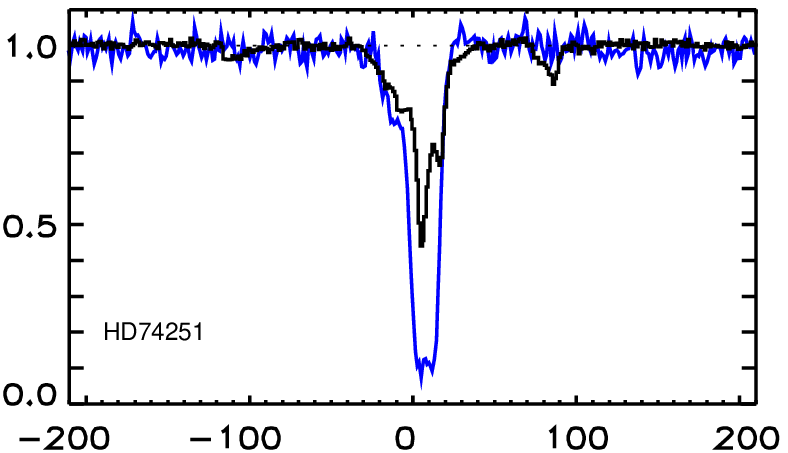}
\hspace{14cm}$V_{\rm LSR}$(km s$^{-1}$) \\
\caption{ Na\,{\sc i} D$_2$ profile of HD 63308, HD 69302, HD 70309, HD 71609,
HD72108, HD 72179, HD 72535, HD 72800, HD 73326, HD73478, HD 74234 and HD 74251  obtained by us in
 2011-12 (blue line)  is superposed on the Ca\,{\sc ii} K profile obtained in 1993-94
(black line) by Cha \& Sembach (2000).}

\end{figure*}

\begin{figure*}
\vspace{0.3cm}
\rotatebox{90}{\hspace{1.2cm}Normalised Intensity}
\includegraphics[width=5.7cm,height=5cm]{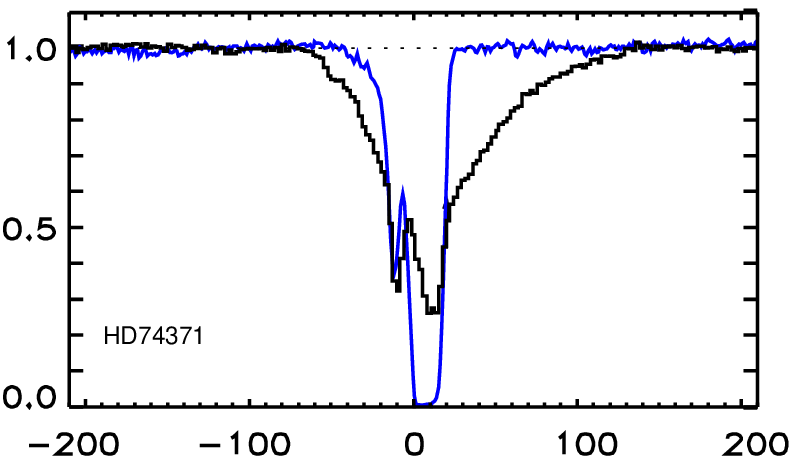}
\includegraphics[width=5.7cm,height=5cm]{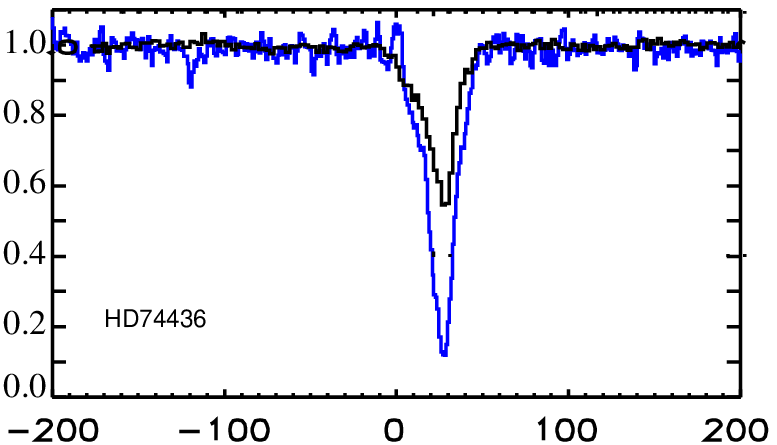}
\includegraphics[width=5.7cm,height=5cm]{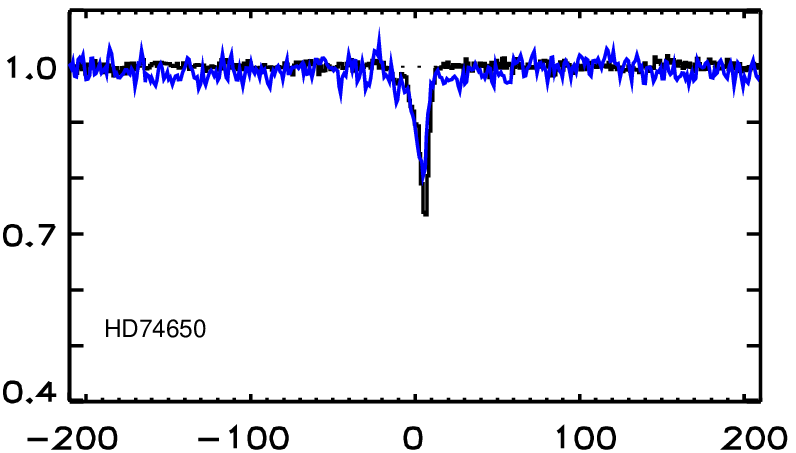}
\hspace{.5cm}
\vspace{.3cm}
\rotatebox{90}{\hspace{1.2cm}Normalised Intensity}
\includegraphics[width=5.7cm,height=5cm]{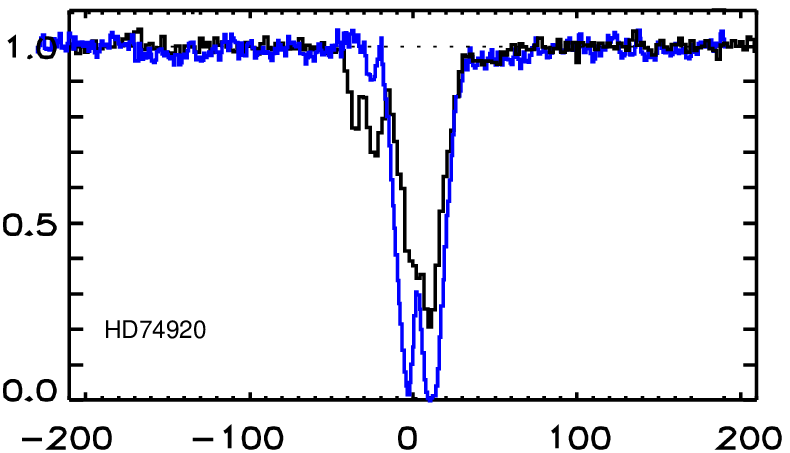}
\includegraphics[width=5.7cm,height=5cm]{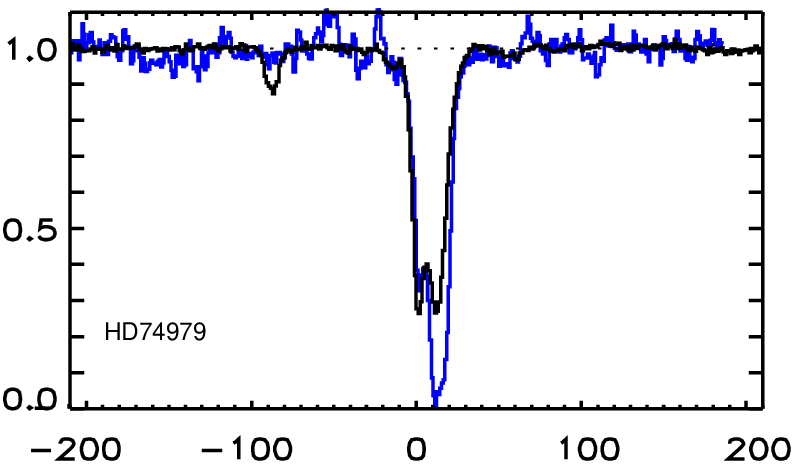}
\includegraphics[width=5.7cm,height=5cm]{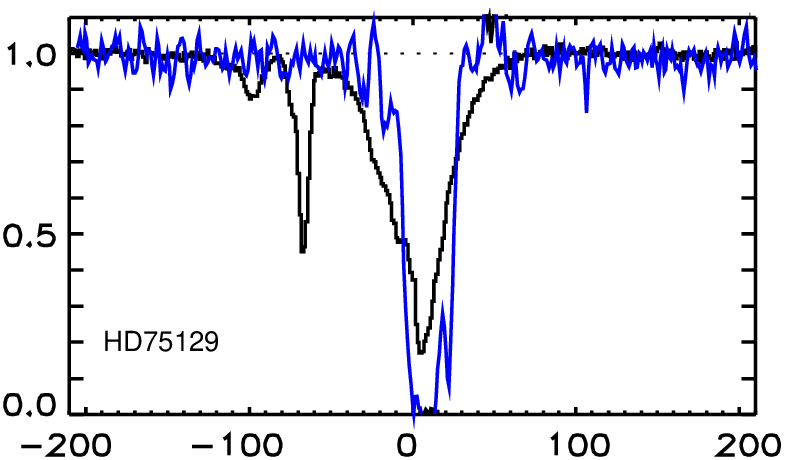}
\hspace{.5cm}
\vspace{.1cm}
\rotatebox{90}{\hspace{1.2cm}Normalised Intensity}
\includegraphics[width=5.7cm,height=5cm]{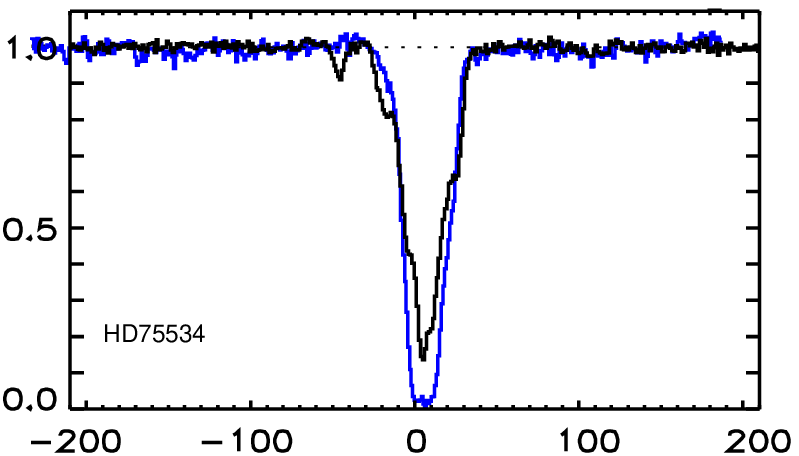}
\includegraphics[width=5.7cm,height=5cm]{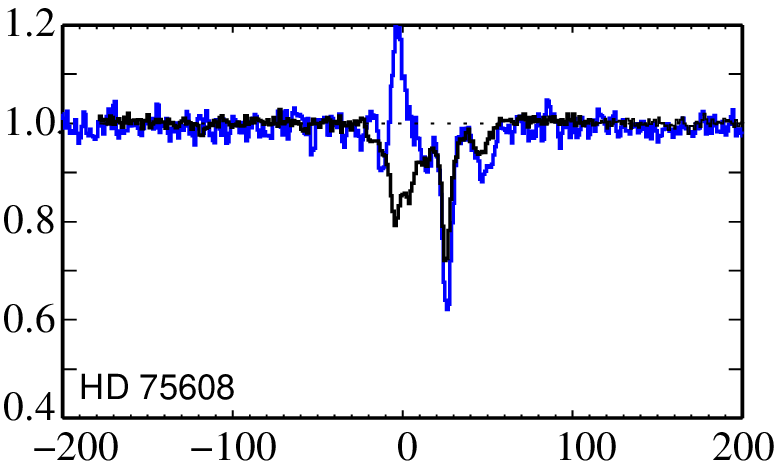}
\includegraphics[width=5.7cm,height=5cm]{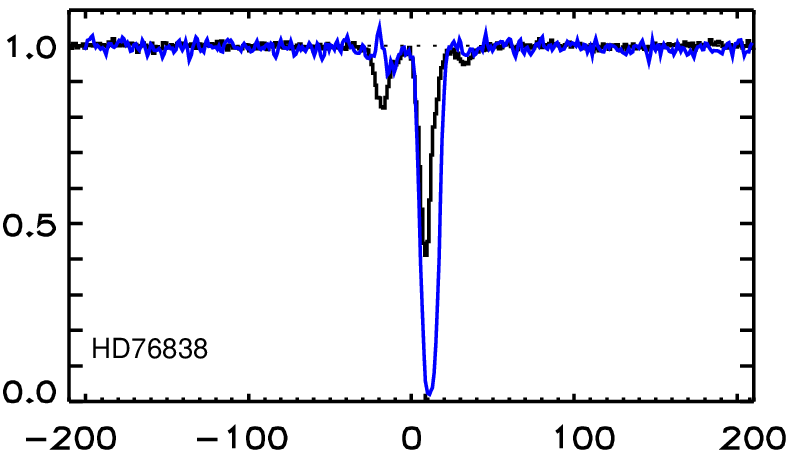}
\hspace{.5cm}
\vspace{-.1cm}
\rotatebox{90}{\hspace{1.2cm}Normalised Intensity}
\includegraphics[width=5.7cm,height=5cm]{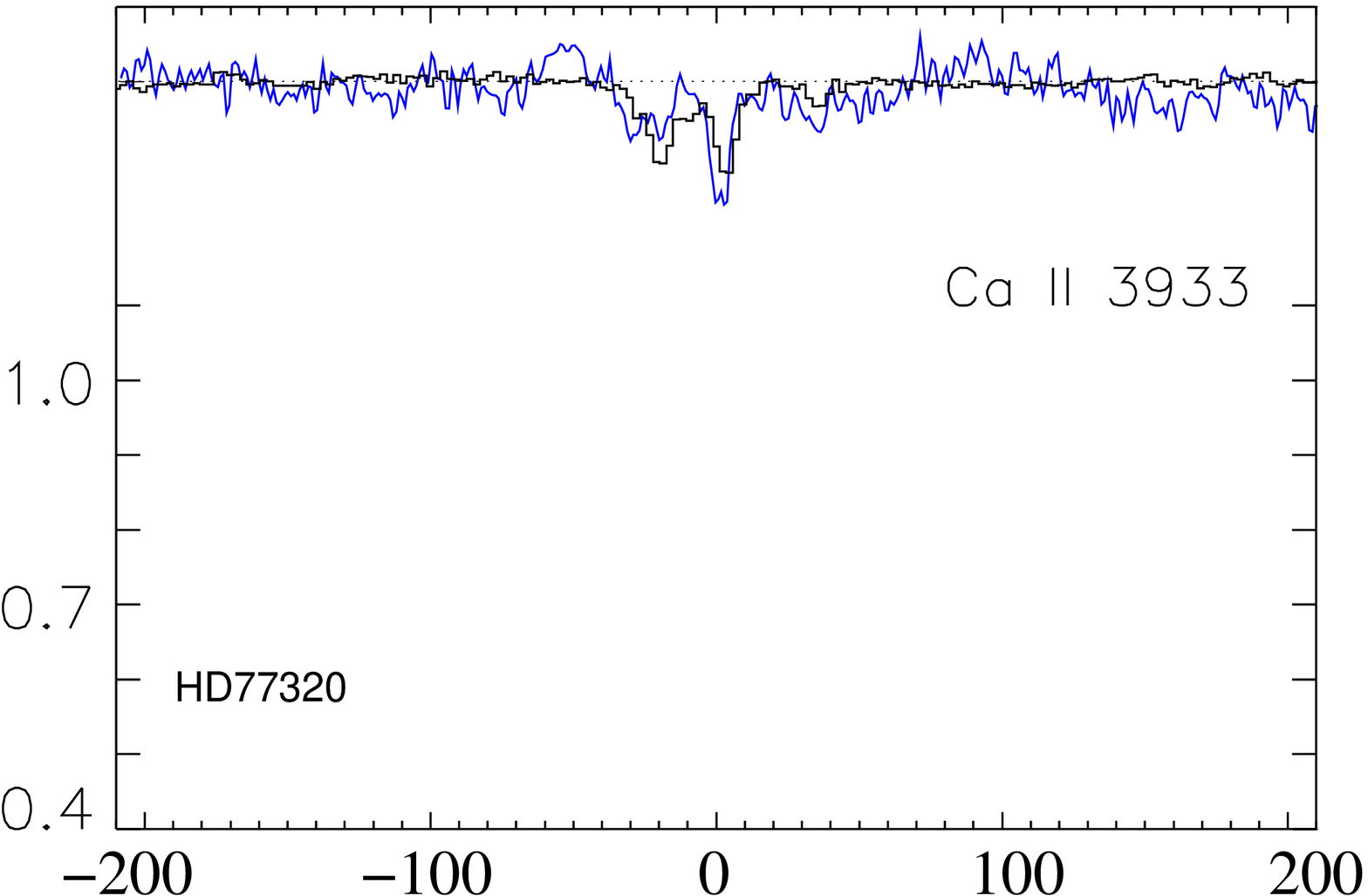}
\hspace{14cm}$V_{\rm LSR}$(km s$^{-1}$) \\
\caption{Na\,{\sc i} D$_2$ profile of  HD 74371, 
 HD 74436, HD 74650, HD 74920 , HD 74979, HD 75129, HD 75534, HD 75608, HD 76838 and
 HD 77320  obtained by us in
 2011-12 (blue line)  is superposed on the Ca\,{\sc ii} K profile obtained in 1993-94
 (black line) by Cha \& Sembach (2000). The prominent emission in the Na D profile of HD 75608 is due to
 terrestrial Na D airglow: weaker emission is seen in some other Na D profiles in this and Figure 2.}
\end{figure*}



\section{Overview of The Na D profiles}

The VBT Na D profiles were obtained primarily to
 define their evolution, if any, since similar quality Na D profiles were obtained in 1993-1996 by Cha \& Sembach (2000).
Cha \& Sembach searched for  evolutionary changes  in both the Ca\,{\sc ii} K and Na D profiles over their three year baseline but only among the 13 stars observed
twice.  Of the 13, seven showed variable profiles with Ca\,{\sc ii} K line variations being generally more prominent than those affecting the Na D lines. Changes were seen both
among the low velocity complex of lines (two cases) and at high velocity (six cases).   Our  Na D observations cover 43 stars out of the 44 for which Cha \& Sembach
provide Na D profiles. (Our Na D spectra include 22 stars for which Cha \& Sembach obtained
Ca K line profiles but not
 Na D profiles.)   Remarkable weakening of low-velocity absorption in Na D between 1993-1996
and 2011-2012 is seen in three cases --  HD 63578, HD 68217 and HD 76161--
and strengthening in the case of HD 68243.   Cha \& Sembach reported changes in
 some high velocity components and our
spectra provide limited information on such components. In a few cases,  cloud
 accelerations are  evident from  the present observations (e.g., HD 75821, HD 74234). For the majority of the stars observed at the VBT, the Na D profile is not distinguishable from the counterpart observed between 1993-1996
by Cha \& Sembach.   

\begin{table*}
\centering
\begin{minipage}{160mm}
\caption{\Large Na\,{\sc i}(now) and Ca\,{\sc ii}(then) Absorption Lines  }
\begin{footnotesize}
\begin{tabular}{lcrrrcrrccrrrccr}
\hline
   &\multicolumn{3}{c}{Cha \& Sembach }& &\multicolumn{4}{c}{Franco/Other}&&
\multicolumn{4}{c}{Present Observations$^a$}&  \\
\cline{2-4} \cline{6-9} \cline{11-14} \\
 Star   &\multicolumn{3}{c}{ Ca\,{\sc ii} K}&&\multicolumn{4}{c}{Na\,{\sc i} D}&&\multicolumn{4}{c}{Na\,{\sc i} D} \\
\cline{2-4} \cline{6-9} \cline{11-14} \\
  &   epoch&$V_{\rm LSR}$& Eq.w & & epoch&$V_{\rm LSR}$& Eq.w &Eq.w& & epoch&$V_{\rm LSR}$& Eq.w &Eq.w&   \\
     &  & km s$^{-1}$ &(mA) & &  & km s$^{-1}$ &(mA) &(mA) & & & km s$^{-1}$ &(mA) &(mA) &  \\
\hline
HD 63308&  1993&-15& 29& &     &    &  &  & &2011-12& -9.7&43&19& \\
       &      &   &   & &     &    &  &  & &     &  0.6   &80&66&  \\
       &      & 5 & 71& &     &    &  &  & &     &6.3  &173&141& \\
       &      & 16&123& &     &    &  &  & &     &16.6 &83 &40 & \\
       &      & 28& 47& &     &    &  &  & &     &29.4 &43 &20 & \\
       &      & 45&  5& &     &    &  &  & &     &45.5 &11 & 5 & \\
HD 69302&  1996&   &   & & 1989&-30.2&9&0.6&&2011 &     &   &   & \\
       &      &-23&  5& &     &-22.3&10&4& &     &-21.1& 5 & 1 & \\
       &      &   &   & &     &-16.8&9 &4& &     &     &   &   & \\
       &      & -9& 14& &     &-8.5&43&18& &     &-8.9 &28 & 14& \\
       &      & -1&  8& &     &-1.2&29&18& &     &     &   &   & \\
       &      &  5& 15& &     & 3.3&52&25& &     & 0.8 &81 & 46& \\
       &      &  6& 60& &     &    &  &  & &     &     &   &   & \\
       &      &   &   & &     &    &  &  & &     & 22.5&8  & 2 & \\
       &      & 40& 6 & &     &    &  &  & &     &     &   &   & \\
HD 70309&  1996&-28& 14& &     &    &  &  & &     &-29.5& 6 & 3 & \\
       &      &-13& 21& &     &    &  &  & &     &-12.0&34 & 17& \\
       &      &  1& 53& &     &    &  &  & &     &  2.0&73 & 40& \\
HD 71609&  1996&-20& 35& &     &    &  &  & &2011 &-25.8&12 & 4 & \\
       &      &- 8& 31& &     &    &  &  & &     &-14.7&45 & 23& \\
       &      &   &   & &     &    &  &  & &     &  3.3&160&145& \\
       &      &  6&107& &     &    &  &  & &     & 11.0&160&135& \\
       &      & 17& 34& &     &    &  &  & &     & 23.8&  8&  4& \\
       &      & 32& 22& &     &    &  &  & &     & 33.7& 24& 6 & \\
       &      & 48& 15& &     &    &  &  & &     & 45.2&  6& 2 & \\
HD 72108&  1996&-111& 4& & 1989&    &  &  & &2012 &-111 &$\leq$1.7&$\leq$1& \\
       &      &-31 &13& &     &-28.4&24&10&&     &-28.3&28 &8.5& \\
       &      &-23 &67& &     &-21.1&162&134&&   &-21.2&146&132& \\
       &      &-12&35 & &     &-11.0&99 &69 &&   &-10.7&87 &60 & \\
       &      &-1 &20 & &     &-0.2&97 &66& &    &-0.1 &91 &71 & \\
       &      & 5 &22 & &     &7.7 &48 &30& &    &7.3  &39 &21 & \\
       &      &   &   & &     &15.9&12 &7 & &    &15.6 &22 &13 & \\
       &      &29 &11 & &     &    &   &  & &    &     &   &   & \\
HD 72179&  1996&-33&6  & &     &    &   &  & &2011&-25.3&6  & 3 & \\
       &      &-12&66 & &     &    &   &  & &    &-12.1&31 &15 & \\
       &      & 1 &11 & &     &    &   &  & &    & 1.3 &72 &33 & \\
       &      &11 &117& &     &    &   &  & &    & 12.0&88 &55 & \\
       &      &29 &10 & &     &    &   &  & &    & 25.6&12 & 6 & \\
       &      &35 &17 & &     &    &   &  & &    & 34.9&12 & 5 & \\
       &      &61 &26 & &     &    &   &  & &    & 57.7& 2 & 1 & \\
       &      &71 &65 & &     &    &   &  & &    & 71.6&87 &54 & \\
       &      &78 &50 & &     &    &   &  & &    & 78.5&14 & 7 & \\
       &      &84 &15 & &     &    &   &  & &    & 88.9&10 & 5 & \\
       &      &127&19 & &     &    &   &  & &    &     &   &   & \\
HD 72535&  1996& -9& 8 & &     &    &   &  & &2011&-12.7&12 & 6 & \\
       &      & 10&142& &     &    &   &  & &    &  7.3&181&164& \\
       &      &   &   & &     &    &   &  & &    & 16.0&172&157& \\
       &      & 21& 44& &     &    &   &  & &    & 25.3& 70& 35& \\
       &      & 29& 42& &     &    &   &  & &    & 35.5& 10&  8& \\
HD 72800&  1996&-35& 26& &     &    &   &  & &2011&     &   &   & \\
       &      &   &   & &     &    &   &  & &    &-20.9& 18& 9 & \\
       &      &-10&106& &     &    &   &  & &    &-10.0& 60&30 & \\
       &      &   &   & &     &    &   &  & &    &  3.0&180&152& \\
       &      &  7&124& &     &    &   &  & &    &     &   &   & \\
       &      &   &   & &     &    &   &  & &    & 11.0&168&147& \\
       &      & 21& 90& &     &    &   &  & &    & 26.9& 32& 16& \\
       &      &   &  & &     &    &   &  & &    & 43.7& 10&  5& \\
\hline
\end{tabular}
\\
\end{footnotesize}
\label{default}
\end{minipage}
\end{table*}

\begin{table*}
\flushleft{\Large Table 2 (continued)}
\centering
\begin{minipage}{160mm}
\begin{footnotesize}
\begin{tabular}{lcrrrcrrccrrrccr}
\hline
   &\multicolumn{3}{c}{Cha \& Sembach }& &\multicolumn{4}{c}{Franco/Other}&&
\multicolumn{4}{c}{Present Observations$^*$}&  \\
\cline{2-4} \cline{6-9} \cline{11-14} \\
 Star   &\multicolumn{3}{c}{ Ca\,{\sc ii} K}&&\multicolumn{4}{c}{Na\,{\sc i} D}&&\multicolumn{4}{c}{Na\,{\sc i} D} \\
\cline{2-4} \cline{6-9} \cline{11-14} \\
  &   epoch&$V_{\rm LSR}$& Eq.w & & epoch&$V_{\rm LSR}$& Eq.w &Eq.w& & epoch&$V_{\rm LSR}$ & Eq.w &Eq.w&   \\
     &  & km s$^{-1}$ &(mA) & &  & km s$^{-1}$ &(mA) &(mA) & & & km s$^{-1}$ &(mA) &(mA) & \\
\hline

HD 73326&  1996&-46& 14& &     &    &   &  & &2011& 45.5& 5& 2 & \\
       &      &-37& 18& &     &    &   &  & &    &     &   &   & \\
       &      &-10& 50& &     &    &   &  & &    &-11.3& 41& 20& \\
       &      &  0&  7& &     &    &   &  & &    & -0.7& 36& 18& \\
       &      &  7& 48& &     &    &   &  & &    & 10.1&164&143& \\
       &      & 17& 36& &     &    &   &  & &    & 20.3&156&133& \\
HD 73478&  1996&-36& 10& &     &    &   &  & &2011&     &   &   & \\
       &      &-22& 44& &     &    &   &  & &    &-19.1& 13& 5 & \\
       &      &-8 & 73& &     &    &   &  & &    &-10.0& 24& 11& \\
       &      &   &   & &     &    &   &  & &    &-2.8 & 57& 30& \\
       &      & 5 & 70& &     &    &   &  & &    & 8.2 &170&122& \\
       &      & 13& 51& &     &    &   &  & &    &     &   &   & \\
       &      &   &   & &     &    &   &  & &    &28.5 & 15& 8 & \\
HD 74234&  1996&-17&27 & &     &    &   &  & &2011&-19.3&14 & 7 & \\
       &      &-6 &18 & &     &    &   &  & &    &-7.6 &36 &18 & \\
       &      & 5 &64 & &     &    &   &  & &    & 5.2 &180&147& \\
       &      &   &   & &     &    &   &  & &    &15.8 &49 &26 & \\
       &      &28 &15 & &     &    &   &  & &    &26.0 &13 & 7 & \\
       &      &69 &47 & &     &    &   &  & &    &71.7 &14 & 7 & \\
       &      &85 &6  & &     &    &   &  & &    &     &   &   & \\
HD 74251&  1996&-108&8 & &     &    &   &  & &2011&     &   &   & \\
       &      &-5 &41 & &     &    &   &  & &    &     &   &   & \\
       &      & 6 &56 & &     &    &   &  & &    & 0.2 &55 &29 & \\
       &      & 16&40 & &     &    &   &  & &    &13.6 &190&122& \\
       &      &   &  &  &     &    &   &  & &    &22.5 &180&132& \\
       &      & 78&5 &  &     &    &   &  & &    &     &   &   & \\
       &      & 86&9 &  &     &    &   &  & &    &     &   &   & \\
HD 74371&  1994&   &  &  &     &    &   &  & &2011-12&-30.5&8&4 & \\
       &      &   &  &  &     &    &   &  & &    &-19.6& 18& 9 & \\
       &      &-10&42&  &     &    &   &  & &    &- 9.7& 98&58 & \\
       &      &  2&286*&&     &    &   &  & &    & 3.8 &$>$250&207&\\
       &      & 11&96&  &     &    &   &  & &    &14.6 &$>$189&190&\\
       &      & 29&301*&&     &    &   &  & &    &     &   &   & \\
HD 74436&  1994&-7 &57 & &     &    &   &  & &2011&-9.0 & 60& 27& \\
       &      & 7 &80 & &     &    &   &  & &    & 6.9 &260&185& \\
       &      &   &   & &     &    &   &  & &    &20.1 & 35& 18& \\
HD 74650&  1996& -1& 8 & &     &    &   &  & &2011&-1.7 & 8 & 4 & \\
       &      &  6& 21& &     &    &   &  & &    & 6.1 & 28& 16& \\
HD 74920&  1993&-36& 23& &     &    &   &  & &2011&     &   &   & \\
       &      &-23& 50& &     &    &   &  & &    &-25.0& 11& 6 & \\
       &      & -9& 34& &     &    &   &  & &    &-11.2& 79& 45& \\
       &      & -3& 42& &     &    &   &  & &    & -3.7&165&150& \\
       &      &  2& 35& &     &    &   &  & &    &  6.9&148&136& \\
       &      & 10&134& &     &    &   &  & &    & 13.7&148&130& \\
       &      & 21& 40& &     &    &   &  & &    & 22.5& 35& 20& \\
       &      & 45& 14& &     &    &   &  & &    &     &   &   & \\
HD 74979&  1996&-87& 15& &     &    &   &  & &2011&-84.5&$<$8&  &  \\
       &      &-13&  9& &     &    &   &  & &    & -8.1& 25& 13& \\
       &      &  1& 73& &     &    &   &  & &    & 0.9 & 54& 35& \\
       &      & 12&146& &     &    &   &  & &    & 12.5&312&280& \\
HD 75129&  1996&-98& 26& &     &    &   &  & &2011&     &   &   & \\
       &      &-67& 62& &     &    &   &  & &    &-67.8& 10&  7& \\
       &      &   &   & &     &    &   &  & &    &-21.4& 20& 11& \\
       &      &   &   & &     &    &   &  & &    & -8.8& 33& 15& \\
       &      &  2&396& &     &    &   &  & &    &  5.7&146&131& \\
       &      &  7&109& &     &    &   &  & &    & 12.7&138&137& \\
       &      &   &   & &     &    &   &  & &    & 19.7&137&130& \\
       &      &   &   & &     &    &   &  & &    & 27.9&116& 97& \\
       &      &   &   & &     &    &   &  & &    & 41.9&  7&  2& \\
\hline
\end{tabular}
\\
\end{footnotesize}
\label{default}
\end{minipage}
\end{table*}

\begin{table*}
\flushleft{\Large Table 2 (continued)  }
\centering
\begin{minipage}{160mm}
\begin{footnotesize}
\begin{tabular}{lcrrrcrrccrrrccr}
\hline
   &\multicolumn{3}{c}{Cha \& Sembach }& &\multicolumn{4}{c}{Franco/Other}&&
\multicolumn{4}{c}{Present Observations$^*$}&  \\
\cline{2-4} \cline{6-9} \cline{11-14} \\
 Star   &\multicolumn{3}{c}{ Ca\,{\sc ii} K}&&\multicolumn{4}{c}{Na\,{\sc i} D}&&\multicolumn{4}{c}{Na\,{\sc i} D} \\
\cline{2-4} \cline{6-9} \cline{11-14} \\
  &   epoch&$V_{\rm LSR}$& Eq.w & & epoch&$V_{\rm LSR}$& Eq.w &Eq.w& & epoch&$V_{\rm LSR}$ & Eq.w &Eq.w&   \\
     &  & km s$^{-1}$ &(mA) & &  & km s$^{-1}$ &(mA) &(mA) & & & km s$^{-1}$ &(mA) &(mA) & \\
\hline
HD 75534&  1996&-46&  8& &     &    &   &  & &2011&     &   &   & \\
       &      &-17& 32& &     &    &   &  & &    &-14.4& 16&  9& \\
       &      & -5& 62& &     &    &   &  & &    & -0.2&145&120& \\
       &      &  5&128& &     &    &   &  & &    &  5.7&132&118& \\
       &      & 13& 72& &     &    &   &  & &    & 13.6&179&164& \\
       &      & 23& 62& &     &    &   &  & &    & 23.0& 85& 50& \\
HD 75608&  1996&-37&  8& &     &    &   &  & &2011&-33.9& 10&  4& \\
       &      &-27& 21& &     &    &   &  & &    &     &   &   & \\
       &      &-19& 14& &     &    &   &  & &    &     &   &   & \\
       &      & -9& 12& &     &    &   &  & &    & -7.8& 15&  8& \\
       &      &  4& 26& &     &    &   &  & &    &  4.2& 54& 34& \\
       &      & 25& 12& &     &    &   &  & &    & 22.9& 13&  5& \\
       &      &   &   & &     &    &   &  & &    & 29.0& 14&  6& \\
HD 76838&  1996&-18& 24& &     &    &   &  & &2011&-22.7&  6&  3& \\
       &      &   &   & &     &    &   &  & &    & -9.3& 11&  6& \\
       &      &  9& 65& &     &    &   &  & &    & 12.8&232&212& \\
       &      & 17&  7& &     &    &   &  & &    &     &   &   & \\
       &      & 33&  7& &     &    &   &  & &    & 35.7&  4&  2& \\
HD 77320&  1994&-20& 18& &     &    &   &  & &2011&-19.2& 13&  7& \\
       &      & -8&  3& &     &    &   &  & &    & -4.2&(3)&(3)& \\
       &      &  3& 16& &     &    &   &  & &    &  3.9& 25& 13& \\
       &      & 36&  4& &     &    &   &  & &    & 36.7& 10&  5& \\
\hline
\end{tabular}
\\
$^a$ :
       The $V_{\rm LSR}$ is an average of  $D_{\rm 2}$ and
      $ D_{\rm 1}$. \\
\end{footnotesize}
\label{default}
\end{minipage}
\end{table*}

             We have checked our velocities with telluric lines, which
  suggest errors around 0.15 km s$^{-1}$. The agreement of our $V_{\rm LSR}$ of the  gaussian
   components for
  many sight lines   with those of Cha \& Sembach is good with a mean
   difference
  of  -0.09$\pm $ 1.25 km s$^{-1}$  for 31 components.
  Moreover , when we mention change in velocity of the components they are essentially differential with respect to the low velocity components that remain unchanged.

\subsection{Na D now and Ca\,{\sc ii} then}

In this section, we compare the Ca\,{\sc ii} K profile from 1993-1996 observations by Cha \& Sembach (2000) with our Na D$_2$ profile for stars for
which Cha \& Sembach did not have a Na D profile.  Figures 2 and  3 show the profiles.  As in other directions through the Galaxy, the Ca\,{\sc ii} K and
Na D interstellar absorption profiles while similar are never exact replicas. 
In addition,  several profiles show high-velocity components, usually exclusively in the Ca\,{\sc ii} K profile: see, for example, HD 72179 (Figure 2) with its strong high-velocity
components at  about $+80$ km s$^{-1}$ in both Ca\,{\sc ii} and Na D and a higher velocity and weaker Ca\,{\sc ii}  component at  about $+130$ km s$^{-1}$ with no
counterpart (after almost two decades) in Na D. 

  Table 2 lists the gaussian components of both  the Ca\,{\sc ii} K profiles obtained by Cha \& Sembach in 1993-96 period and the  Na D profiles obtained by us
 mainly during 2011-12. Most of the components in both
 lines have the same $V_{\rm LSR} $ suggesting that the same clouds are 
  present even after about 20 years .  Two of the stars in the current  sample were observed by Franco (2012) in 1989 April for their Na D profiles at high resolution with the telescope-spectrometer
combination used by Cha \& Sembach (2000): HD 63902 and HD 72108.  Franco's and our Na D profiles are compared below. Gaussian fits to the profiles and Cha \& Sembach's fit to their Ca\,{\sc ii} profile are summarized in Table 2.

Most of the Na D profiles  are unchanged in equivalent width between 1993-1996 and 2011-2012. If it is assumed that the
 Na D components are unchanged, the ratio of column densities of Ca\,{\sc ii}  to Na\,{\sc i} could be estimated from the ratio of equivalent widths. 
  As emphasized by Wallerstein et al.
 (1980) a  high ratio of column densities of Ca\,{\sc ii} to Na\,{\sc i}  for high-velocity components is
 a sign of shocked gas, presumably resulting from the SNR interacting with interstellar clouds. This conclusion is supported by our larger sample.
Stars in the direction of the Vela SNR offer the opportunity to examine
the number density ratio  $R$(Ca,Na) = $N$(Ca\,{\sc ii})/$N$(Na\,{\sc i}) in
clouds of higher velocity than those in the undisturbed diffuse interstellar
low-velocity 
medium where an increase in $R$(Ca,Na) with radial velocity was long ago
noted by Routly \& Spitzer (1952) over the approximate velocity range of
$\pm$60 km s$^{-1}$. For the highest velocity clouds in the Vela survey
spanning $\pm$190 km s$^{-1}$, 
one might expect differences in physical conditions -- densities and
ionizing radiation -- as well as destruction of grains with appreciable
release of Ca (relative to Na); Ca in quiescent interstellar gas is much more
severely depleted onto/into grains than Na. An analysis of $R$(Ca,Na) in 
high-velocity components along seven Vela SNR sight lines was reported by
Danks \& Sembach (1995) using a subset of the line profiles assembled and
discussed by Cha \& Sembach (2000). Our extension of their analysis
uses the full dataset provided by Cha \& Sembach (2000) with new information
on the Na D profiles for stars observed by them at the K line but not the Na D
line. Additional information on high-velocity gas is modest because 
high-velocity components are rarely observed in the Na D profiles.

Our evaluation of $R$(Ca,Na) is made for about 40 stars for which
equivalent widths are available for both the Ca\,{\sc ii} K and the
Na\,{\sc i} D lines. All of the K line profiles are taken from
Cha \& Sembach (2000). For those K line profiles for which Cha \& Sembach
did not provide a D line profile that profile is taken from the
VBT observations with the assumption that the profile has not changed in
the 15-18 year interval. We limit the study to components with LSR peculiar
velocities greater than $\pm$20 km s$^{-1}$ to avoid the strong (saturated)
foreground clouds, a limit close to that suggested by Cha et al. (1999) as
a representative (minimum) velocity for clouds shocked by the SNR. For the
stronger Na D lines (equivalent width greater than 100 m\AA) we used the doublet
method to derive column densities. Almost all the K line equivalent
widths are less than 100 m\AA\ and were assumed to be optically thin, the
same assumption was adopted by Cha \& Sembach (2000). The plot of
$R$(Ca,Na) versus $|V_{\rm LSR}|$ from 20  to 200 km s$^{-1}$ not
surprisingly is similar to that (their Figure 3) provided by Danks \&
Sembach (1995). 

At  the `low' velocities from $|V_{\rm LSR}|$ of 20 to about 50 km s$^{-1}$,
$R$(Ca,Na) increases steeply from about unity to about 20 with a large
scatter all velocities. At higher velocities, $R$(Ca,Na) appears to attain an
approximately constant value of 10 with a large scatter and with many lower
limits arising from the non-detection of the Na D lines. The observed
scatter at low and high velocities most probably exceeds the
measurement errors.  With the possible
exception of the $R$(Ca,Na) values at the lowest considered velocities, all
values are greater than values reported by Sembach \& Danks (1994) for
diffuse interstellar clouds for which $R$(Ca,Na) does not attain a value of
10 until peculiar velocities exceed about 50 km s$^{-1}$. 
Danks \& Sembach (1995) suggest that although (incomplete)
dust destruction in shocked
gas may explain high $R$(Ca,Na) values, clouds with low $R$(Ca,Na) values
may be found in gas which is in process of being compressed and accelerated.
Testing this suggestion will require high-resolution ultraviolet spectra and
thus access to more diagnostic lines.

Interstellar absorption line studies towards a few other SNRs have been
reported in the literature but no study approaches the richness achieved
for the Vela SNR and, in particular, no other SNR has been exposed to
precision spectroscopy at more than a single epoch.
The studied  SNRs with a range of
ages are all older than the 11000 year old Vela SNR and, thus, 
comparisons may shed light on the evolution of the
interaction between a SNR and its interstellar environment. The 
SNRs for which
Na D and Ca K observations are available include the Cygnus Loop at 20000
years (Welsh et al. 2002), IC 443 at 30000 years (Welsh \& Sallmen 2003),
Shajn 147 at 100000 years (Silk \& Wallerstein 1973; Sallmen \& Welsh 2004),
and the Monoceros Loop at 150000 years (Wallerstein \& Jacobsen 1976;
Welsh et al. 2001) where the ages are those given in the cited references.

In the case of the Cygnus Loop, Welsh et al. (2002) observed nine OB stars
with distances of 2500 to 2300 pc with the SNR at a distance of about 440 pc.
There is strong interstellar absorption at low velocity ($V_{\rm LSR}
\simeq 1$ km s$^{-1}$ with the negative $R$(Ca,Na) ratio representative
of the diffuse interstellar medium. Components at 9 and 20 km s$^{-1}$ are 
present for  a majority of the sight lines and a component at 30 km s$^{-1}$
is seen at the Na D lines but not the Ca K line for one star. These
latter components appear for stars at distances beyond the Cygnus Loop.
However, Welsh et al. are `unable to definitely associate these components
with an interaction between the expansion of the SN shock wave and the
ambient interstellar medium' and suggest the alternative origin with
the `an old pre-cursor SN neutral gas shell'.

For the IC 443 SNR, Welsh \& Sallmen (2003) describe Na D and Ca K absorption
profiles towards four early-type stars, three probably beyond the SNR and one in
the foreground. Two of the more distant stars show high-velocity 
components extending to $-100$ km s$^{-1}$ and $+50$ km s$^{-1}$ with
$R$(Ca,Na) values which, 
as Welsh \& Sallmen write, are `consistent with appreciable levels of dust
grain destruction due to interstellar shocks caused by interaction of the
expanding SNR blast-wave with the ambient interstellar medium.' 

Interstellar gas in the direction of the Shajn 147 SNR has been subject to
limited exploration: Na D and Ca K lines  were measured off
photographic spectra for seven stars by Silk \& Wallerstein (1973) and
Na D and Ca K line profiles from CCD spectra of three stars (plus {\it FUSE}
spectra of two stars) were obtained  by Sallmen \& Welsh (2004). High-velocity gas
in one star - HD 36665 - first noted by Silk \& Wallerstein is discussed
in detail by Sallmen \& Welsh. {\it FUSE} spectra of a second star 
- HD 37318 -  show a high-velocity red-shifted component in ions
such as N$^+$ and Fe$^+$ which is not seen in Na D and Ca K. The
various high-velocity components `can be associated with the expansion of 
the SNR that has disrupted the surrounding interstellar gas' (Sallmen 
\& Welsh 2004).

A photographic spectroscopic survey of 25 sight lines 
towards the Monoceros Loop by Wallerstein \& Jacobsen (1976) yielded
one star -- HD 47240 --  showing high-velocity components. This star
was subsequently reobserved at Na D and Ca K
by Welsh et al. (2001) who also discuss 
{\it FUSE} spectra. Components at $V_{\rm LSR}$ $\simeq$ $+65$ km s $^{-1}$ show
$R$(Ca,Na) values indicative of some release of Ca from grain destruction,
a result in line with measurements of Fe, Si and Al column densities from
{\it FUSE}.    

Studies of these four SNRs, all older than the Vela SNR, show that 
high-velocity gas linked to the SNR can persist for at least 150000 years. 
An impression is suggested that gas at the highest velocities seen in the
Vela observations is less common among the older SNRs. No information is
available about the evolution of the
high-velocity components. Reobservation at Na D and Ca K would seem
now warranted in order to constrain evolution over about a decade.

\begin{figure*}
\vspace{0.3cm}
\rotatebox{90}{\hspace{1.2cm}Normalised Intensity}
\includegraphics[width=5.7cm,height=5cm]{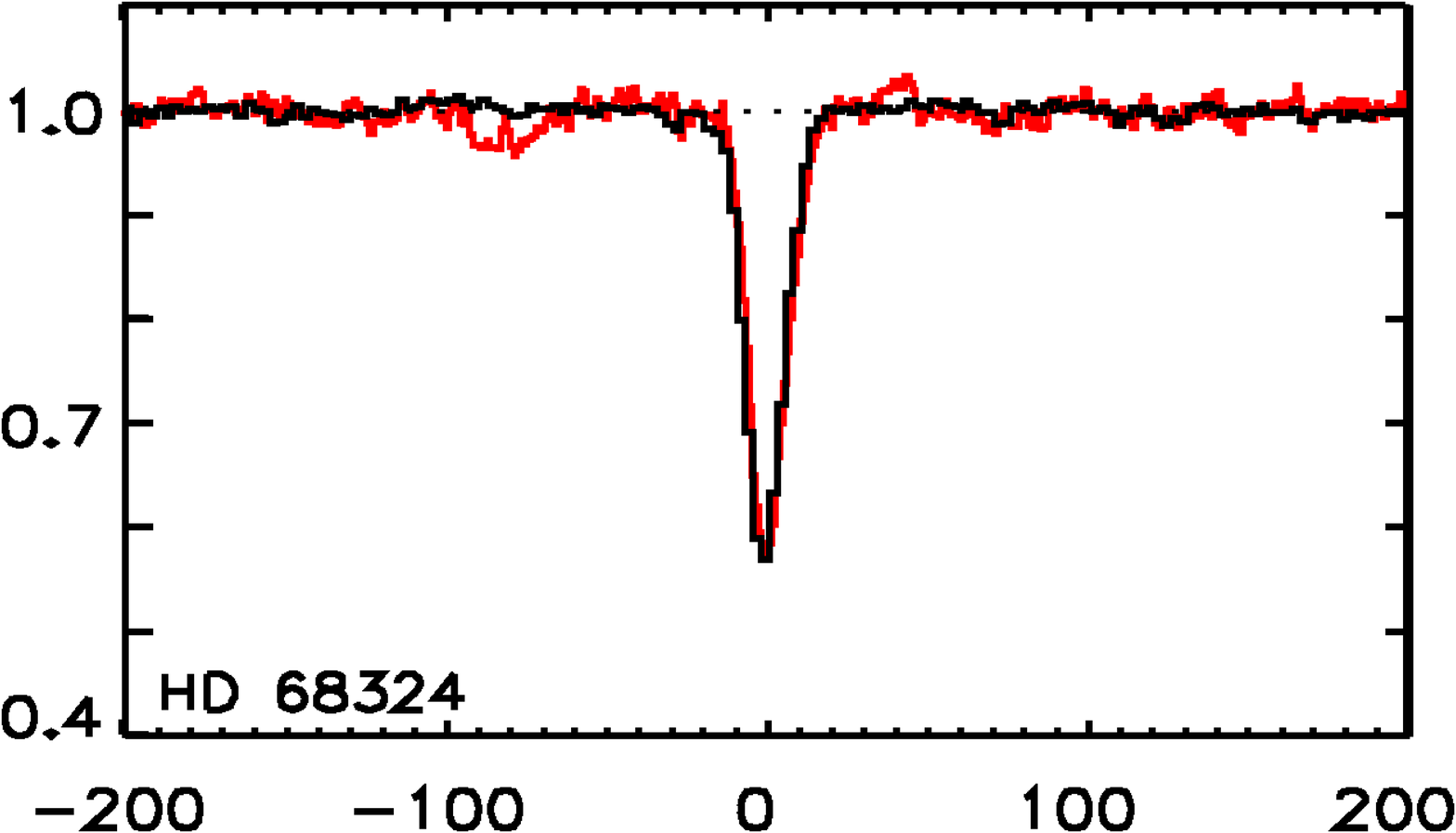}
\includegraphics[width=5.7cm,height=5cm]{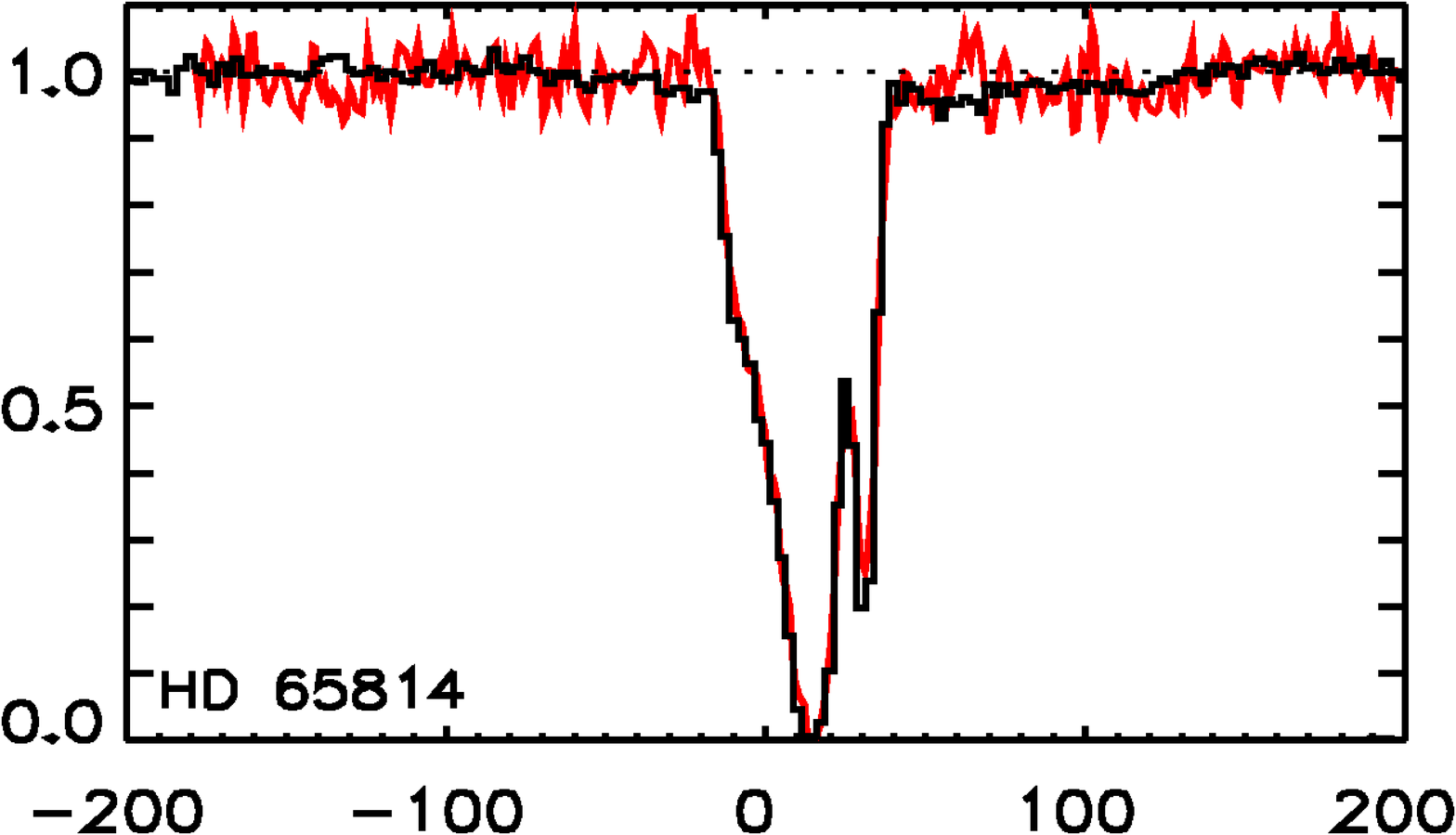}
\includegraphics[width=5.7cm,height=5cm]{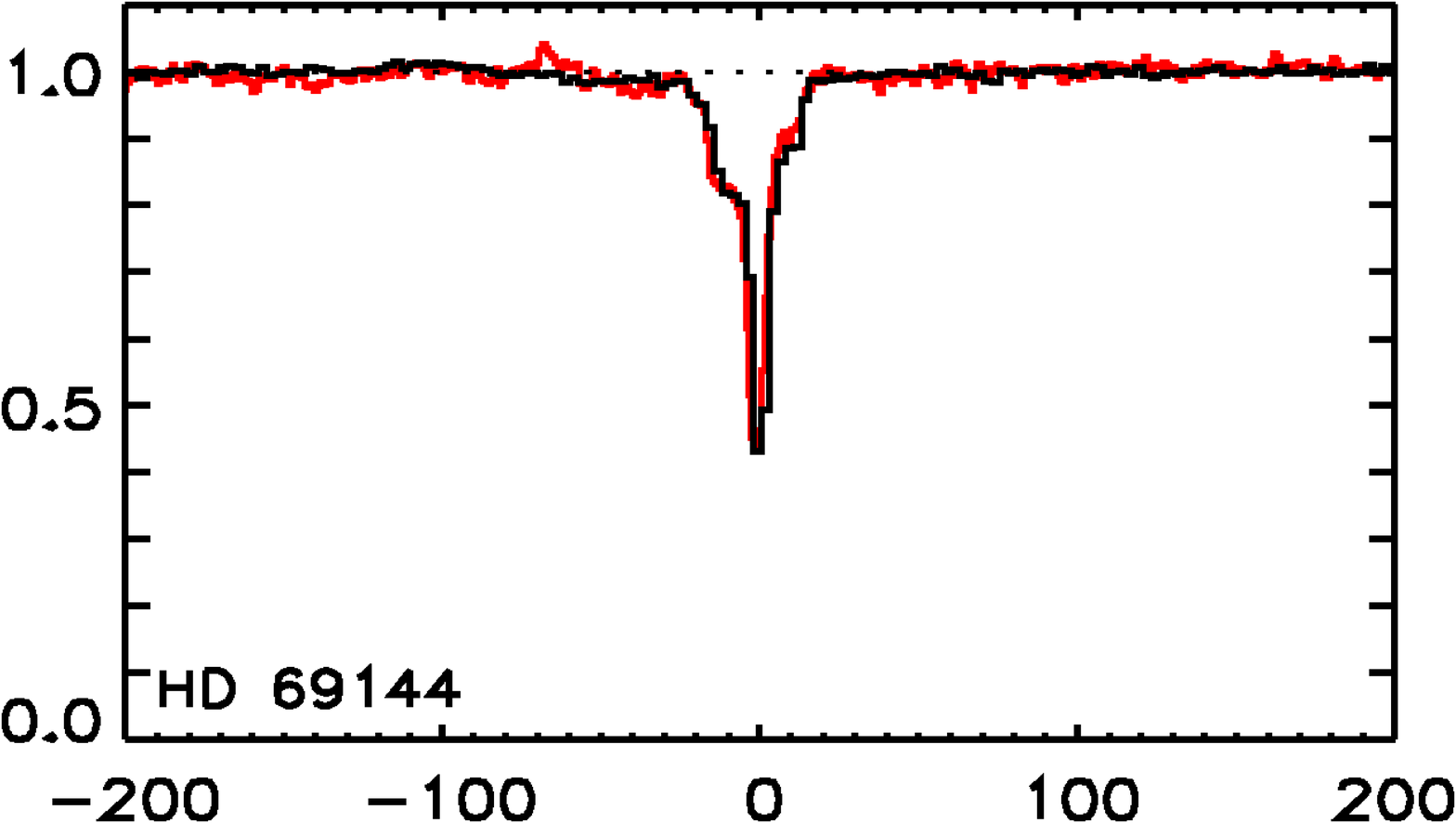}
\hspace{.5cm}
\vspace{.3cm}
\rotatebox{90}{\hspace{1.2cm}Normalised Intensity}
\includegraphics[width=5.7cm,height=5cm]{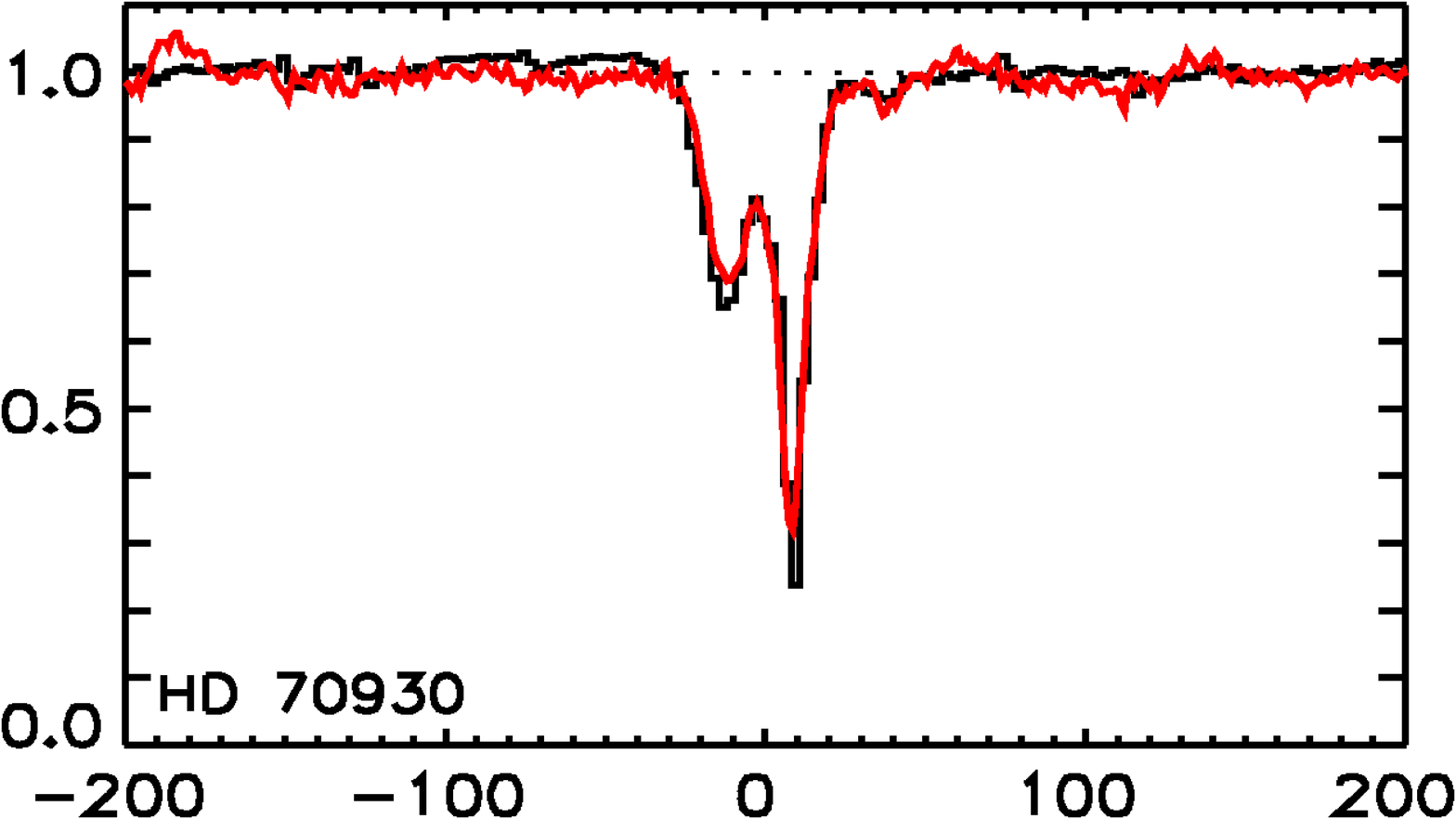}
\includegraphics[width=5.7cm,height=5cm]{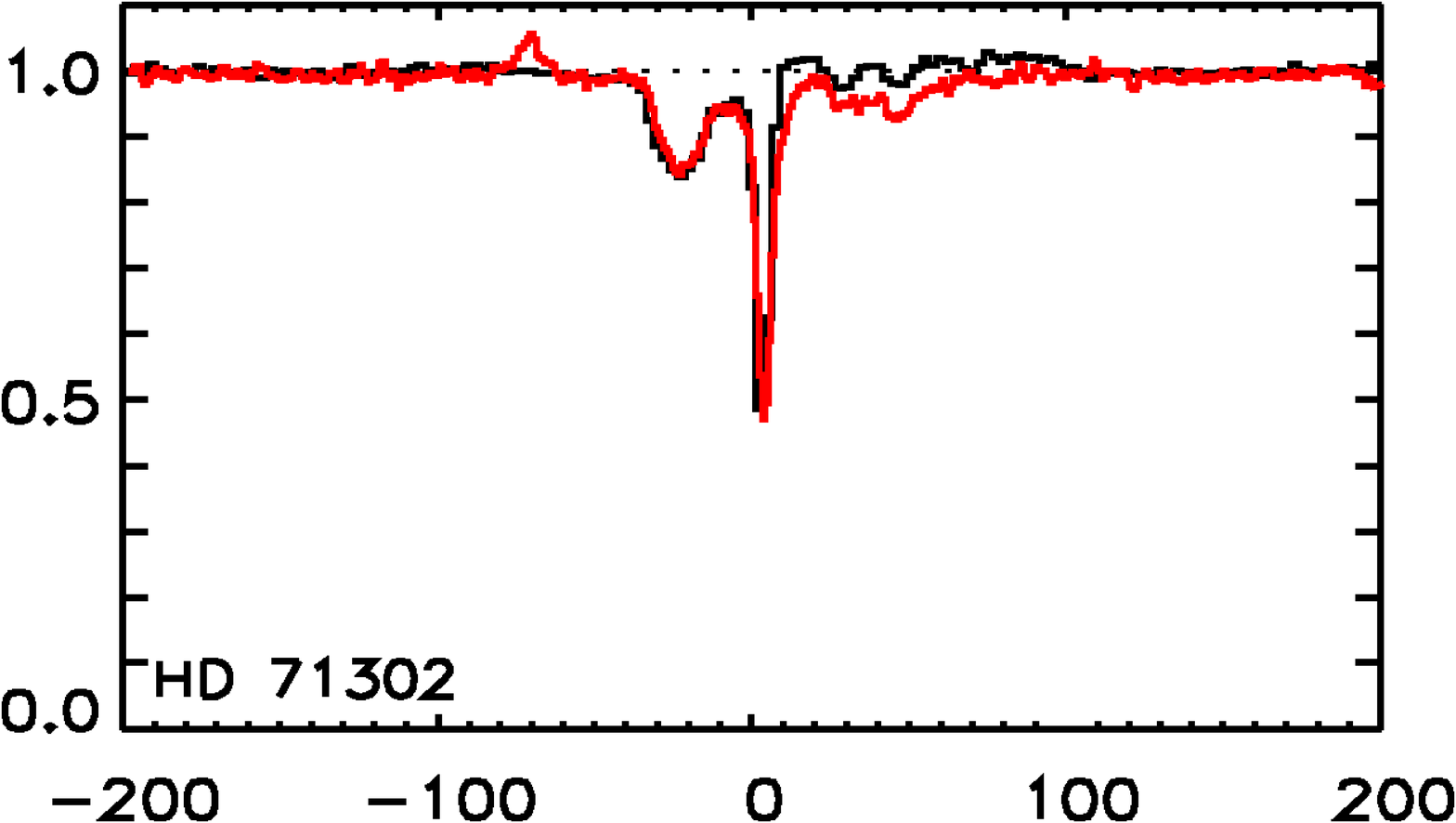}
\includegraphics[width=5.7cm,height=5cm]{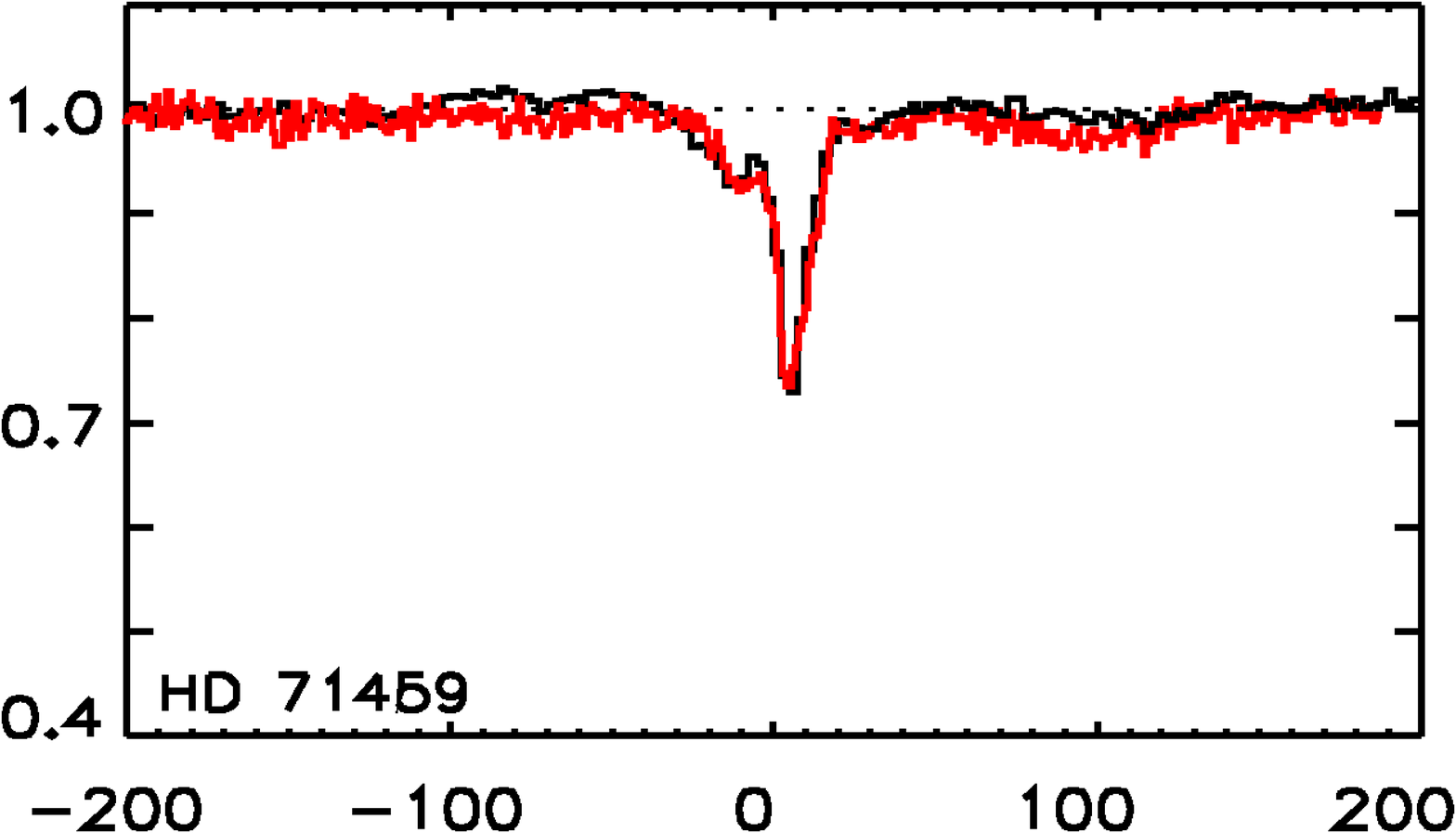}
\hspace{.5cm}
\vspace{.1cm}
\rotatebox{90}{\hspace{1.2cm}Normalised Intensity}
\includegraphics[width=5.7cm,height=5cm]{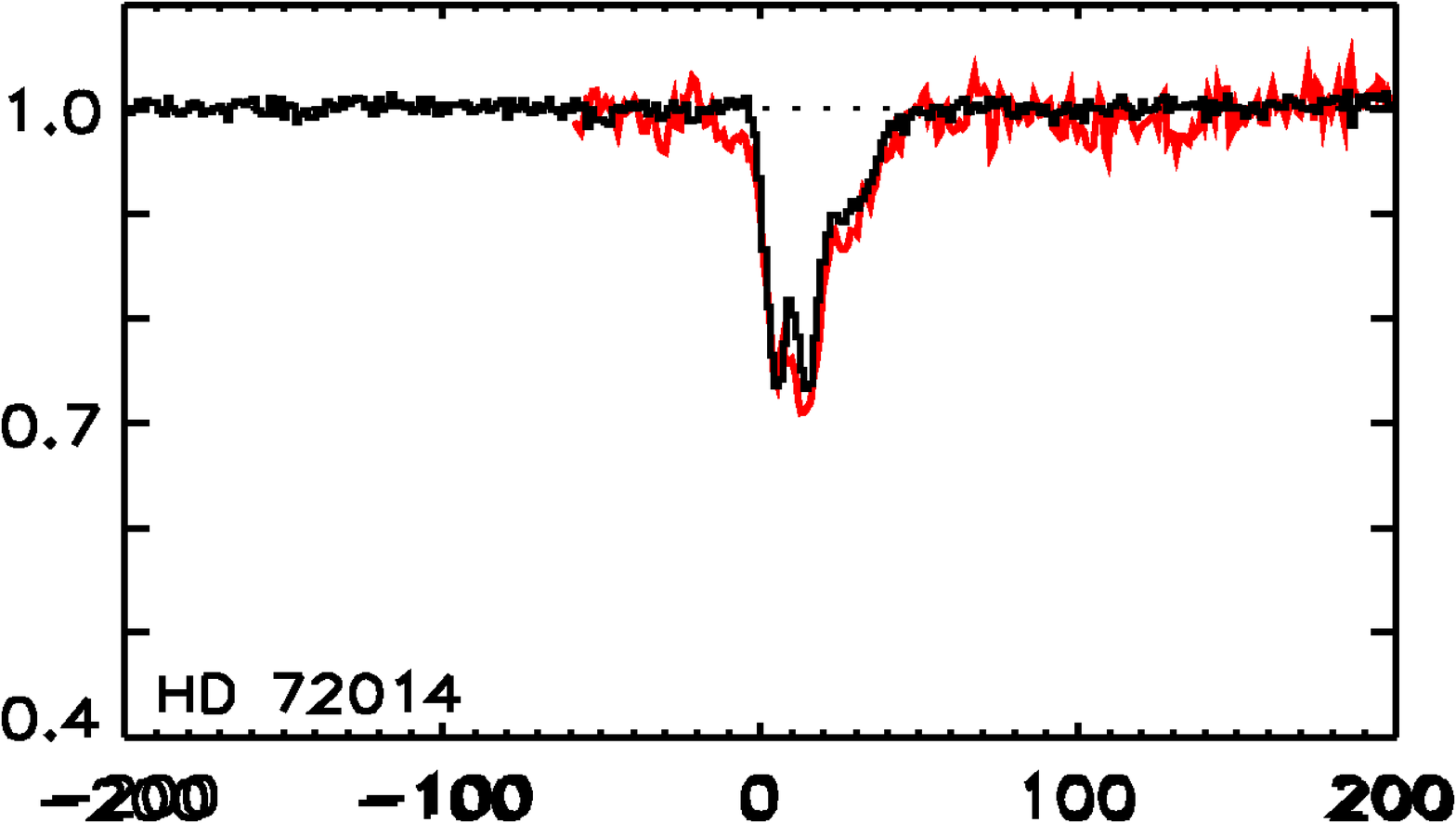}
\includegraphics[width=5.7cm,height=5cm]{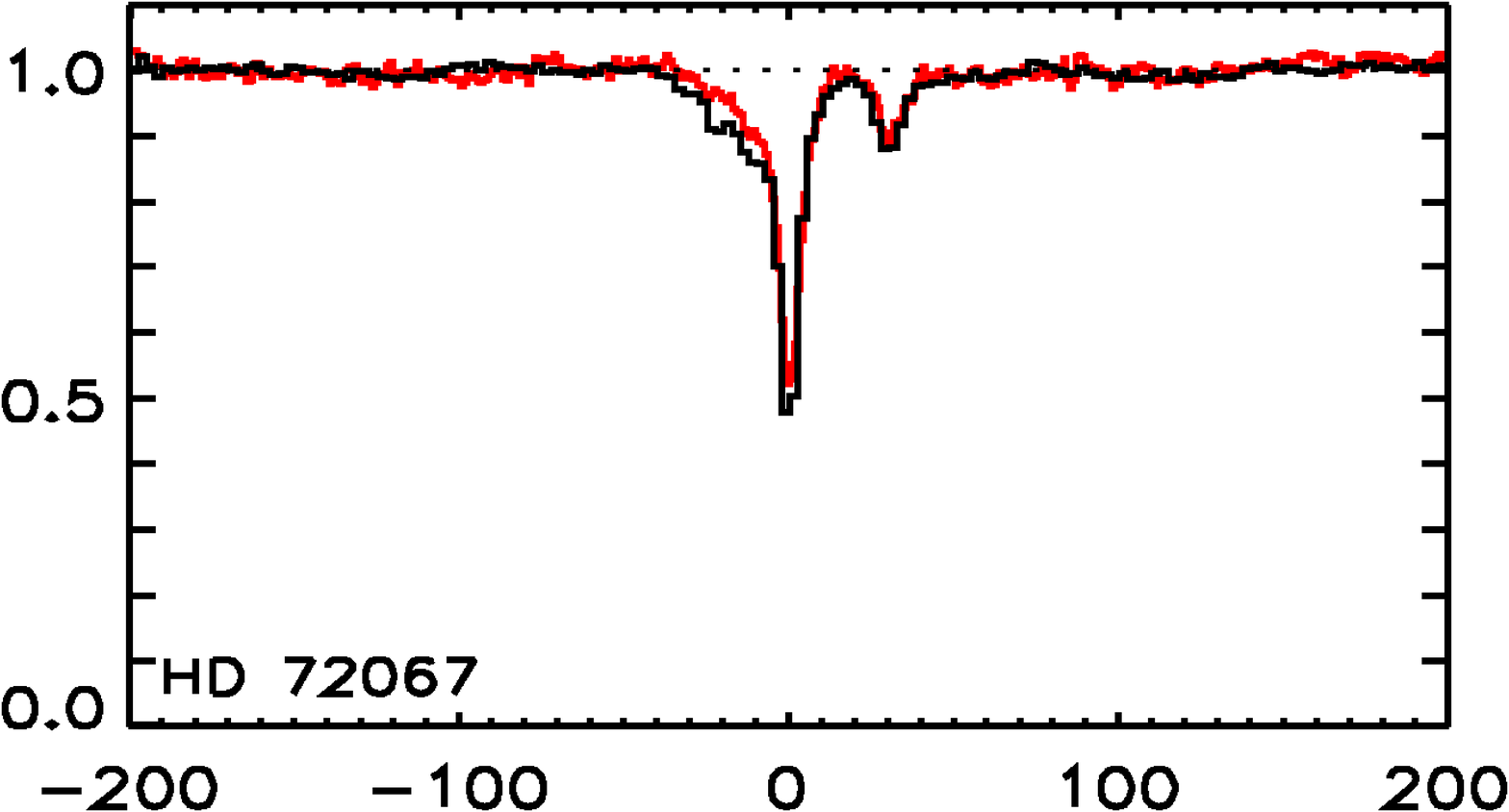}
\includegraphics[width=5.7cm,height=5cm]{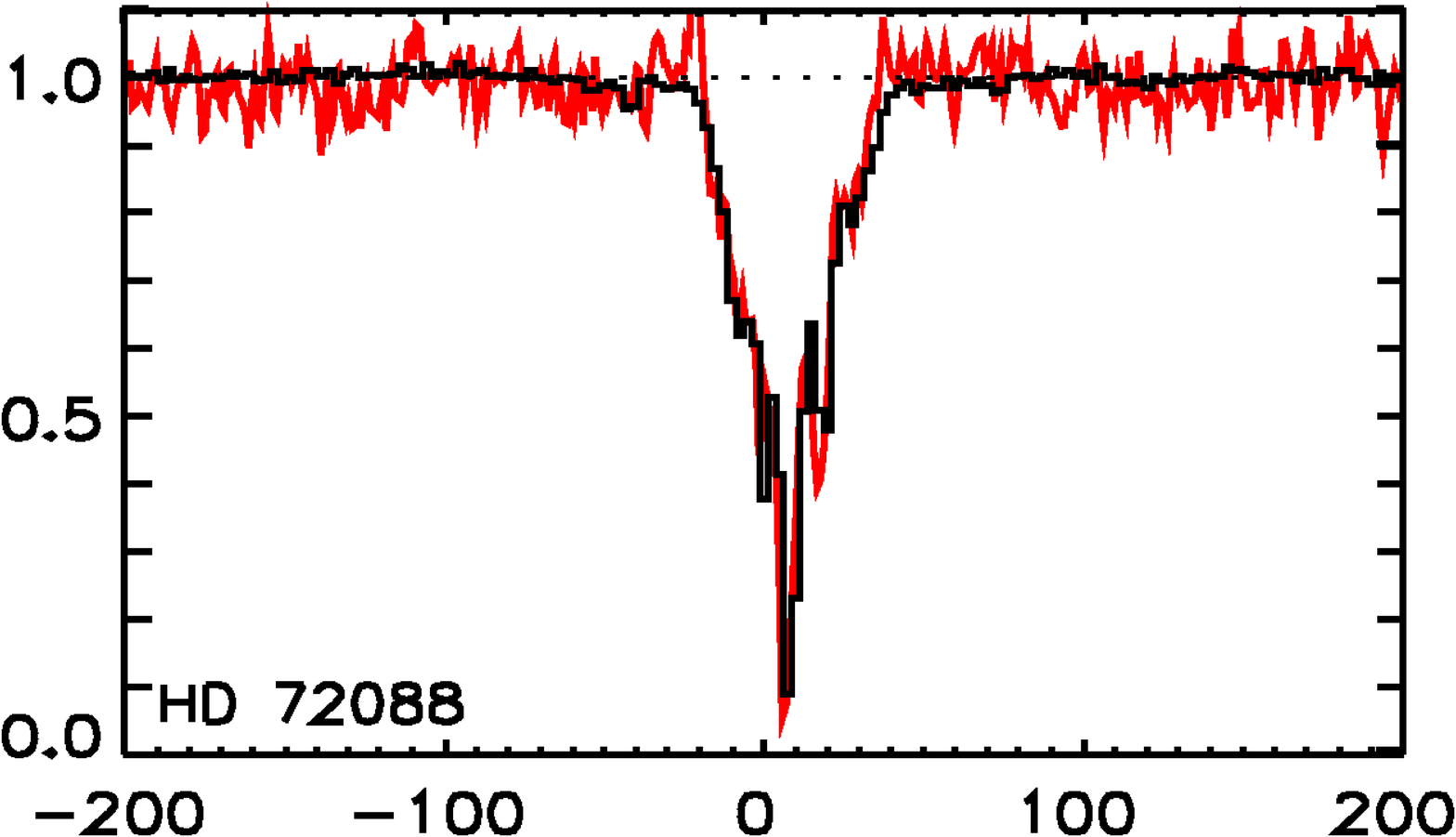}
\hspace{.5cm}
\vspace{-.1cm}
\rotatebox{90}{\hspace{1.2cm}Normalised Intensity}
\includegraphics[width=5.7cm,height=5cm]{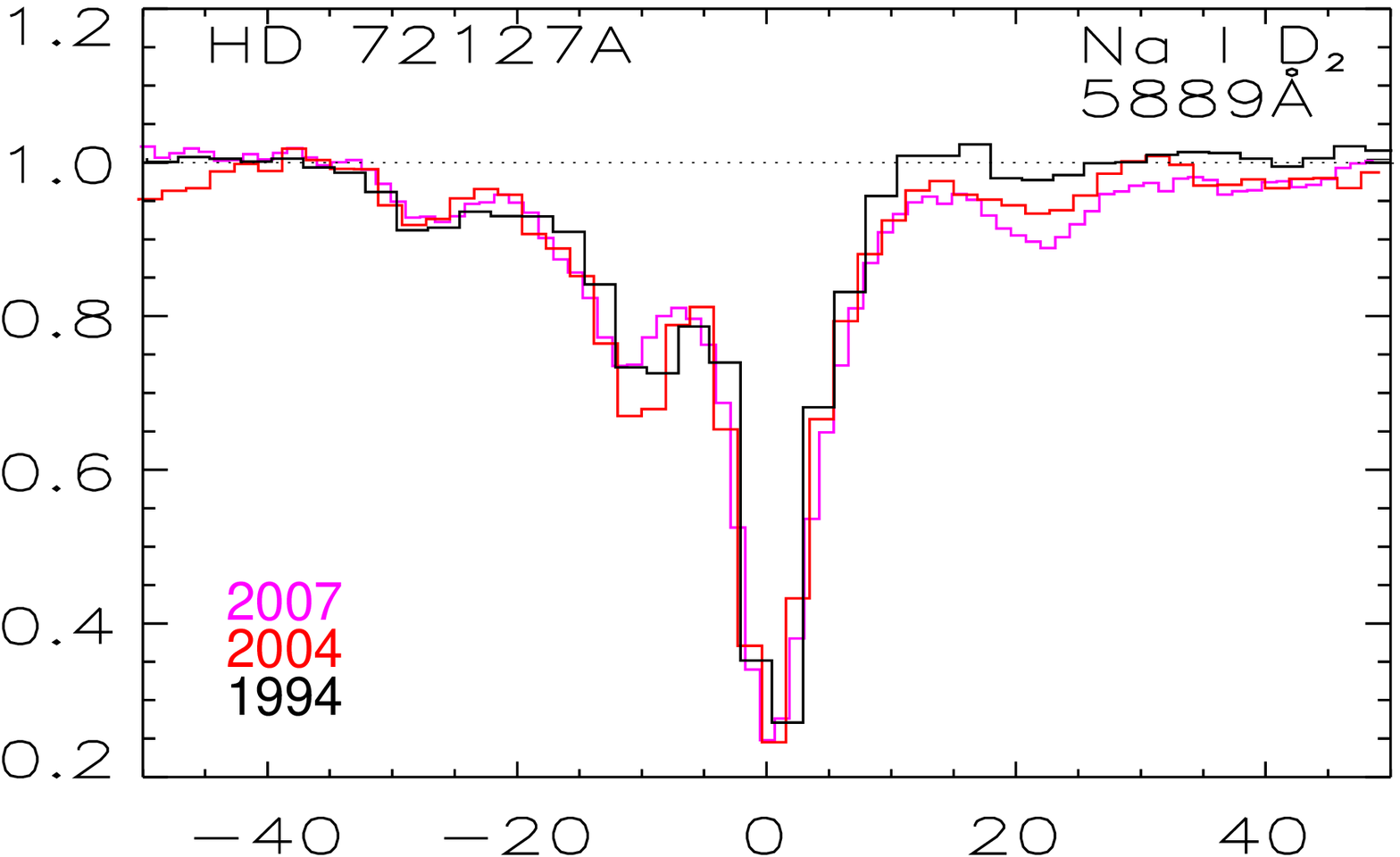}
\includegraphics[width=5.7cm,height=5cm]{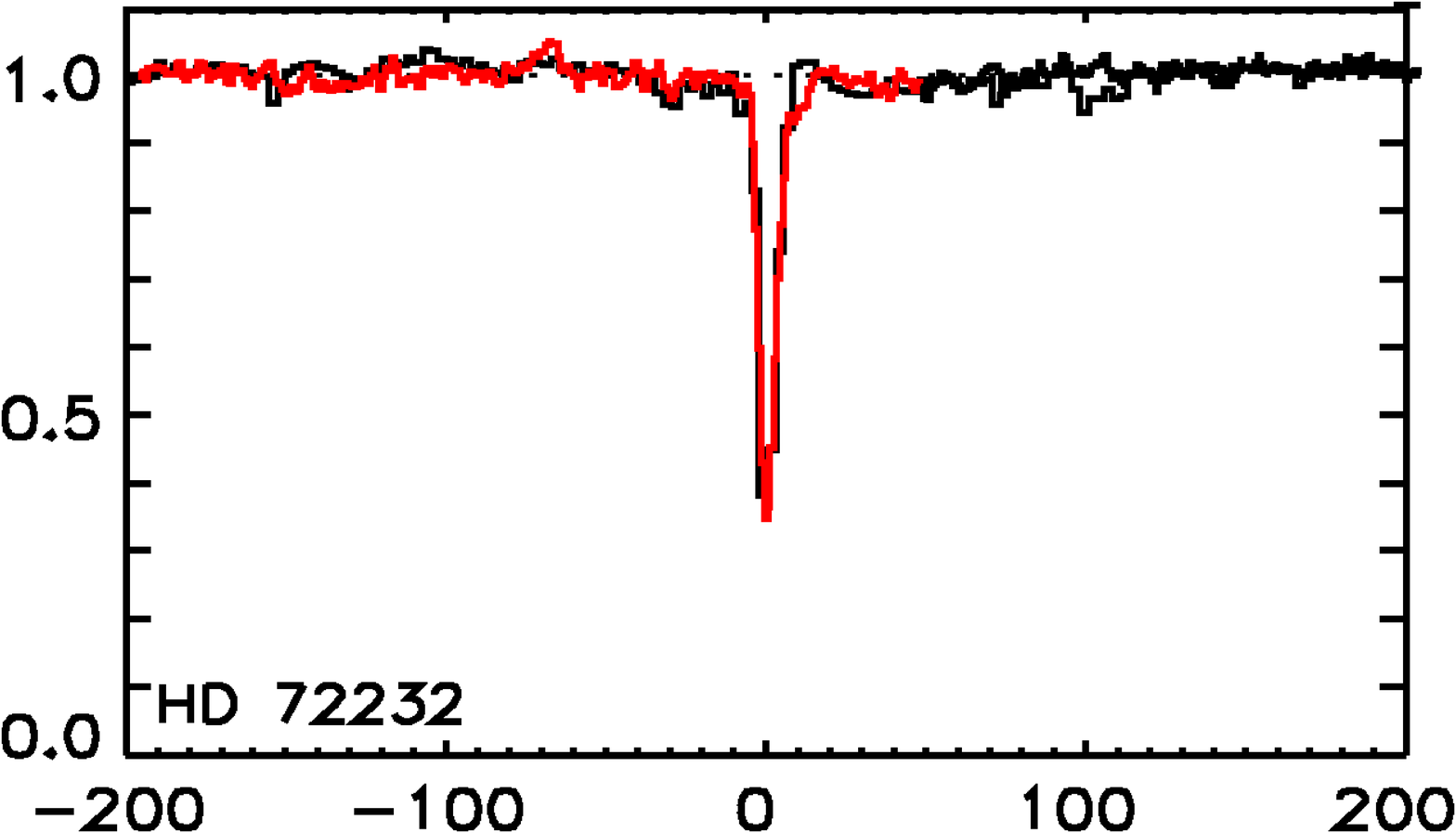}
\includegraphics[width=5.7cm,height=5cm]{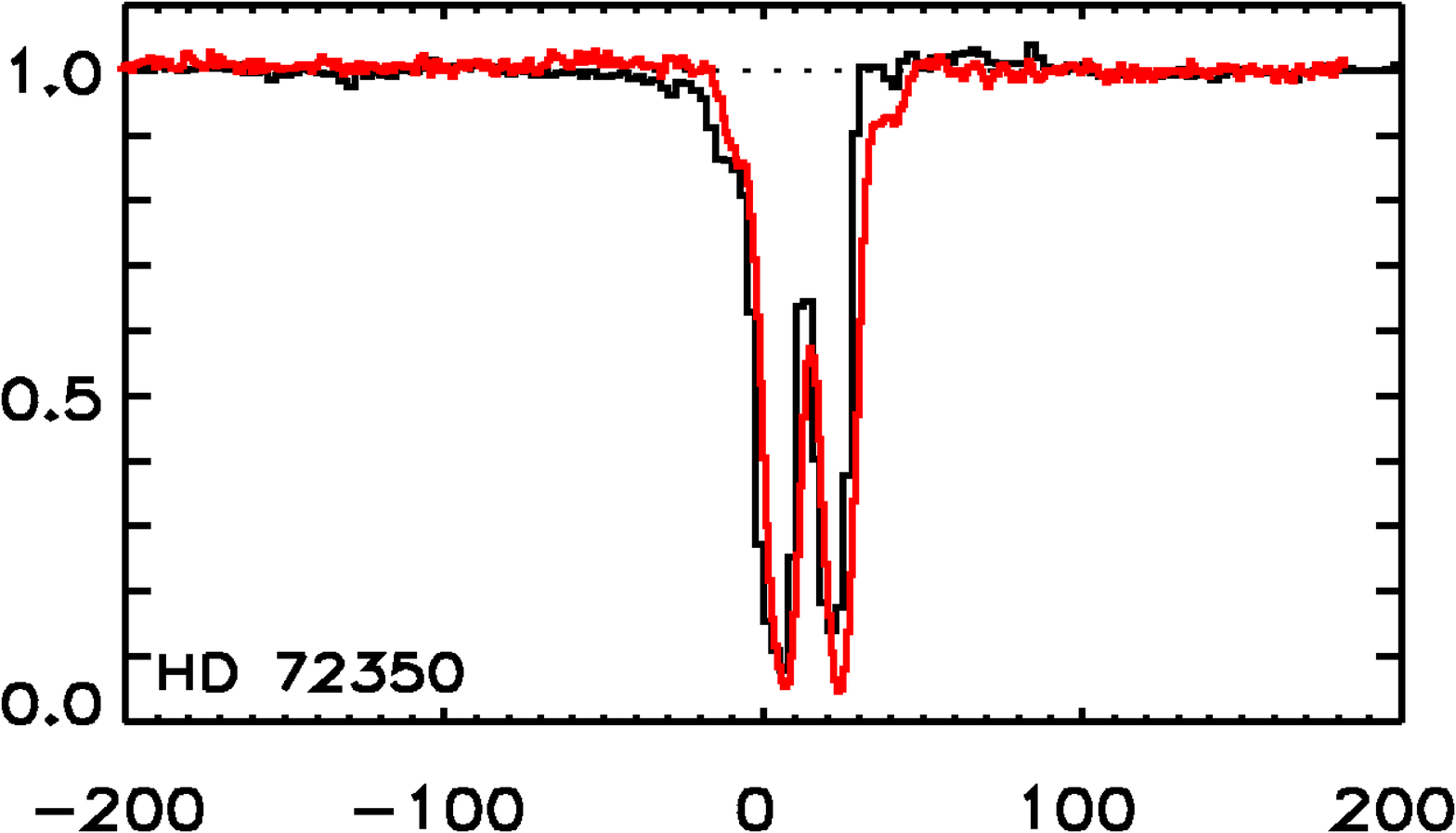}
\hspace{14cm}$V_{\rm LSR}$(km s$^{-1}$) \\
\caption{The Na\,{\sc i} D$_2$ profile of HD 68324, HD 65814, HD 69144, HD 70930,
 HD 71302, HD 71459, HD 72014, HD 72067, HD 72088, HD 72127A, HD 72232, and
 HD 72350  obtained by us in
 2011 (red line)  is superposed on the Na\,{\sc i} D$_2$ profile obtained in 1993
 (black line) by Cha \& Sembach (2000). For HD 72127A, VBT profiles from 2004 and 2007 are shown with Cha \& Sembach's profile from 1994.}
\end{figure*}

\begin{figure*}
\vspace{0.3cm}
\rotatebox{90}{\hspace{1.2cm}Normalised Intensity}
\includegraphics[width=5.7cm,height=5cm]{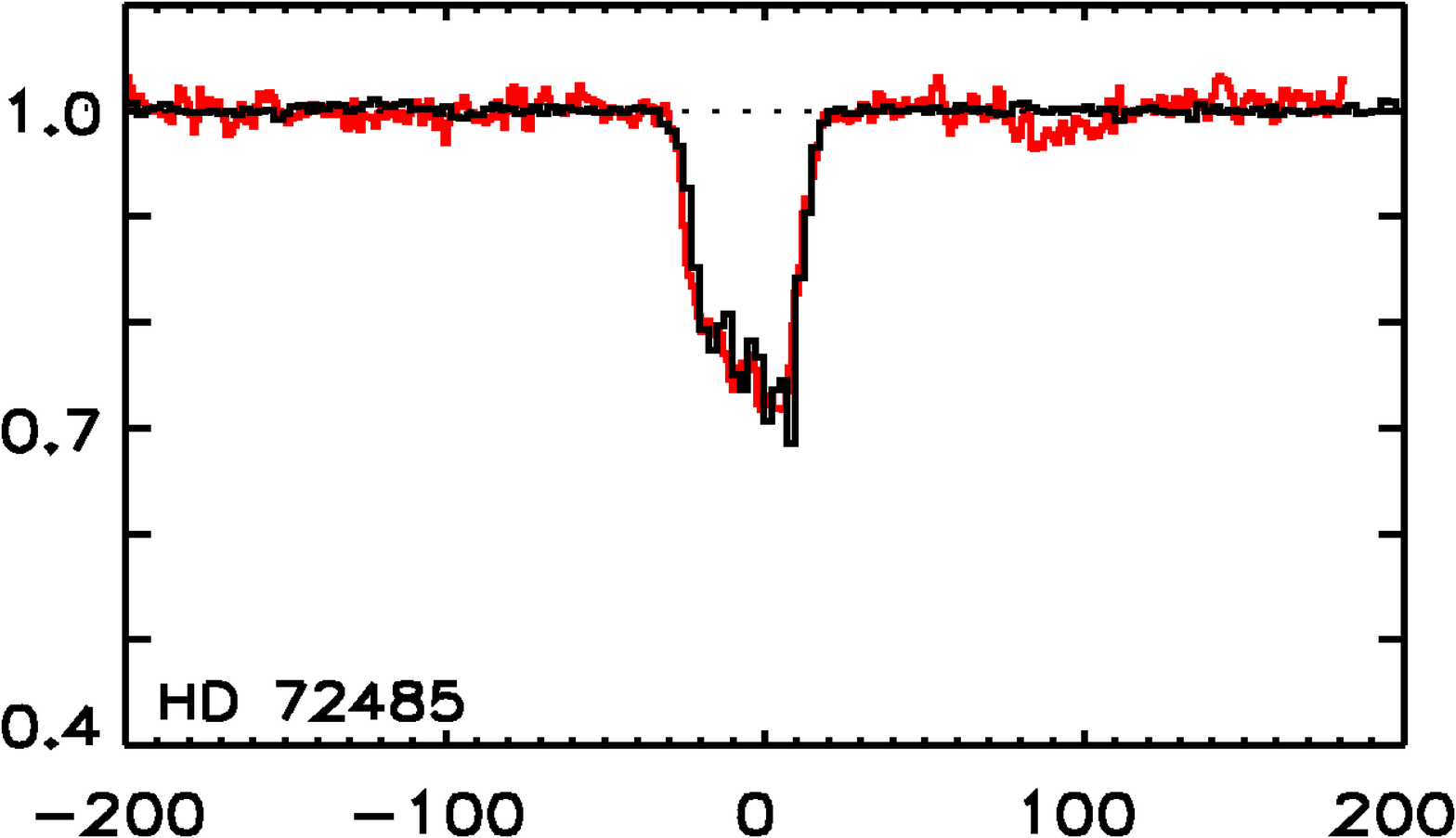}
\includegraphics[width=5.7cm,height=5cm]{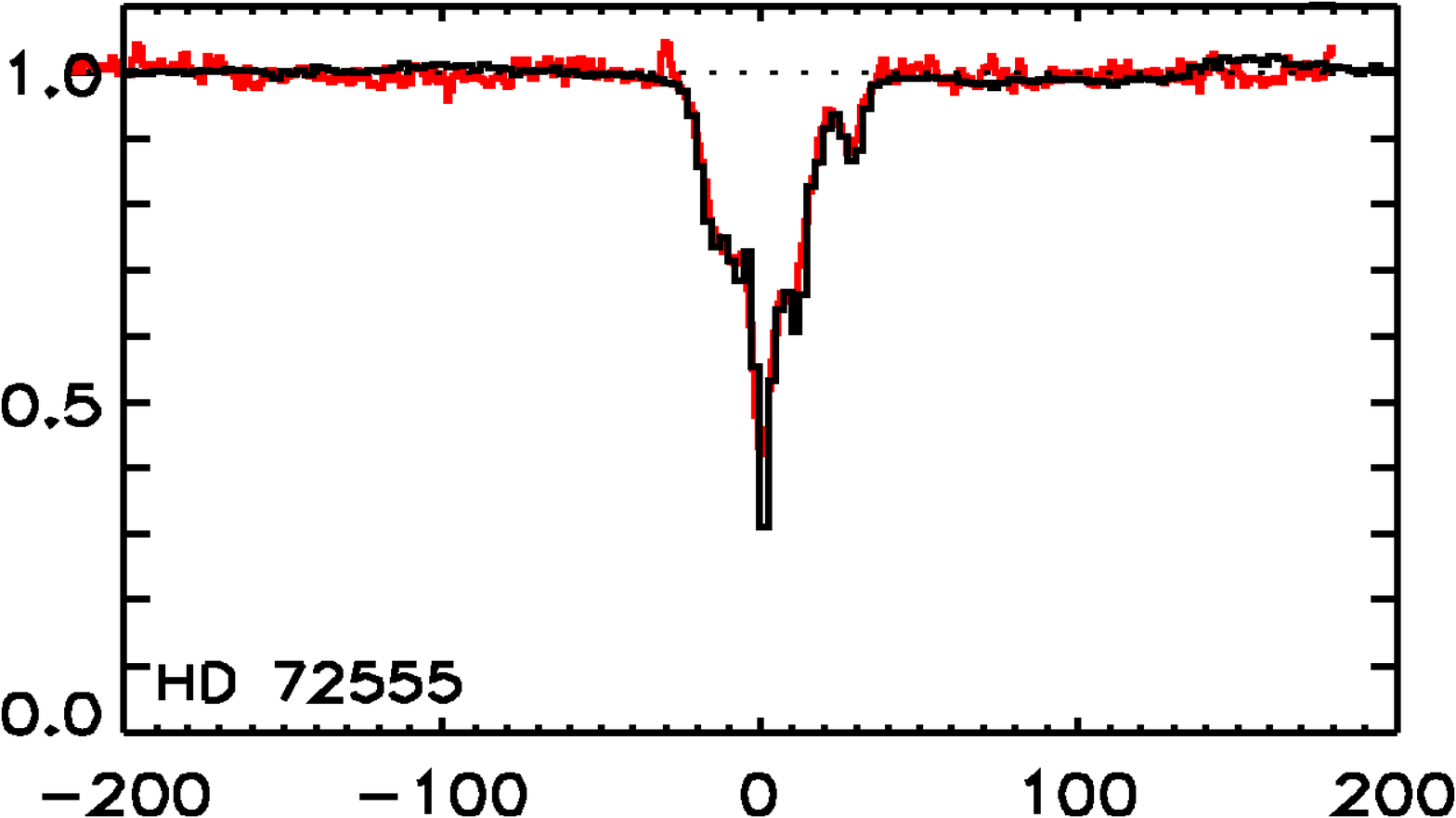}
\includegraphics[width=5.7cm,height=5cm]{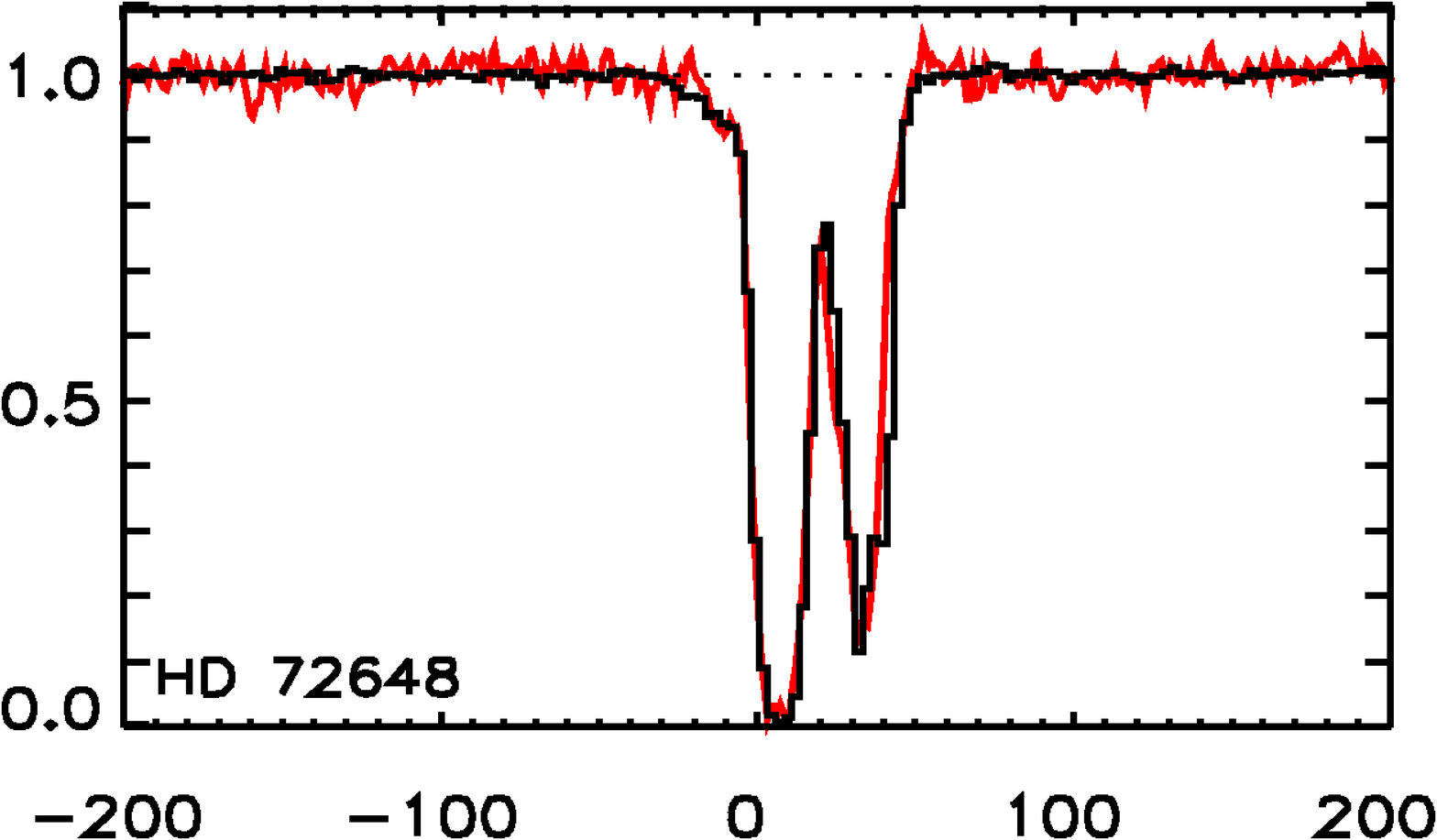}
\hspace{.5cm}
\vspace{.3cm}
\rotatebox{90}{\hspace{1.2cm}Normalised Intensity}
\includegraphics[width=5.7cm,height=5cm]{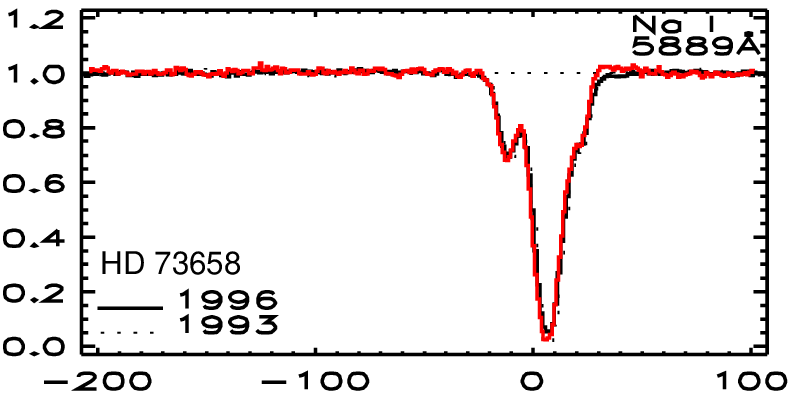}
\includegraphics[width=5.7cm,height=5cm]{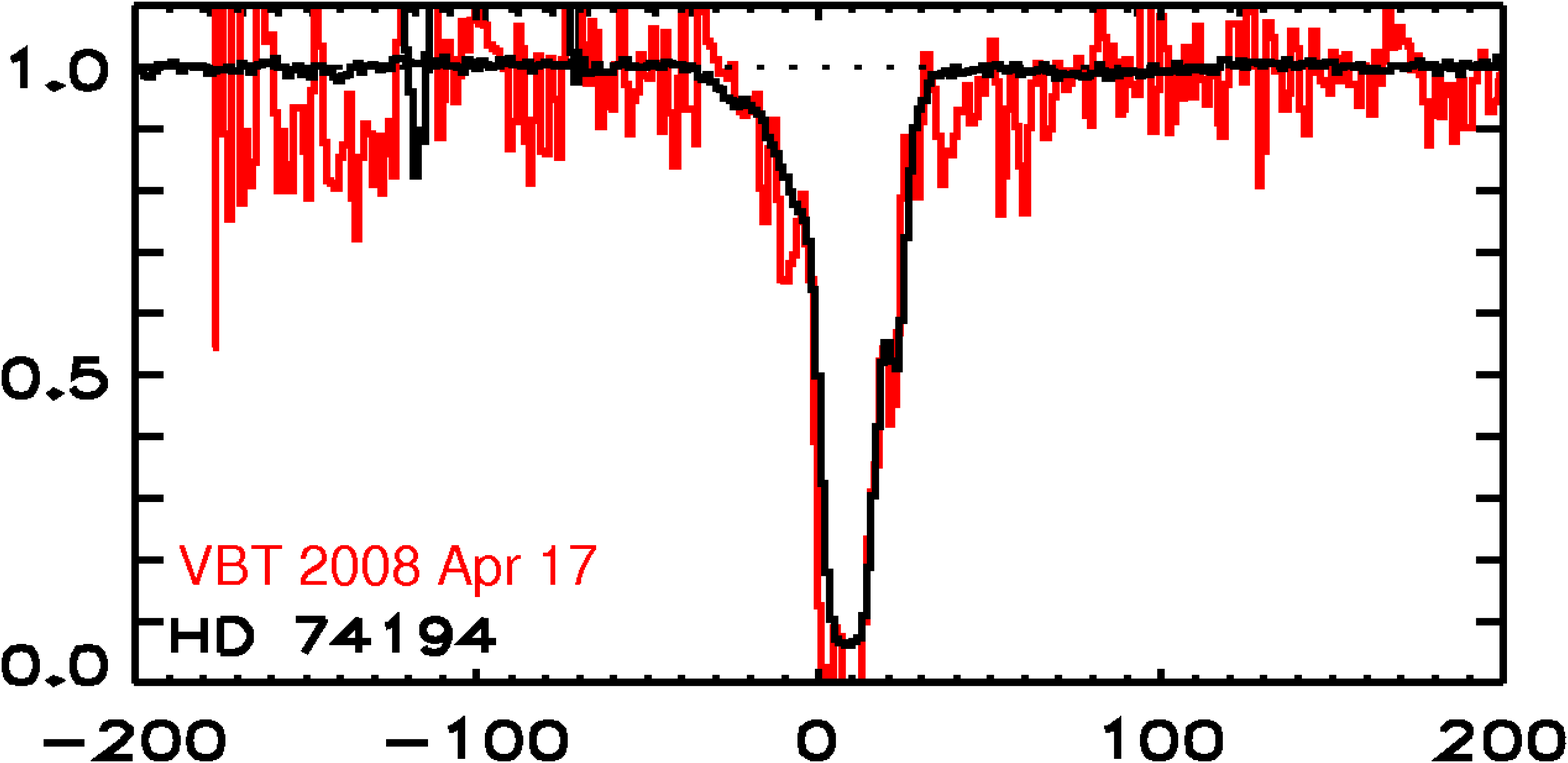}
\includegraphics[width=5.7cm,height=5cm]{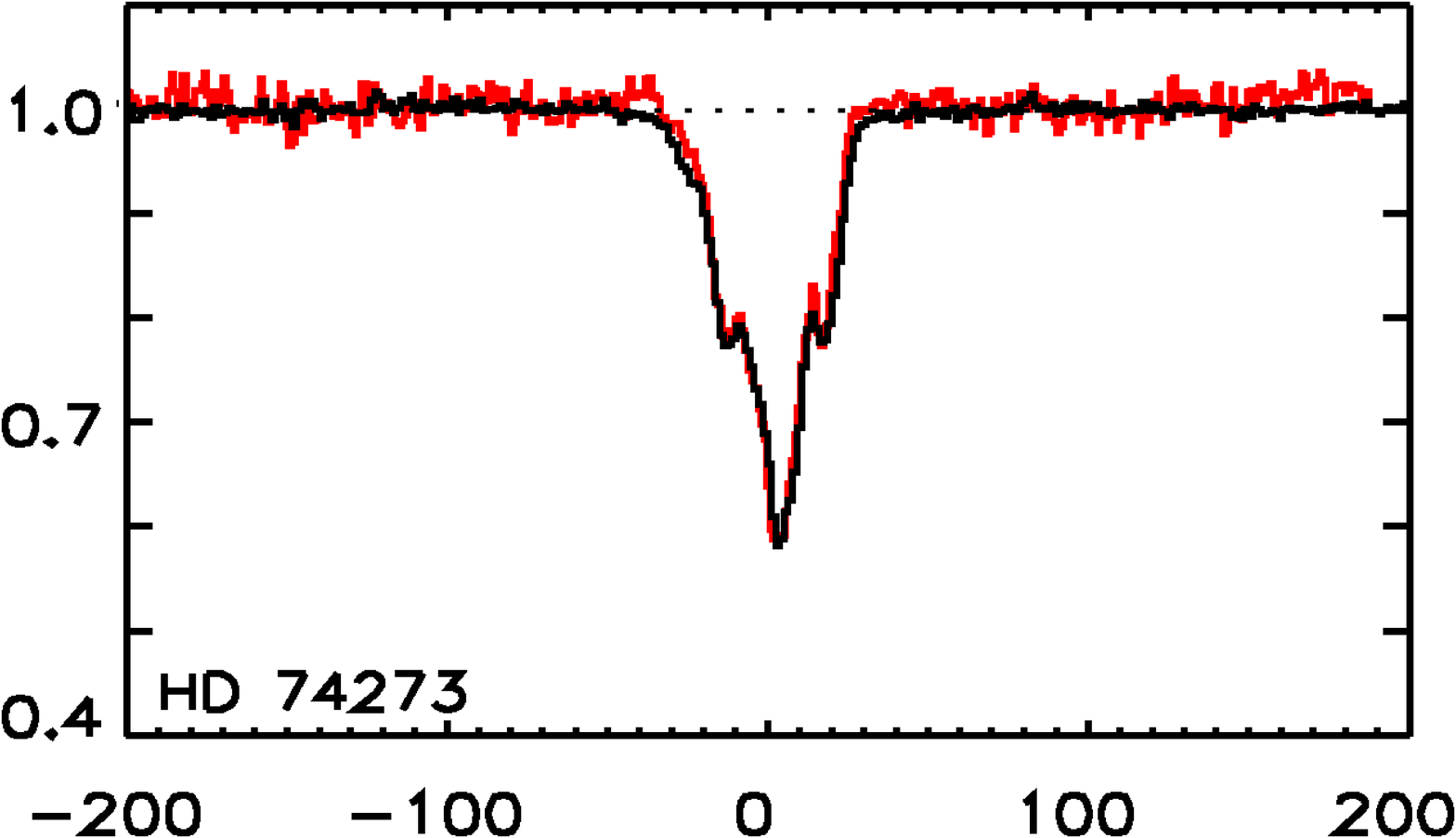}
\hspace{.5cm}
\vspace{.1cm}
\rotatebox{90}{\hspace{1.2cm}Normalised Intensity}
\includegraphics[width=5.7cm,height=5cm]{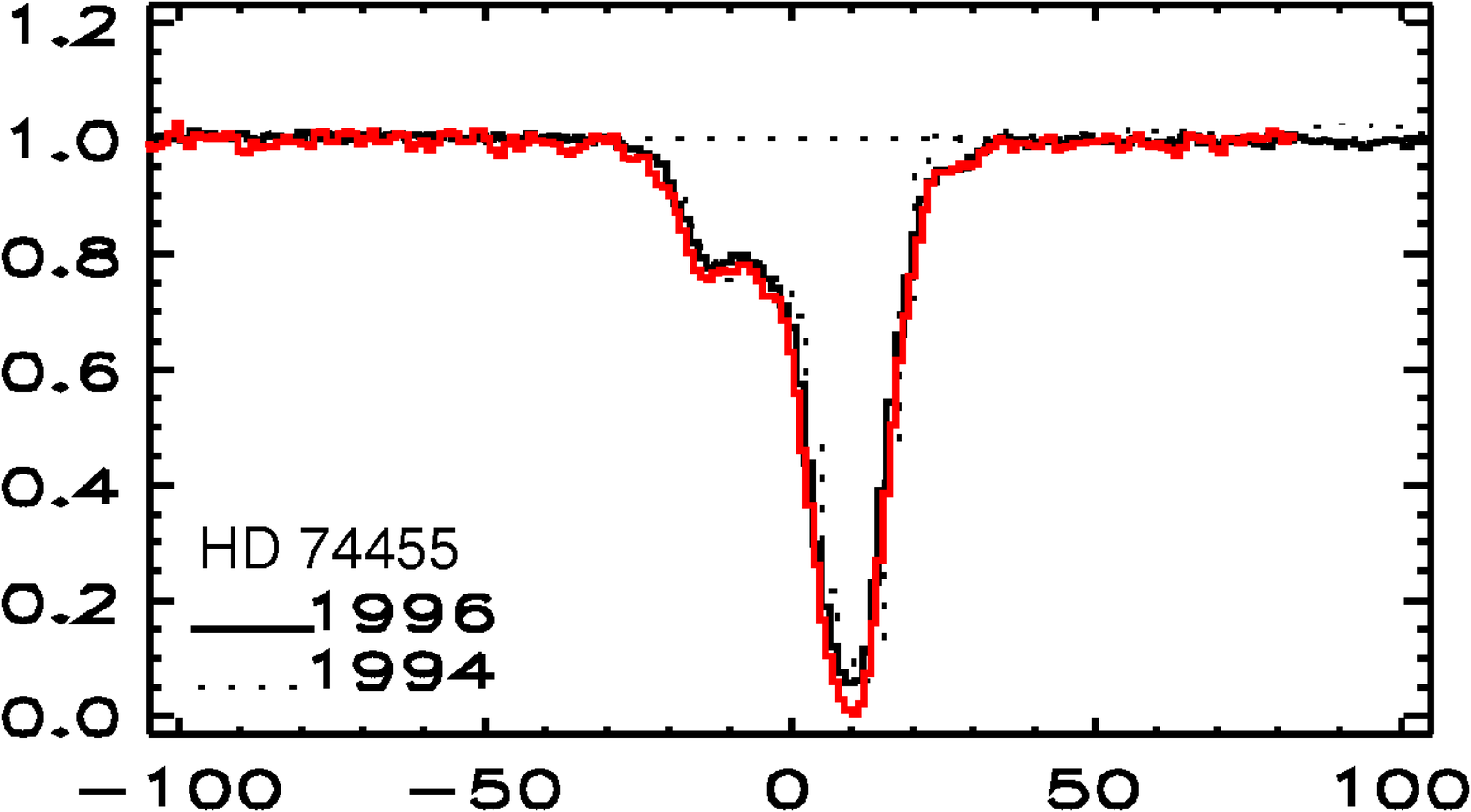}
\includegraphics[width=5.7cm,height=5cm]{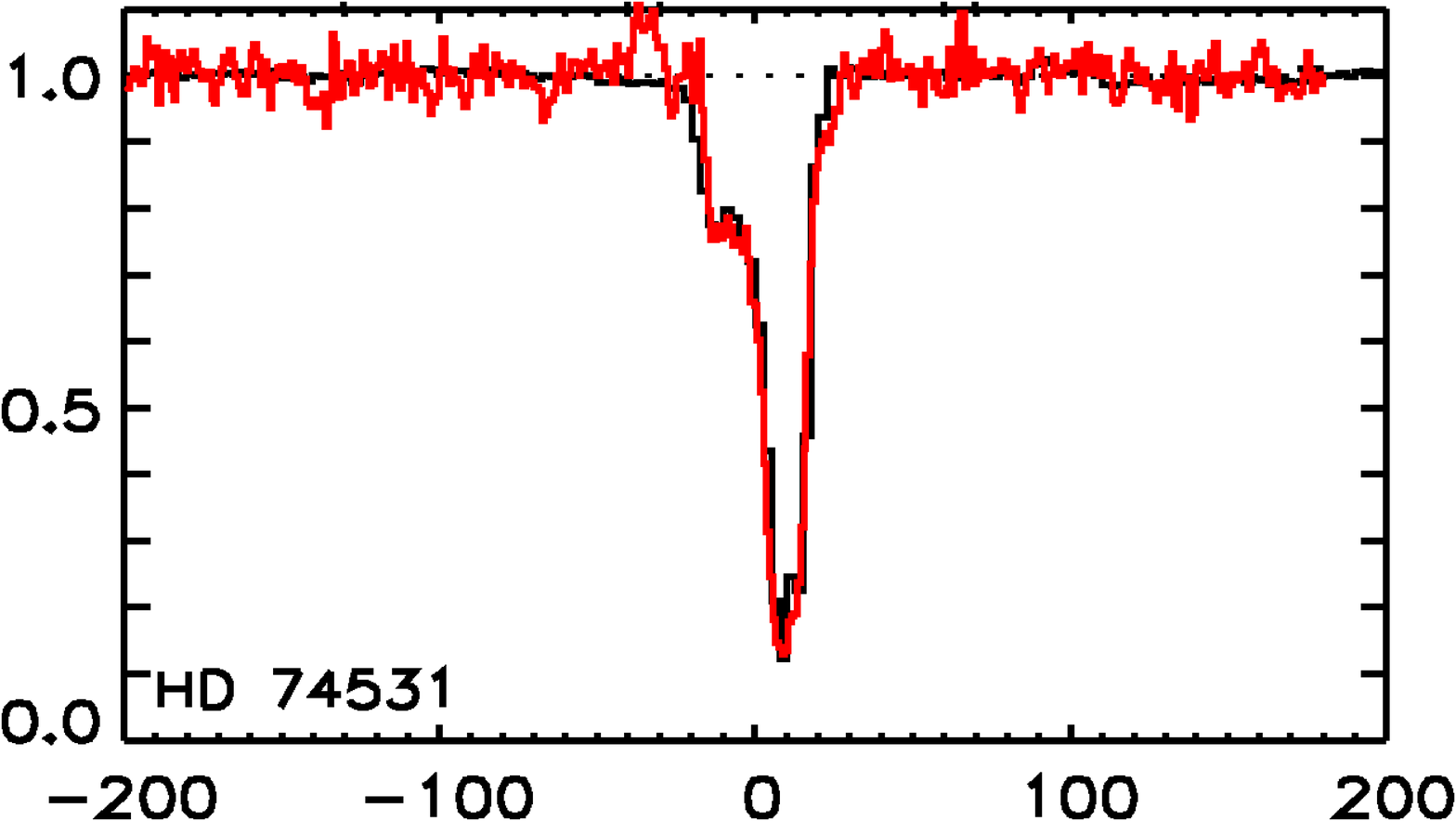}
\includegraphics[width=5.7cm,height=5cm]{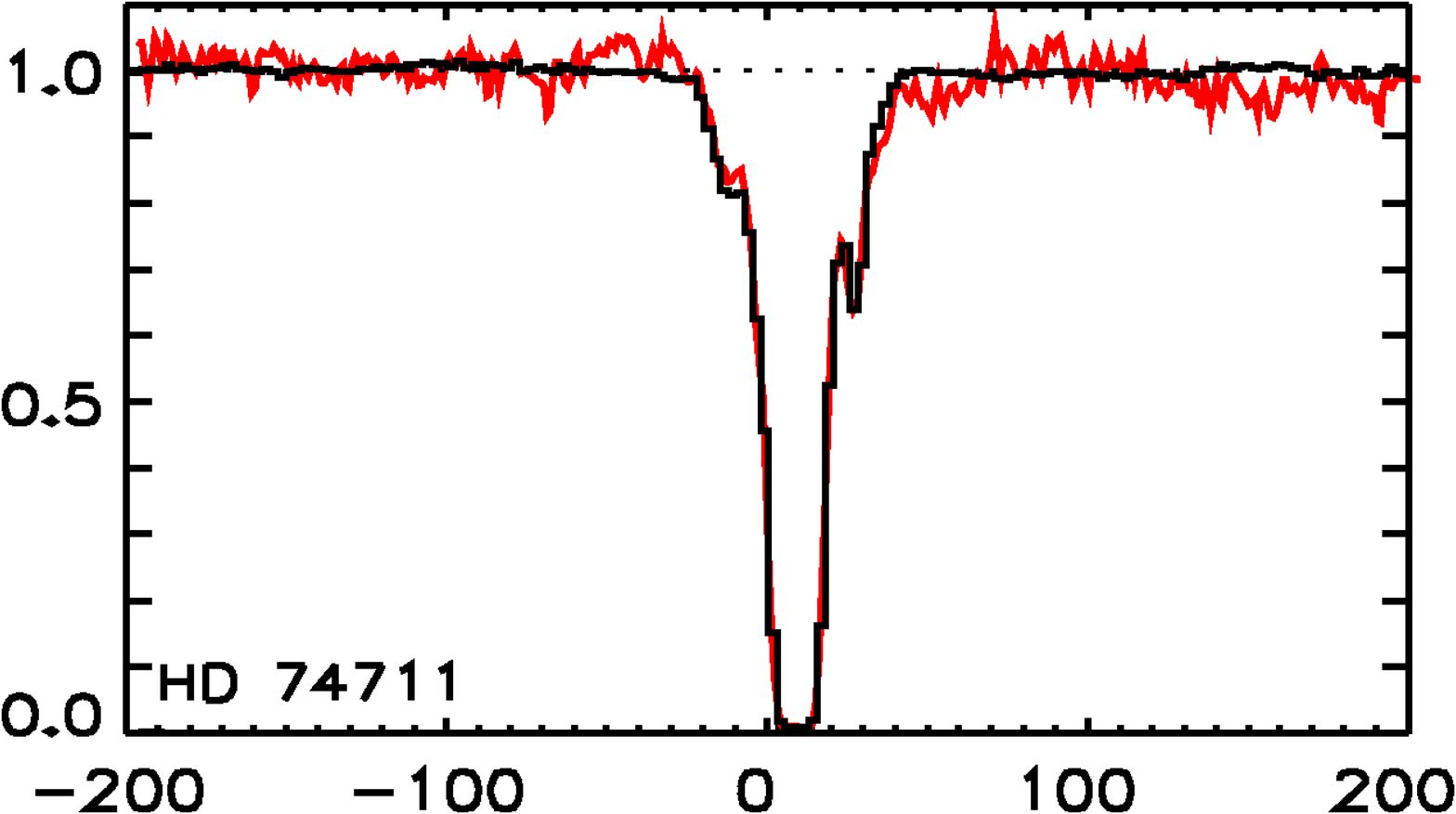}
\hspace{.5cm}
\vspace{-.1cm}
\rotatebox{90}{\hspace{1.2cm}Normalised Intensity}
\includegraphics[width=5.7cm,height=5cm]{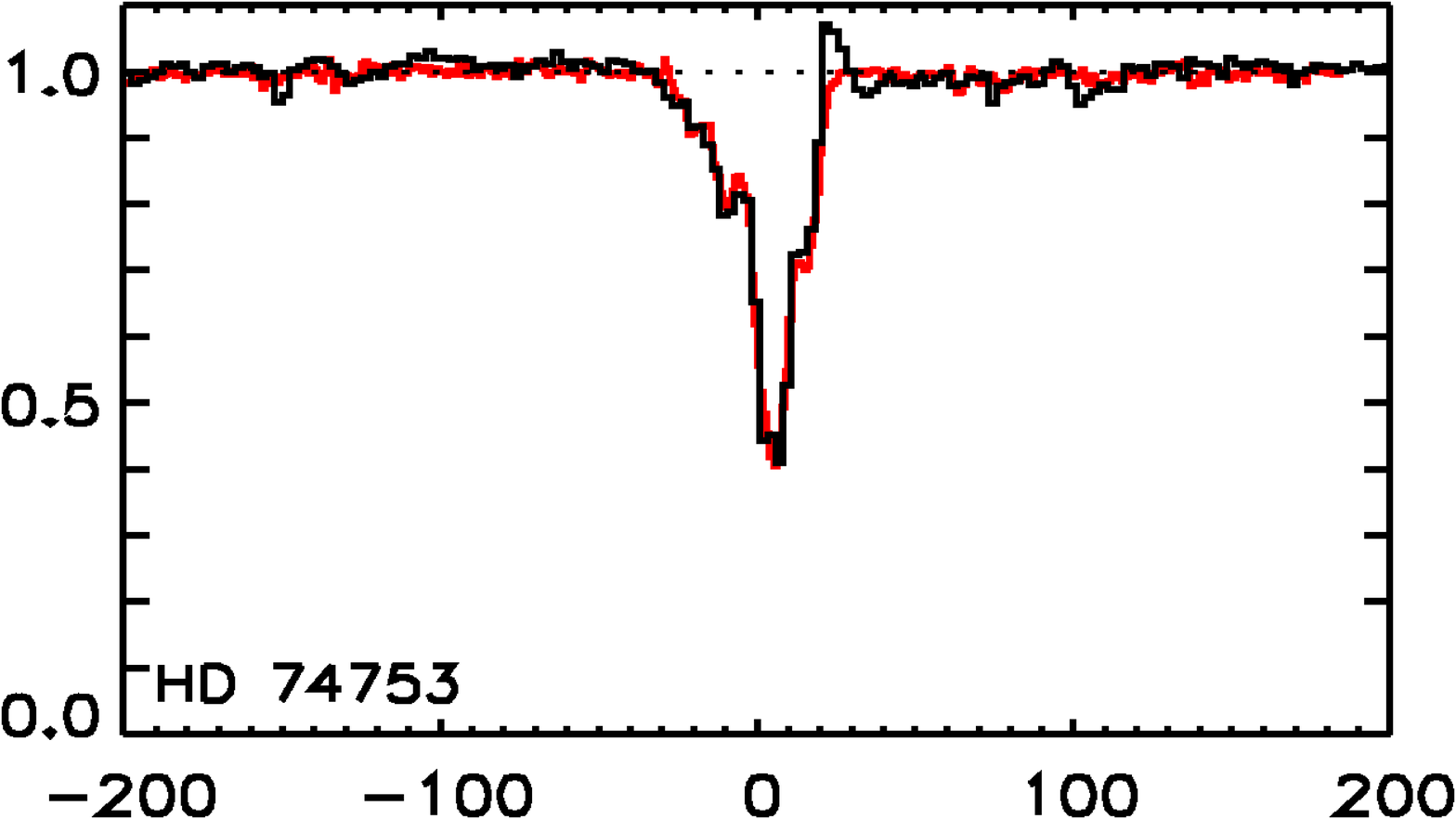}
\includegraphics[width=5.7cm,height=5cm]{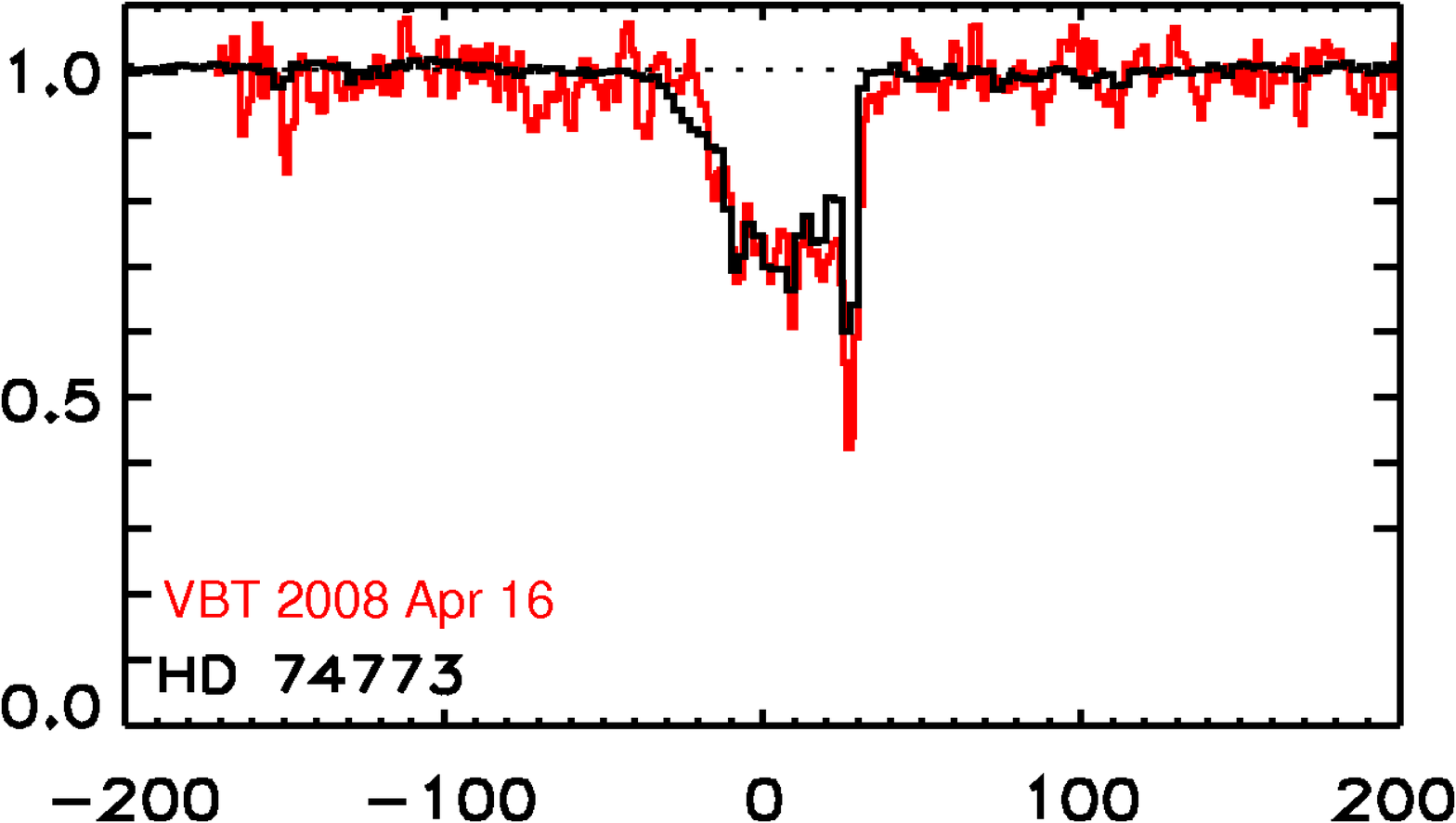}
\includegraphics[width=5.7cm,height=5cm]{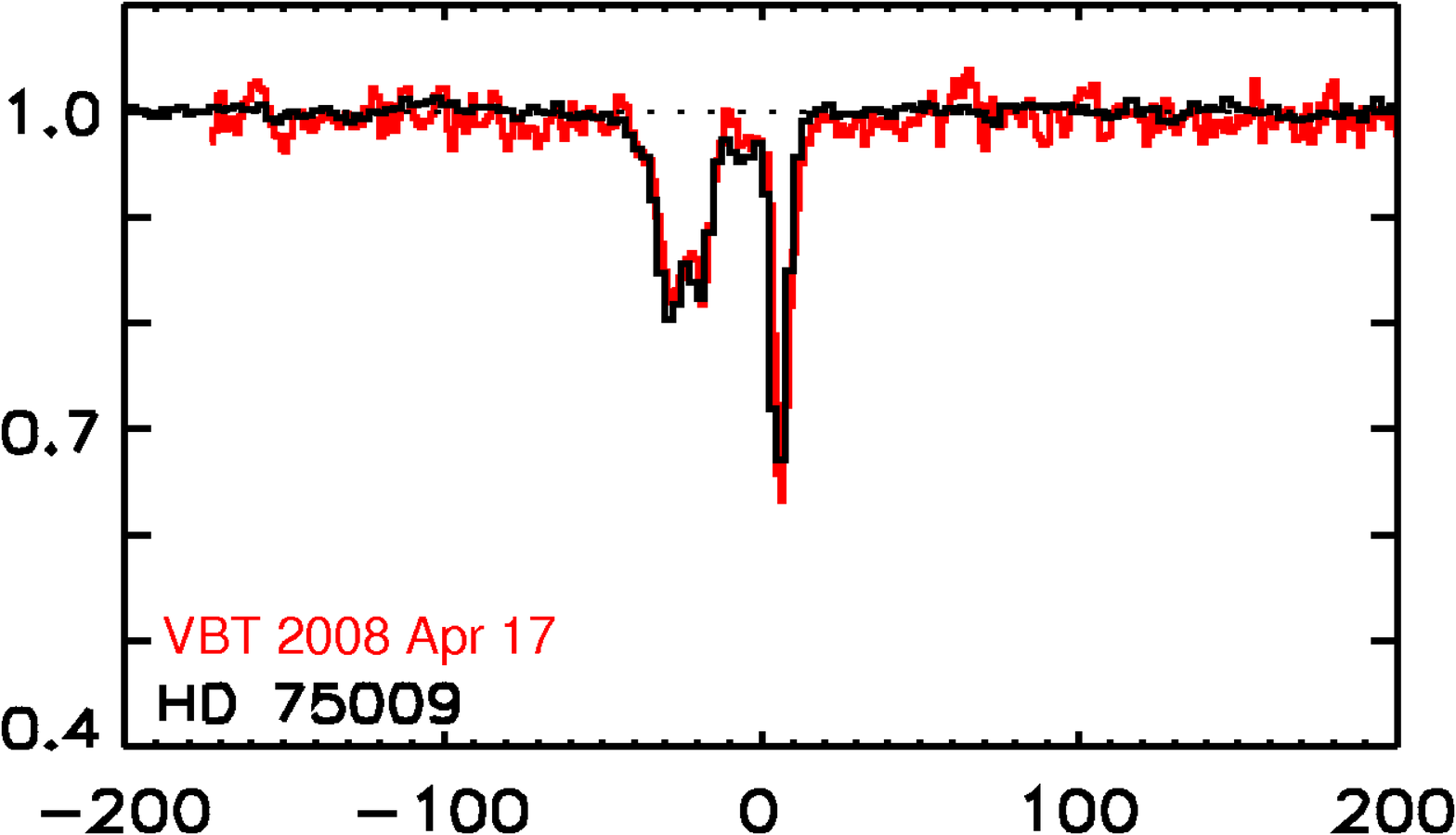}
\hspace{14cm}$V_{\rm LSR}$(km s$^{-1}$) \\
\caption{The Na\,{\sc i} D$_2$ profile of HD 72485, HD 72555, HD 72648, HD 73658, HD 74194, HD 74273,
HD 74455, HD 74531, HD 74711, HD 74753, HD 74773 and HD 75009  obtained by us in 2008 or
 2011 (red line)  is superposed on the Na\,{\sc i} D$_2$ profile obtained in 1993, 1994 or 1996
 (black line) by Cha \& Sembach (2000).}
\end{figure*}

\begin{figure*}
\vspace{0.3cm}
\rotatebox{90}{\hspace{1.2cm}Normalised Intensity}
\includegraphics[width=5.7cm,height=5cm]{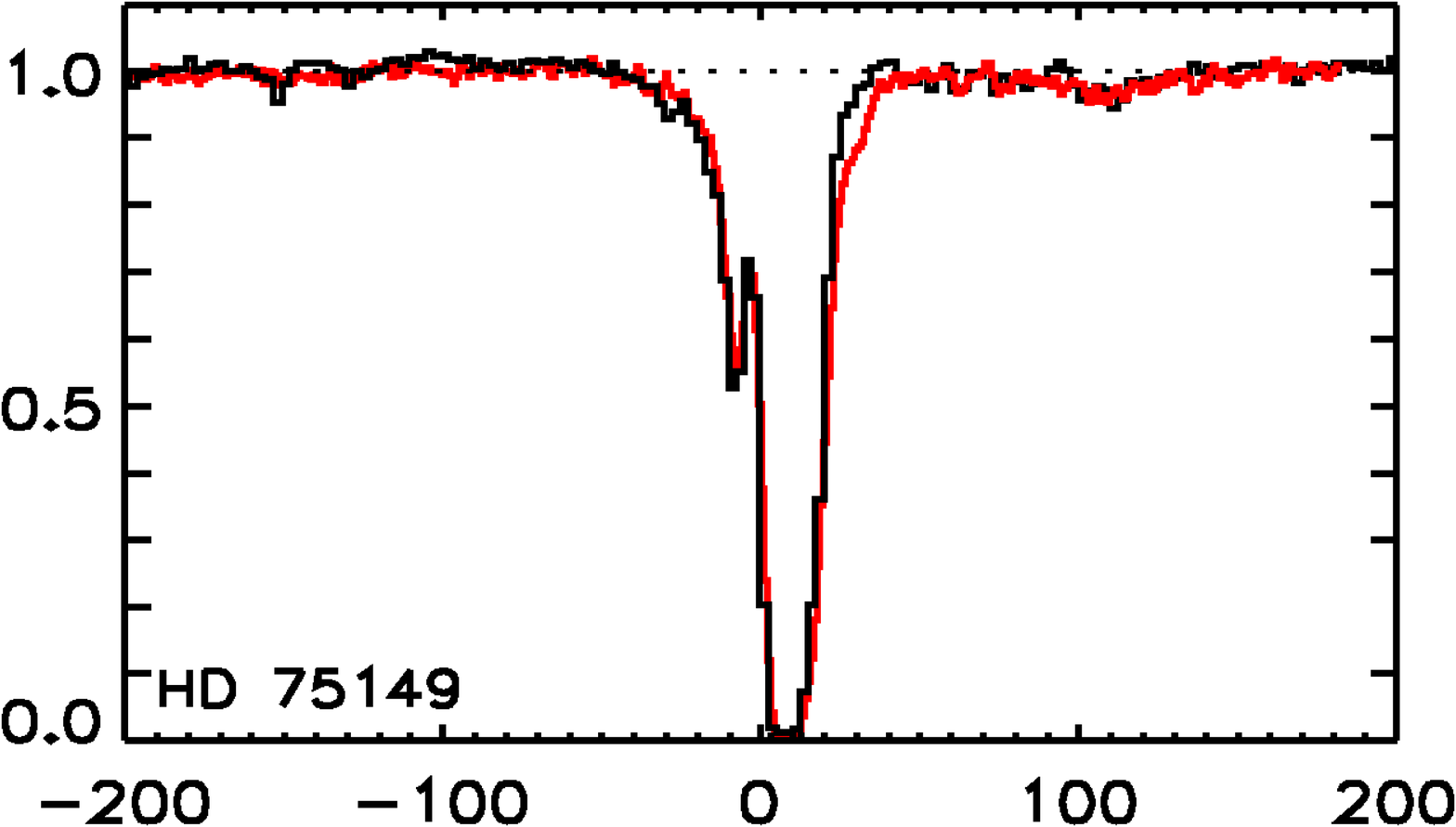}
\includegraphics[width=5.7cm,height=5cm]{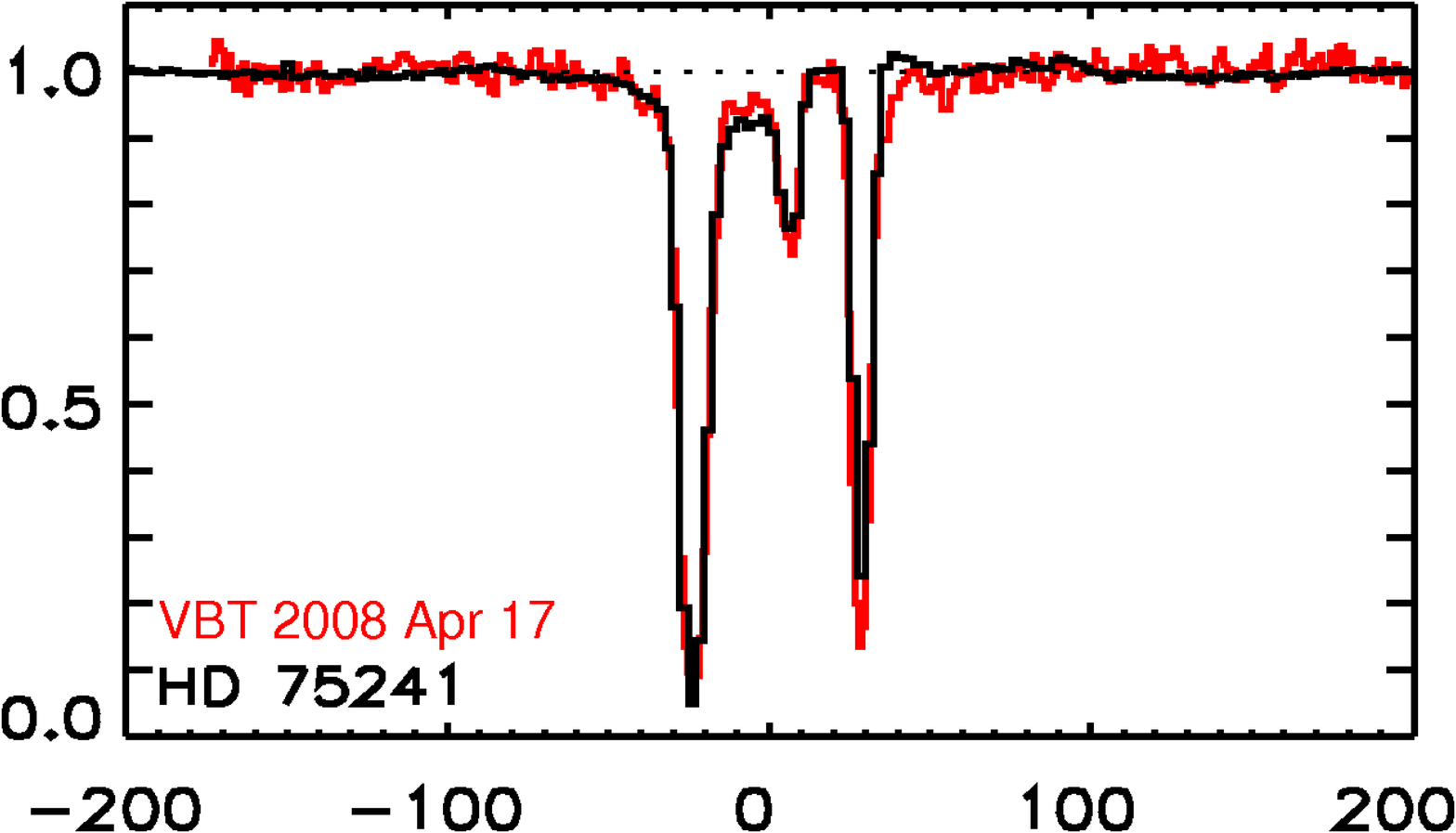}
\includegraphics[width=5.7cm,height=5cm]{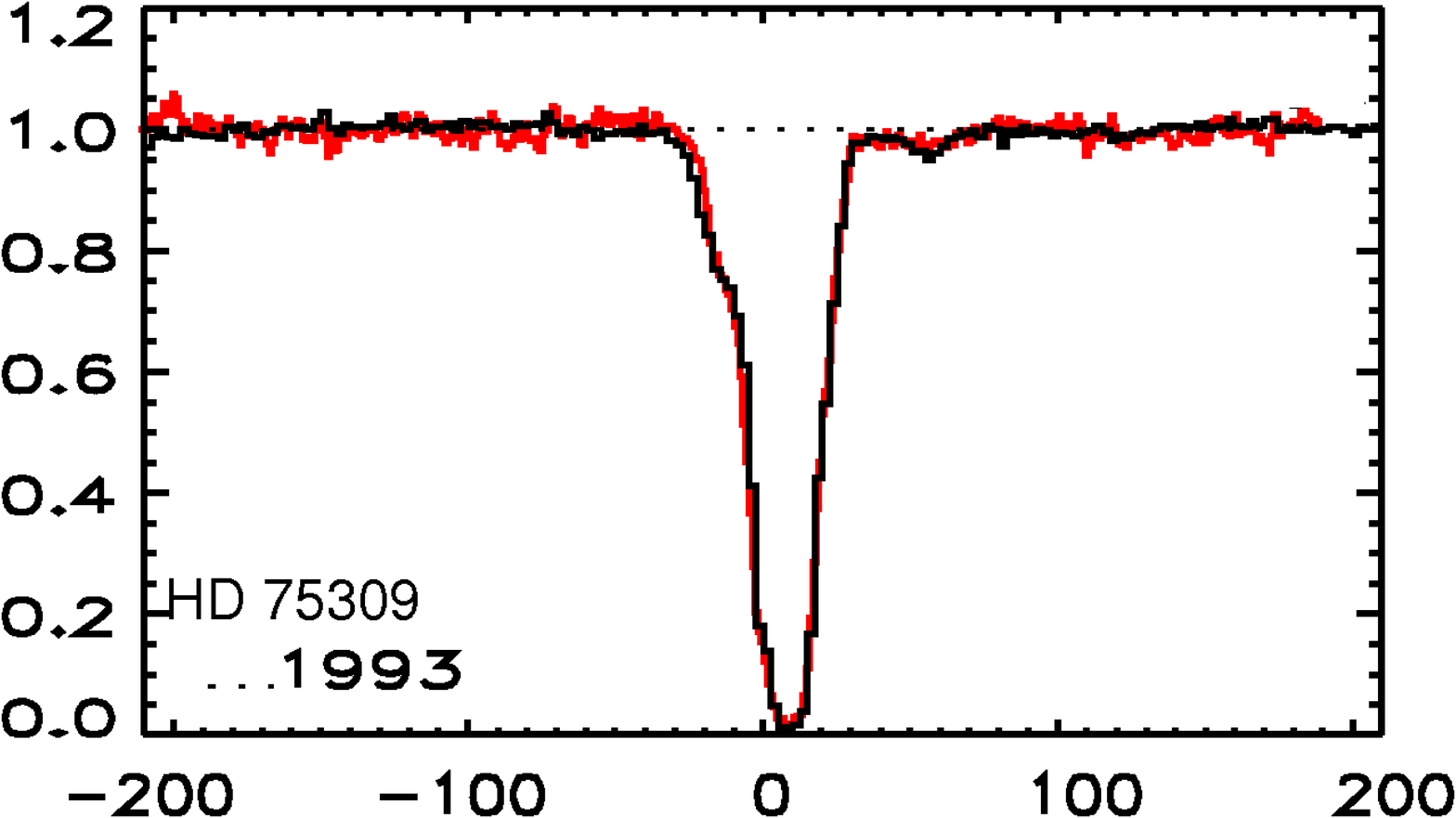}
\hspace{.5cm}
\vspace{.3cm}
\rotatebox{90}{\hspace{1.2cm}Normalised Intensity}
\includegraphics[width=5.7cm,height=5cm]{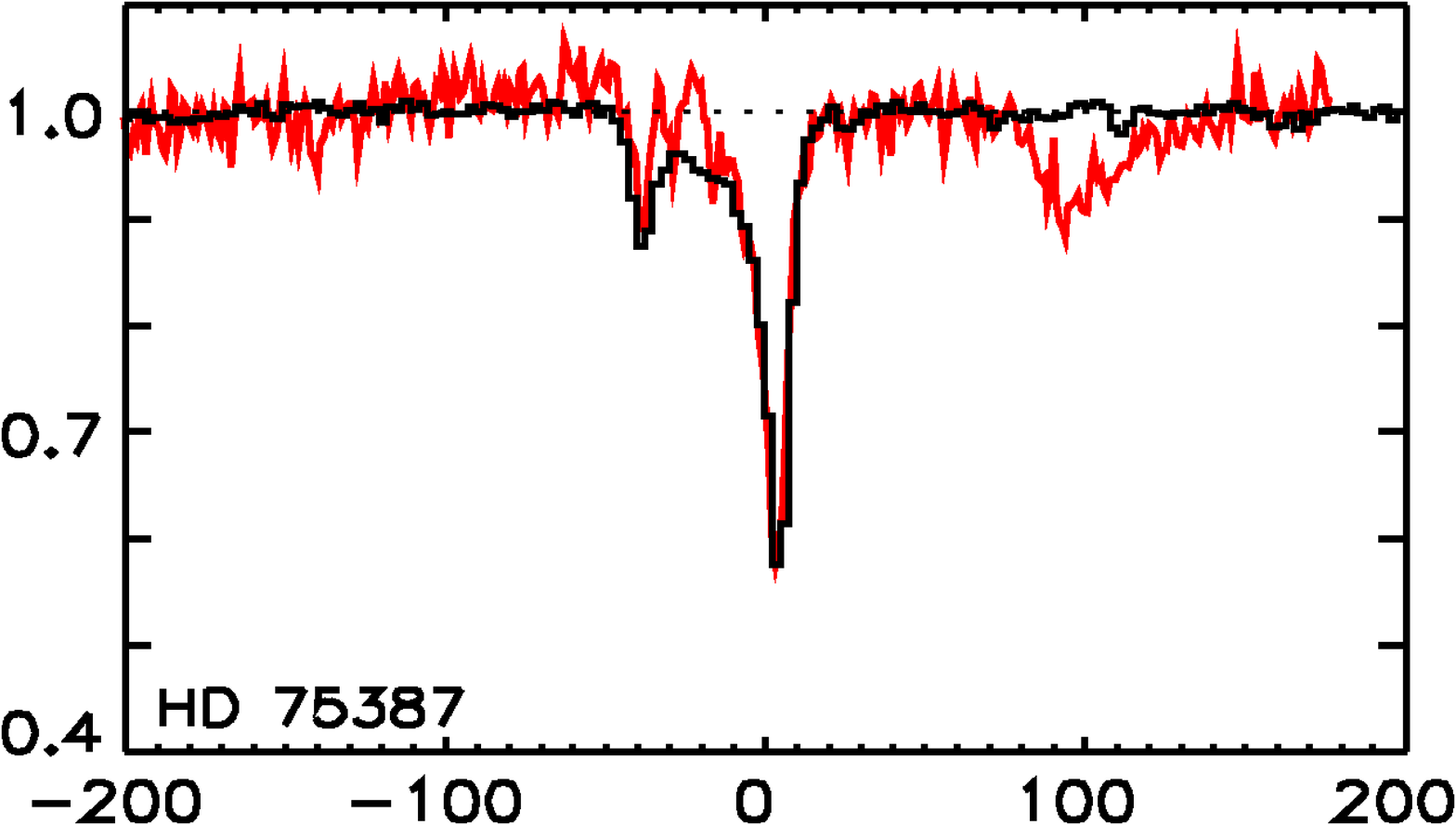}
\includegraphics[width=5.7cm,height=5cm]{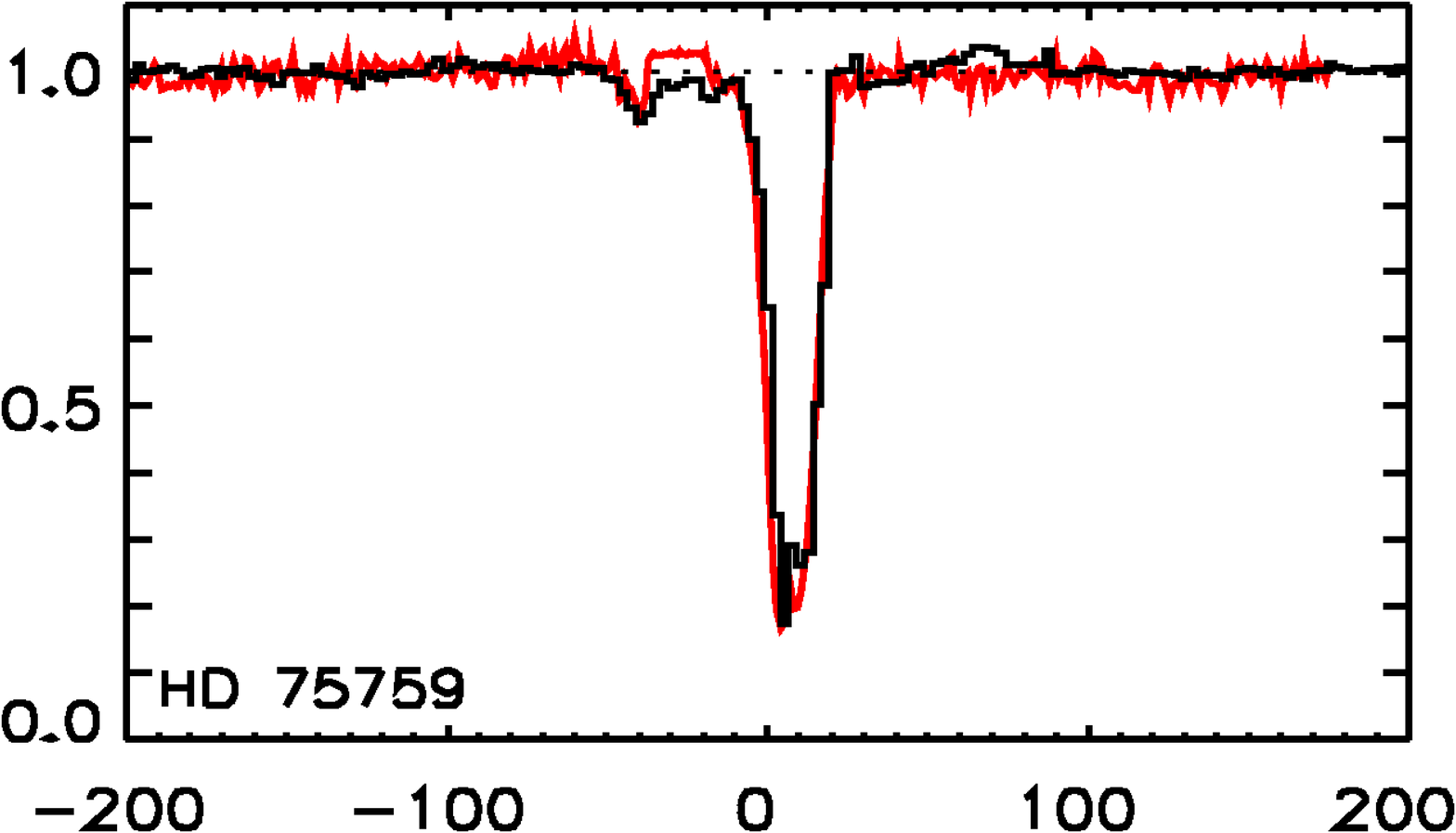}
\includegraphics[width=5.7cm,height=5cm]{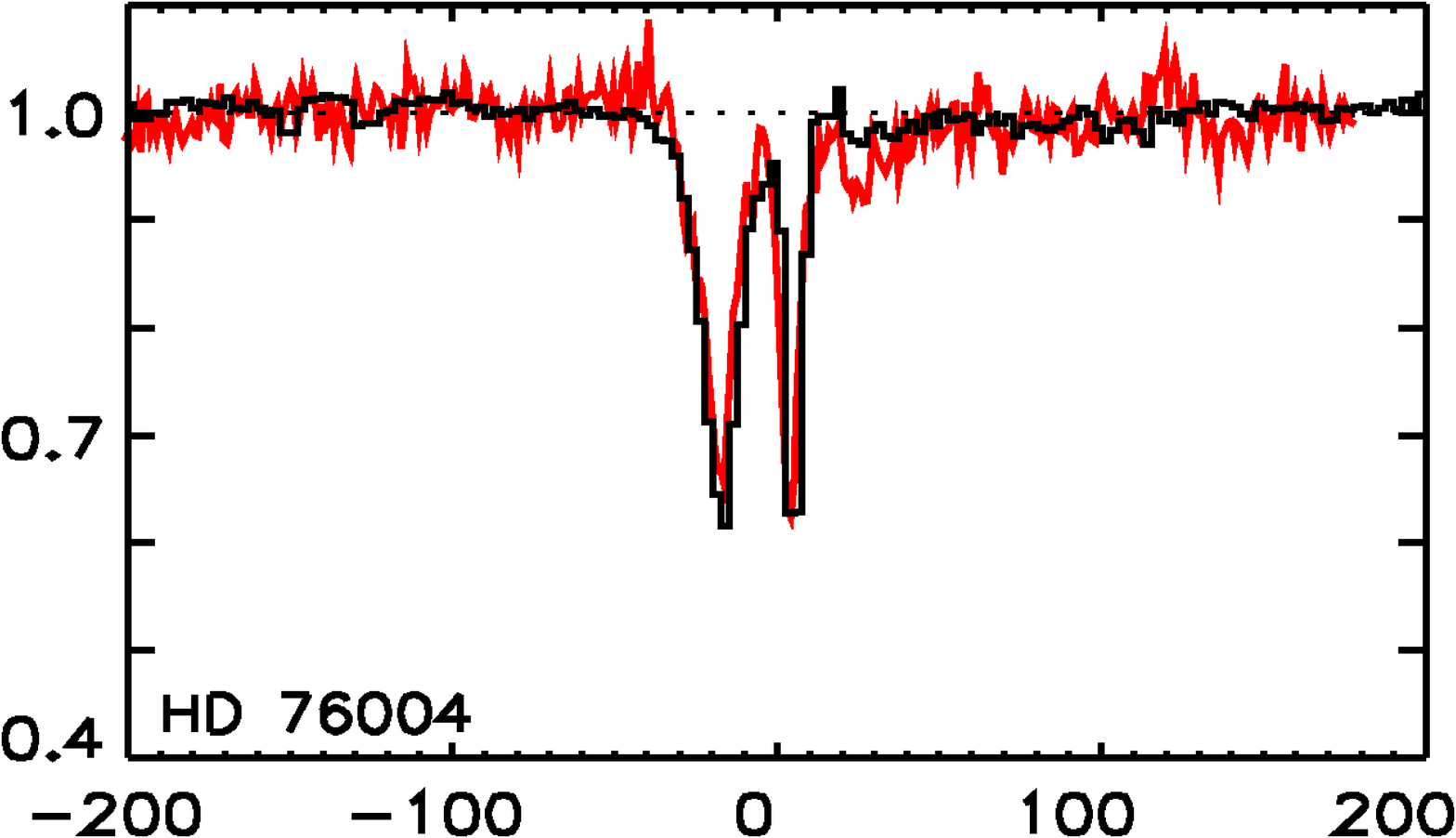}
\hspace{.5cm}
\vspace{-.1cm}
\rotatebox{90}{\hspace{1.2cm}Normalised Intensity}
\includegraphics[width=5.7cm,height=5cm]{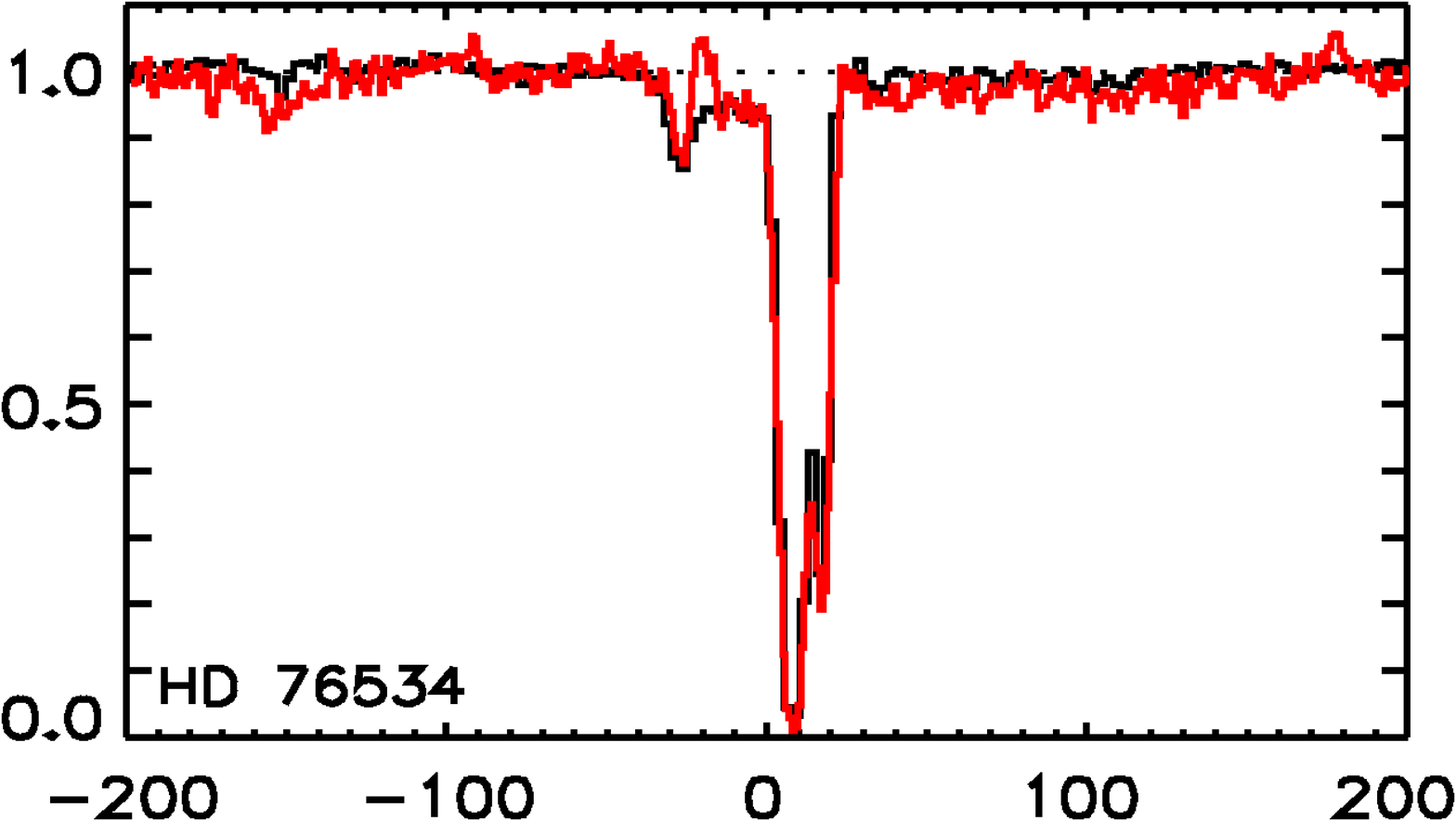}
\includegraphics[width=5.7cm,height=5cm]{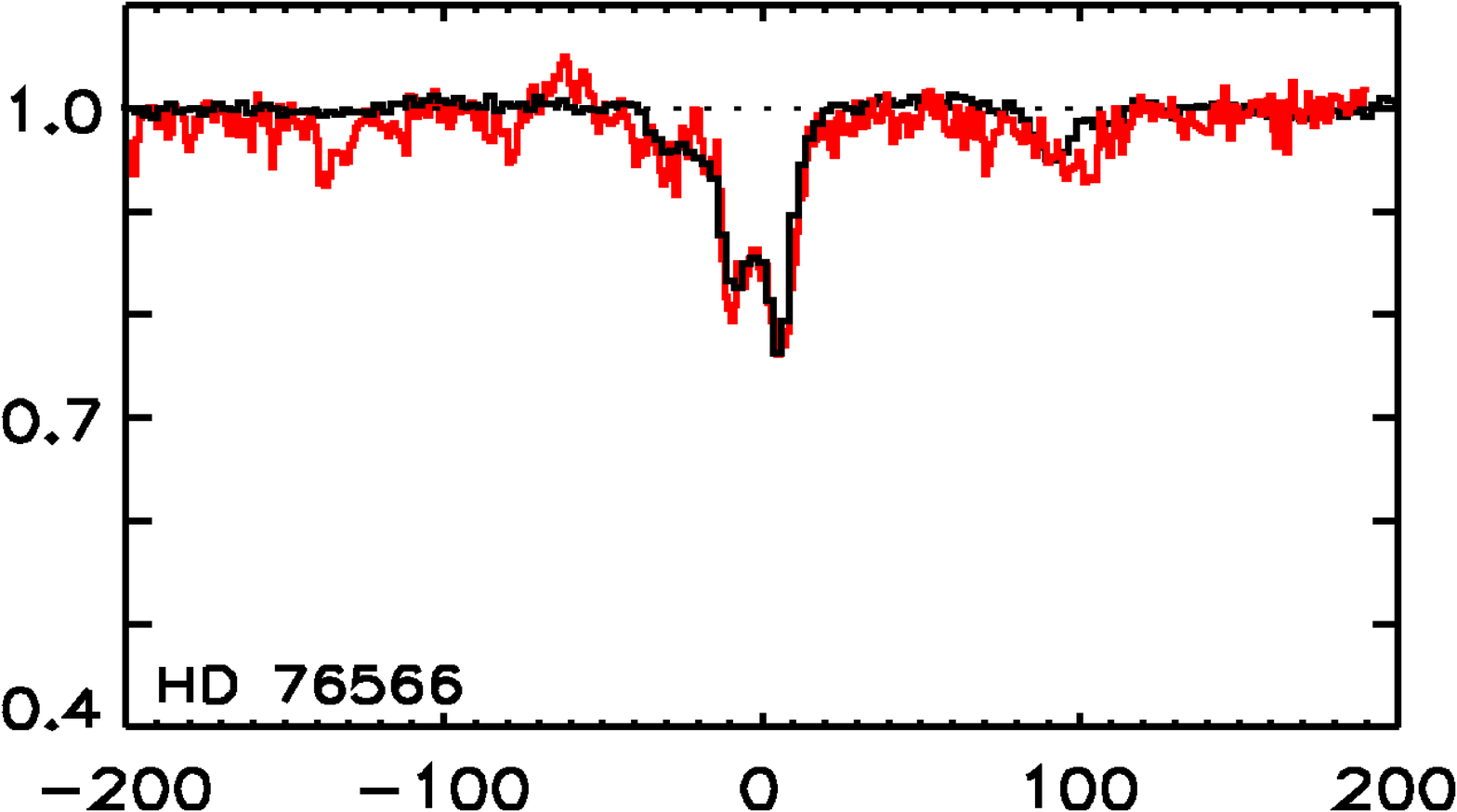}
\includegraphics[width=5.7cm,height=5cm]{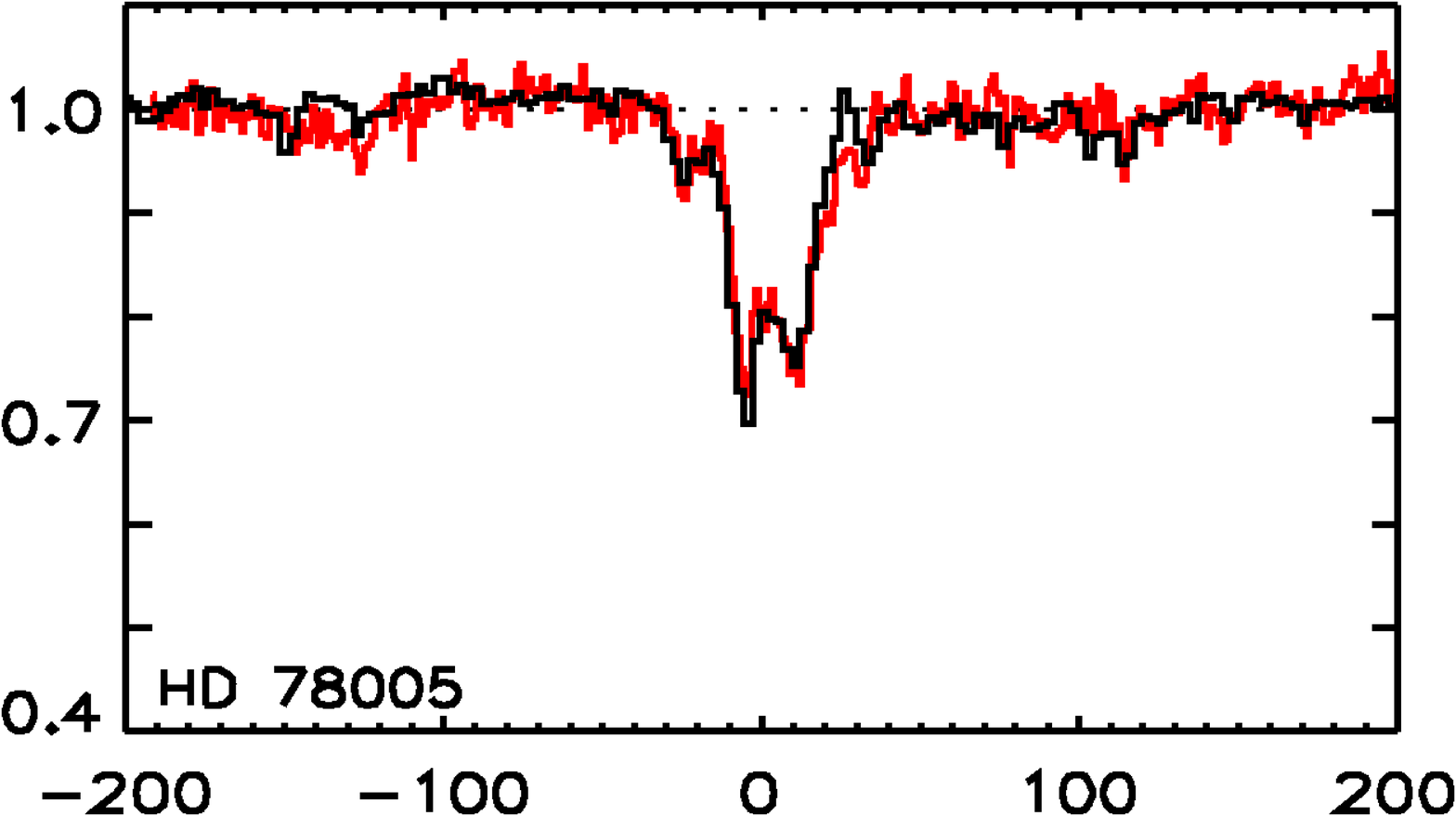}
\hspace{14cm}$V_{\rm LSR}$(km s$^{-1}$) \\
\caption{The Na\,{\sc i} D$_2$ profile of HD 75149, HD 75241, HD 75309, HD 75387, HD 75759, HD 76004,
 HD 76534, HD 76566, and HD 78005  obtained by us in
 2011-12 (red line)  is superposed on the Na\,{\sc i} D$_2$ profile obtained in 1993-94
 (black line) by Cha \& Sembach (2000). The VBT spectrum of HD 75241 was obtained in 2008.}
\end{figure*}

\begin{figure*}
\vspace{-.1cm}
\rotatebox{90}{\hspace{1.2cm}Normalised Intensity}
\includegraphics[width=7cm,height=6cm]{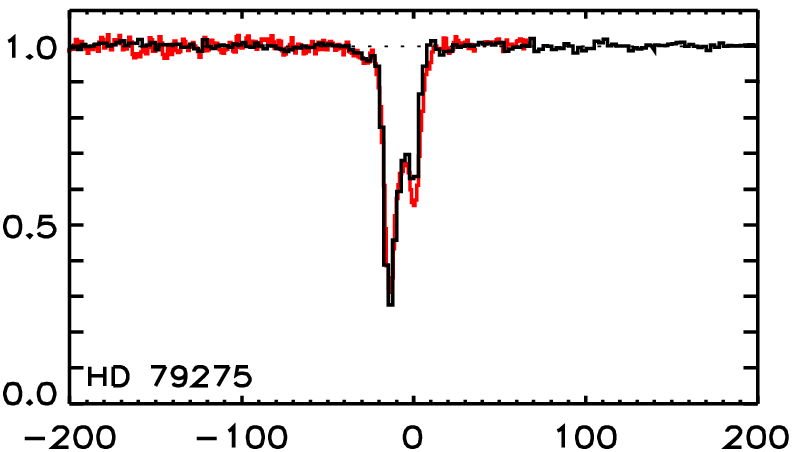}
\hspace{16cm}$V_{\rm LSR}$(km s$^{-1}$) 
\caption{The Na\,{\sc i} D$_2$ profile of HD 79275 obtained by us 
  (red line)  is superposed on the Na\,{\sc i} D$_2$ profile obtained in 1993-94
 (black line) by Cha \& Sembach (2000).}
\end{figure*}

\begin{figure*}
\vspace{0.3cm}
\rotatebox{90}{\hspace{1.2cm}Normalised Intensity}
\includegraphics[width=5.7cm,height=5cm]{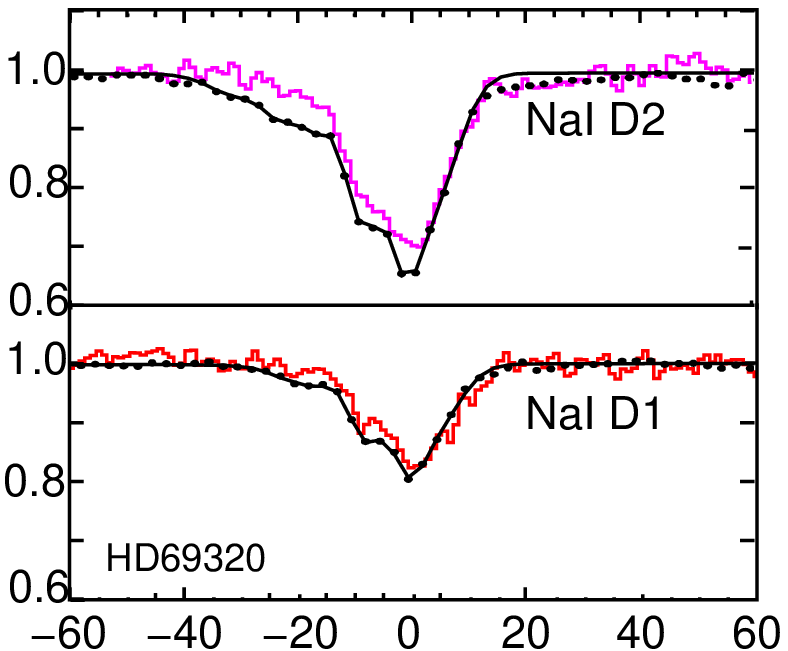}
\includegraphics[width=5.7cm,height=5cm]{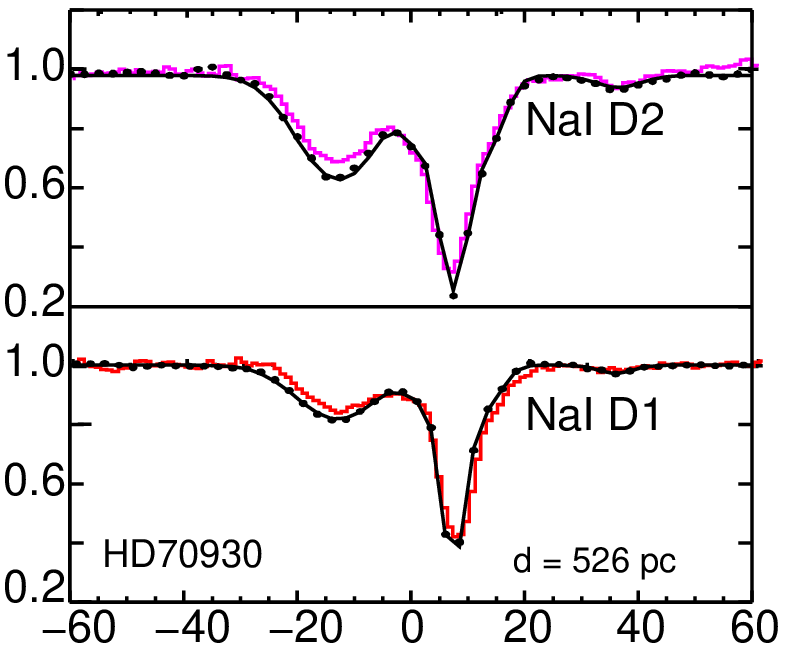}
\includegraphics[width=5.7cm,height=5cm]{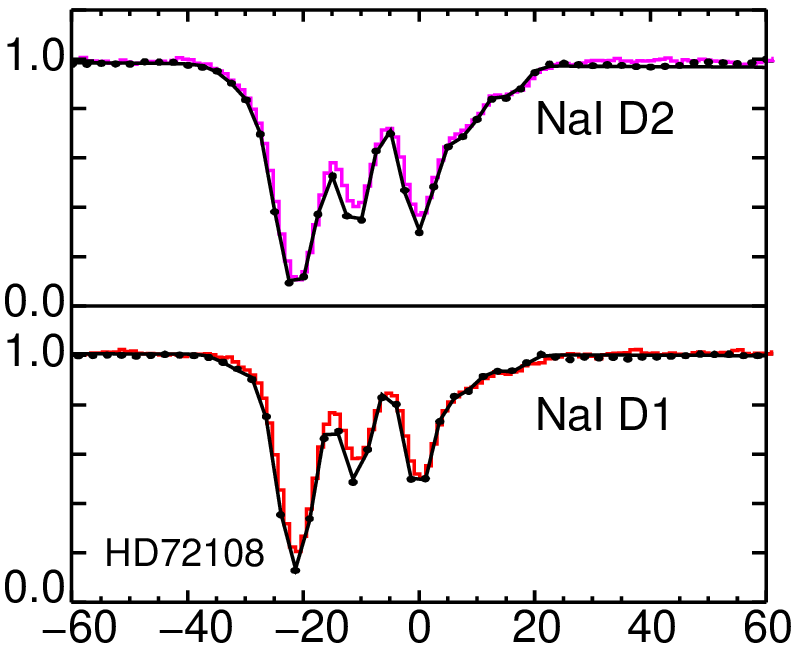}
\hspace{6cm}$V_{\rm LSR}$(km s$^{-1}$) \hspace{8cm}$V_{\rm LSR}$(km s$^{-1}$ \hspace{10cm}$V_{\rm LSR}$(km s$^{-1}$)\\
\caption{The  Na\,{\sc i} D$_1$ and D$_2$ profiles of HD 69302, HD 70930, and HD 72108
 obtained by us in
 2011-12 (red line) are superposed on the Na\,{\sc i} D$_2$ profile obtained in 1989 by Franco (2012). }
\end{figure*}

\subsection{ Na D then and now}
               
Thirty three stars  are judged to have Na D$_2$ profiles which appear to be unchanged between 1993-1996 and 2011-2012 (Figures 4, 5, 6, and 7).
Na D profiles for six stars observed in 1989 with the same telescope-spectrometer
combination  used several years later by Cha \& Sembach were published by Franco (2012).  Two of these six stars were observed neither by Cha \& Sembach
nor by us.  Two were observed by Cha \& Sembach at Na D and one of the pair (HD 68217) shows  changes in the low velocity Na D
absorption (see next subsection).  Three  of the six and  also HD 68217 were observed with the VBT.  Comparison of Franco's  (2012) and the VBT profiles for the three common
stars is given in Figure 8.
 The bulk of the Na D absorption is at low velocity and  likely produced by the
 intervening spiral arms:  here, low velocity corresponds to within $\pm25$ km s$^{-1}$ of $V_{\rm LSR}  = 0$ km s$^{-1}$ (Cha, Sembach \& Danks  1999).  
Of the illustrated stars, eight have high velocity Ca\,{\sc ii} components.  Of this octet, three were judged to have variable high-velocity components by Cha \& Sembach.  Similar 
profiles in 1993-1996 and 2011-2012 should not be taken to  mean that  differences at high velocity did not occur  in the intervening years; several of the observed changes in high-velocity
components occur with a short half-life (see below).

\begin{table*}
\centering
\begin{minipage}{160mm}
\caption{\Large Na\,{\sc i}  Absorption gaussian components towards some sight lines }
\begin{footnotesize}
\begin{tabular}{lcrrrcrrccrrrccrr}
\hline
   &\multicolumn{4}{c}{Cha \& Sembach }& &\multicolumn{4}{c}{Franco/Other}&&
\multicolumn{4}{c}{Present Observations$^a$}&  \\
\cline{2-5} \cline{7-10} \cline{12-15} \\
 Star   &\multicolumn{4}{c}{ Na\,{\sc i} D}&&\multicolumn{4}{c}{Na\,{\sc i} D}&&\multicolumn{4}{c}{Na\,{\sc i} D} \\
\cline{2-5} \cline{7-10} \cline{12-15} \\
  &   epoch&$V_{\rm LSR}$& Eq.w&Eq.w & & epoch&$V_{\rm LSR}$& Eq.w &Eq.w& & epoch&$V_{\rm LSR}$ & Eq.w &Eq.w&   \\
     &  & km s$^{-1}$ &(mA) &(mA) & &  & km s$^{-1}$ &(mA) &(mA) & & & km s$^{-1}$ &(mA) &(mA) & \\
\hline
HD 65814&  1993&1&267&138 & &    &    &  &  & &2011&0.2 &258&138&  \\
       &      &14&354&304& &    &    &  &  & &     &15.7&340&307&  \\
       &      &31&128&104& &    &    &  &  & &     &31.5&90&90& \\
HD 68243&  1994&-8&23 &10 & &    &    &  &  & &2011-12&-11.5&24&10& \\
       &      &-2&33 &25 & &    &    &  &  & &    &-2  &56 &33& \\
       &      &  &   &   & &    &    &  &  & &    &7.6 &12 & 7& \\
HD 70930&  1993&-10&131&66& & 1989&-12.9&129&65&&2011&-9.4&105&48& \\
       &      &   &   &  & &    &0.8&17&11& &     &1.9& 8 & 7 & \\
       &      & 9 &138&84& &    &7.3&102&75& &    &10.1&129&85& \\
       &      &  &   & &   &    &13.4&29&19& &    &18.0 &24 & 17& \\
       &      &  &   & &   &    &36.1&8&5  & &    &39.5 &11 & 4 & \\
HD 71302&  1994&-23&61&24&  &    &    & &   & &2011&-23.4&50 &27& \\
       &      &   &  &  &  &    &    & &   & &   &-7.1 &10 &6 & \\
       &      &  3&58&46&  &    &    & &   & &   & 2.9 &63 &46& \\
       &      &   &  &  &  &    &    & &   & &   &10.4 &10 &5 & \\
       &      &   &  &  &  &    &    & &   & &   &30.1 &12 &6 & \\
       &      &   &  &  &  &    &    & &  & &    &45.4 &17 & 7& \\
HD 71459&  1993&-14&24&$\leq$6&& &    &   &  &  & 2011-12&-11.5&18&8& \\
       &      &  6&60&32&  &    &    &   &  & &     &5.6 &60 &33& \\
HD 72067&  1994&-23&17&$\leq$6&&2001&-22&13&6& & 2011&-23.2& 8& 5& \\
       &      &-10&27&16 & &     &-11&25& 12& &     &-11.2&20&10& \\
       &      &  1&100&46& &     &-0 &85& 50& &     &  0.1&89&44& \\
       &      &   &   &  & &     & 8 &12& 6 & &     &  8  &12&6 & \\
       &      & 31& 23&13& &     &30 &21& 11& &     & 31 & 23&12& \\
HD 72089&  1993&-10&127&76& & 1996&-10&118&70& & 2011&-10.0&121&76& \\
       &      &  2& 89&47& &     &  2& 68&42& &     &  2.0& 90&50&\\
       &      & 11& 51&34& &     & 11& 52&33& &     & 11.0& 50&28&\\
       &      & 20& 77&43& &     & 20& 77&46& &     & 20.4& 65&39&\\
       &      & 75& 12& 6& &     & 75& 14& 6& &     & 71.6&  8&$\leq$6&\\
       &      &104& 56&30& &     &106& 23& 9& &     &107.9&  7&$\leq$4&\\
HD 72232&  1994&-29&11&$\leq$6&& &    &   &  & &2011&-29 & 4&2 & \\
       &      &  1&84&58&  &    &    &   &  & &   &0.1  &80&62& \\
       &      &   &  &  &  &    &    &   &  & &   & 8.7 &11&6 & \\
HD 74773&  1994&-17&43&16&  &    &    &   &  & &2008&-16.9&20&8& \\
       &      & -8&37&18&  &    &    &   &  & &    &- 9.4&36&18&\\
       &      &  2&63&39&  &    &    &   &  & &    &- 0.5&61&40&\\
       &      &  9&31&16&  &    &    &   &  & &    &  7.7&38&22& \\
       &      & 18&53&36&  &    &    &   &  & &    & 16.9&60&36& \\
       &      & 27&47&38&  &    &    &   &  & &    & 26.3&66&45& \\
HD 79275&  1994&-29&18&$\leq$6&& &    &   &  & &2011-12&-31.4&11&$\leq$3&\\
       &      &-14&129&72& &    &    &   &  & &    &-15.2&125&76& \\
       &      &   &   &  & &    &    &   &  & &    & -8.0&   &  & \\
       &      & -2& 95&50& &    &    &   &  & &    & -1.7&100&45& \\
       &      &102& 12&$\leq$6&&&    &   &  & &    & 97.9$^{b}$&10 &$\leq$2&\\
\hline
\end{tabular}
\\
$^a$       The $V_{\rm LSR}$ is an average of  $D_{\rm 2}$ and
      $ D_{\rm 1}$. \\
$^{b}$The feature seems to be superposed on a stellar line of C\,{\sc ii} at 5891.59\AA. The
stellar as well as the ISM features are assumed to have gaussian profiles for
 estimating the equivalent widths.  \\

\end{footnotesize}
\label{default}
\end{minipage}
\end{table*}

\begin{figure*}
\includegraphics[width=7cm,height=7cm]{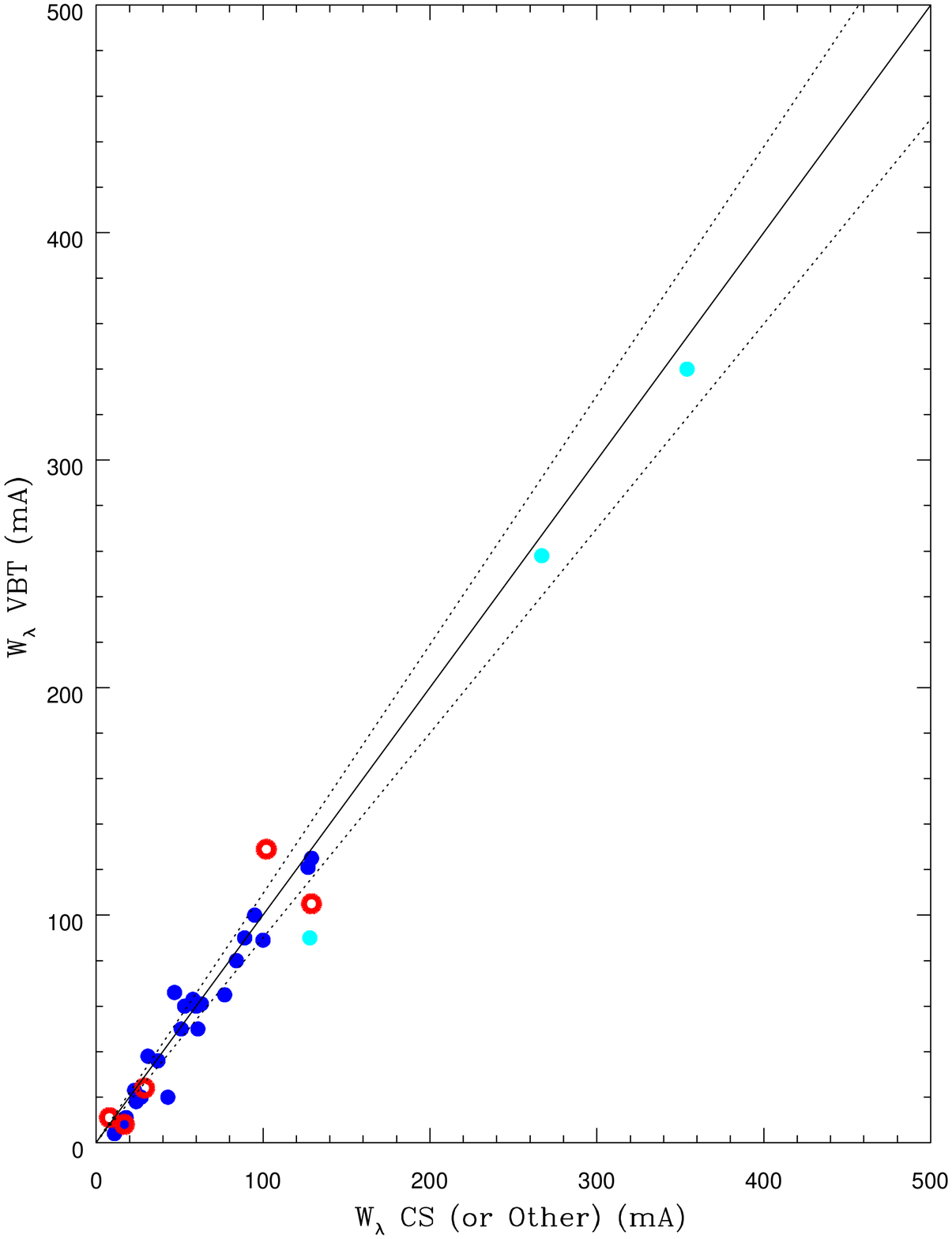}
\caption{ Equivalent widths of low-velocity components of the Na\,{\sc i} D$_2$ profiles  obtained by Cha \& Sembach (2000) or Franco (2012) versus the equivalent widths for the same
components provided by VBT spectra (see Table 3). The line corresponds to equal equivalent widths. Dashed lines represent a 10 percent variation.  Red dots refer to HD 70930 (Franco 2012), 
 and cyan dots refer to HD 65814.}
\end{figure*}

Inspection of Figures 4-8 shows examples where the profiles from then and now are essentially indistinguishable pixel by pixel across the low velocity
complex of blends.  In some cases, small differences occur in the deep line cores and/or wings.  Other apparent differences occur at
high velocity but none are appreciably greater than expected from signal-to-noise ratio differences; in general, the VBT spectra are of a lower
(but high) S/N ratio than Cha \& Sembach's spectra.  Additionally, subtle differences in spectra will arise from differences in the
instrumental profiles, reduction techniques and correction for the telluric H$_2$O absorption lines.  Table 3 summarizes the breakdown of the Na D$_2$
profiles into gaussian components.
In Figure 9, equivalent widths of low velocity
components measured by Cha \& Sembach or Franco are compared with values from the VBT spectra.  With the exception of the low velocity
components for the four stars discussed below (and not plotted in Figure 3) ,   Figure 9 shows that components  have maintained a constant equivalent width over the
multi-year baseline.
 The not unexpected constancy of the majority of the Na D
absorption profiles  from 1993-1994 to
 2011-2012 contrasts sharply with the dramatic and rare examples of  strong low velocity absorption
which disappeared almost completely between 1993-1994 and 2011-2012, as discussed next.  Changes in high-velocity absorption components
were discovered by Cha \& Sembach and are  commented upon below

\subsection{Sight lines with a large change of neutral sodium column density }

Interstellar clouds providing the low velocity Ca\,{\sc ii} and Na D components  are part of the diffuse interstellar medium and generally
associated with the spiral arms between us and the target star.   Clouds close to the Vela supernova  may be expected to have their
physical conditions altered by the supernova and its remnant by radiation or collisions.  At Na D, Cha \& Sembach found no  example
of a star exhibiting large changes at low velocities and only two examples  (HD 72127A and HD 73658) of  modest changes in the Ca\,{\sc ii} K line profiles.
Now, four examples of  unprecedented change in the  interstellar Na D profiles in less than two decades  may be drawn from our sample.  This quartet  -- three
examples of a dramatic weakening and one of a modest strengthening of a Na column density between 1993-1994 and 2011-2012 -- surely
indicate the  affected clouds' close proximity to the supernova or another source of an energetic outflow.

{\bf HD 63578:}  The Na D profile in 2011-2012 is strikingly different from that obtained by Cha \& Sembach in 1993 (Figure 10 ):   a component at $-4$ km s$^{-1}$
and a second and weaker component at $-25$ km s$^{-1}$ in 1993 are considerably weakened by 2011.  Four observations between 2011 January 8 and 2012 November 14
provide identical Na D profiles.  The 1993 Ca\,{\sc ii} K profile (Cha \& Sembach 2000) consists of two components with the stronger one coincident in
velocity with the Na D component at $+6$ km s$^{-1}$ and the other at $-14$ km s$^{-1}$  may be coincident with the Na D components which decreased
greatly in strength between 1993 and 2011 (Table 4).  The Ca\,{\sc ii} K line appears not to have been reobserved since 1993.
 The 1993 Ca\,{\sc ii} K profile is quite similar to that reported by Wallerstein, et al.  (1980) from
observations between 1971 and 1977.  Na D observations of Wallerstein et al.  include the stronger component in the 1993 and 2011-2012 spectra but are probably
of inadequate quality to determine if the blue-shifted  1993 component  was present between 1971 and 1977.   In summary, HD 63578 is an example
where the largest change in Na D profile occurs among the low velocity clouds.  Neither the  available Ca\,{\sc ii} K nor the Na D profiles have shown high-velocity
components.

\begin{figure*}
\includegraphics[width=7cm,height=6cm]{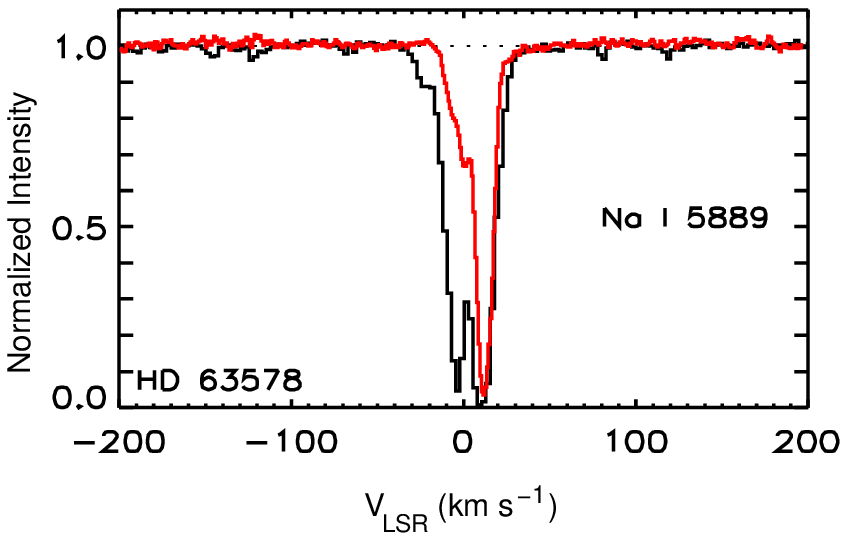}
\caption{ The Na\,{\sc i} D$_2$ profile of HD63578 obtained by us in
 2011 (red line)  is superposed on the Na\,{\sc i} D$_2$ profile obtained in 1993
 (black line) by Cha \& Sembach (2000). Note the absence of the blue-shifted
 absorption components in the 2011 spectrum (red line) that were present (marked   with arrows) in the 1993 spectrum.}
\end{figure*}



\begin{table*}
\centering
\begin{minipage}{170mm}
\caption{\Large HD 63578 Ca K and Na\,{\sc i} Absorption Lines  }
\begin{footnotesize}
\begin{tabular}{lcrrrcrrccrrrccr}
\hline
   \multicolumn{7}{c}{Cha \& Sembach }&& &
\multicolumn{7}{c}{Present Observations} \\
\cline{1-9}  \cline{11-15}  \\
    &  & Ca\,{\sc ii} K &    &  & & &$D_{\rm 2}$&$ D_{\rm 1}$ & & & &$ D_{\rm 2}$& $D_{\rm 1}$ \\
\cline{1-4} \cline{6-9} \cline{11-14}  \\
  epoch&$V_{\rm LSR}$& Eq.w & & &epoch &$V_{\rm LSR}$&Eq.w&Eq.w&  &
epoch& $V_{\rm LSR}$& Eq.w &Eq.w& & \\
     & km s$^{-1}$ &(mA) &  & & &km s$^{-1}$ &(mA)& (mA)&  &   &km s$^{-1}$& (mA)&(mA)&&   \\
\hline
 1993 &-14 & 24 &   &  & 1993 &    &    &   &  & 2011-12&-13.2 &31   &15 &  \\
      &    &    &   &  &      & -4 &223 &178&  &        &- 6.4 &39   &24 &   \\
      & 6  & 34 &   &  &      &    &    &   &  &        &  5.8 &232  &191 & \\
      &    &    &   &  &      & 11 &325 &270&  &        &      &     &    & \\
      &    &    &   &  &      &    &    &   &  &        & 22.0 & 5   &    & \\
\hline
\end{tabular}
\\
\end{footnotesize}
\label{default}
\end{minipage}
\end{table*}

{\bf HD 68217:}   This star 's profiles of interstellar Ca\,{\sc ii} K and Na D, as illustrated by Cha \& Sembach (2000), resemble those of  HD 63578 and, in 
addition, there is a close parallel in the Na D profile's changes between 1993-1994 and 2011-2012.  Figure 11  compares the Na D$_2$ profiles from the
two epochs.  An additional data point on the Na D variations is provided by the 1989 April Na D profiles shown by Franco (2012) which shows that the
blue-shifted components at $-18$ and $-8$ km s$^{-1}$ were slightly weaker in 1989 than in 1994.  A VBT spectrum of lower quality obtained in 2007 February is similar to those
from 2011 and 2012. Thus, the  prominent $-8$ km s$^{-1}$ component   which had strengthened between 1989 and 1994 had largely disappeared by 2007. 
 Table 5 summarizes the results of  Gaussian decomposition of the Na D
profiles.  The Ca\,{\sc ii} K profile closely resembles that for HD 63578 except that the blue-shifted component for HD 68217 is the stronger than the
red-shifted component but the reverse is found for HD 63578.  For both stars, there is an absence of high-velocity components in the Ca\,{\sc ii} K and Na D
profiles.


\begin{figure*}
\vspace{0.0cm}
\includegraphics[width=6cm,height=6cm]{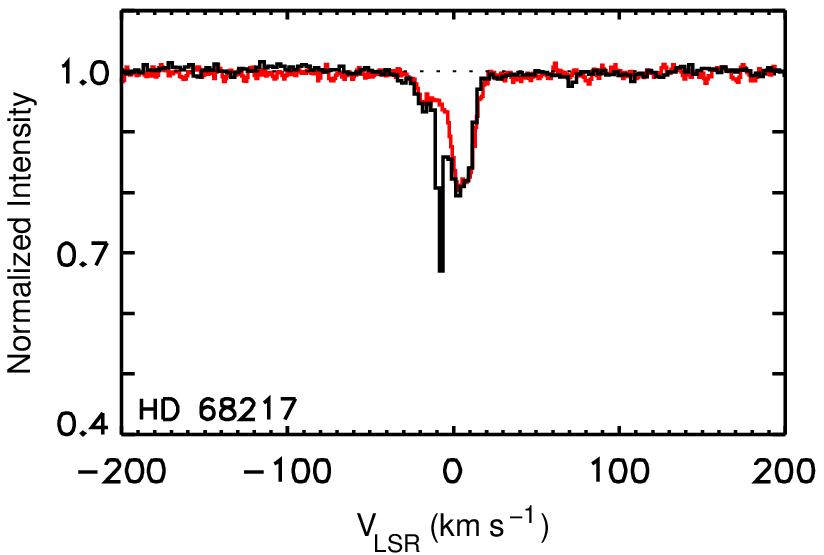}
\vspace{0.3cm}
\includegraphics[width=6cm,height=7cm]{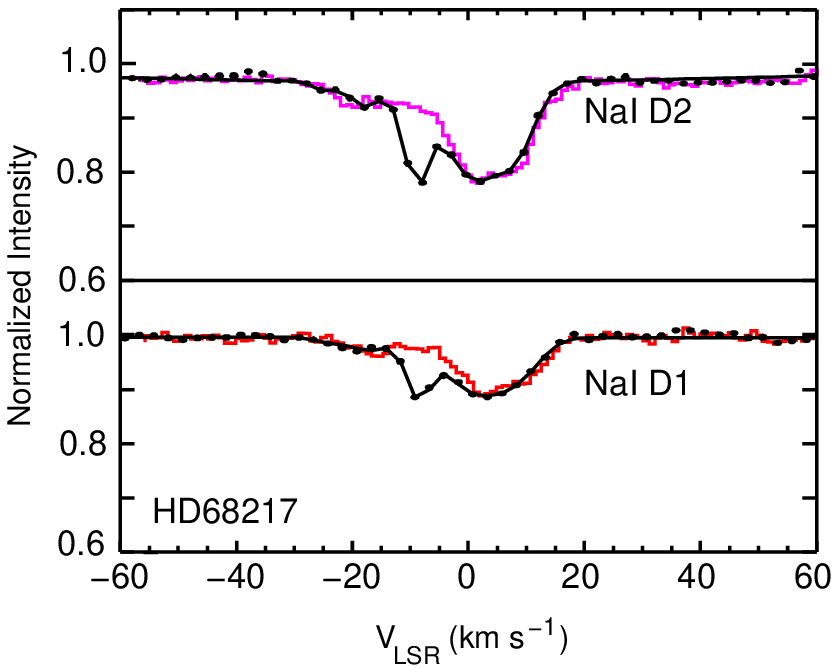}
\caption{The  Na\,{\sc i} D$_2$ profile of HD 68217 obtained by Cha \& Sembach
 in 1994 (black line)   superposed on the profile of D$_2$ obtained 
 by us on  2012 January 16 (red line). Note the strong absorption 
 component at $V_{\rm LSR}$ -9 km s$^{-1}$ present in 1994  is 
greatly weakened in our 2012 profile.
In the righthand two panels,  Na\,{\sc i} D$_2$,and D$_1$   profiles
 obtained in 1989 April by Franco (black lines) are compared with our
 2012 January 16 profiles (magenta -D$_2$ ; red =D$_1$ ). 
 Note the presence of a strong absorption component  at -8 km s$^{-1}$ in the1989  and 1994 profiles that is
 missing in the 2012 profiles. }
\end{figure*}

\begin{table*}
\centering
\begin{minipage}{170mm}
\caption{\Large HD68217 Na\,{\sc i} Absorption Lines  }
\begin{footnotesize}
\begin{tabular}{lcrrrcrrccrrrccr}
\hline
\multicolumn{1}{c}{}&\multicolumn{3}{c}{Franco}&\multicolumn{1}{c}{}&\multicolumn{1}{c}{}&
\multicolumn{3}{c}{Cha \& Sembach} &\multicolumn{1}{c}{} &\multicolumn{3}{c}{Present Observations$^a$}&\multicolumn{1}{c}{}\\
\cline{1-4} \cline{6-9} \cline{11-15}  \\
    &  &  $D_{\rm 2}$&$ D_{\rm 1}$ &  & & &$D_{\rm 2}$&$ D_{\rm 1}$ & & & &$ D_{\rm 2}$& $D_{\rm 1}$ \\
\cline{1-4} \cline{6-9} \cline{11-14}  \\
  epoch&$V_{\rm LSR}$& Eq.w &Eq.w&&epoch &$V_{\rm LSR}$&Eq.w&Eq.w&  &
epoch& $V_{\rm LSR}$ & Eq.w &Eq.w& & \\
     & km s$^{-1}$ &(mA) &(mA) & & &km s$^{-1}$ &(mA)& (mA)&  &   &km s$^{-1}$& (mA)&(mA)&&   \\
\hline
 1989 &-17.7 &6.2 &4.4 &  & 1994 &-18 &17&$\leq$ 6 &  &2011-12&-18.9 &9 &5 &comp 4 \\
 April&-9.0  &19.9&10.9 & &      &-8 &30 &20 &  &             &-8.4 &9.5&4.6&comp 3   \\
      & 0.8 &42.1&24.5 &  &      & 2 &43 &12 &  &             &1.5 &41 &21.5 &comp 2 \\
      & 8.5 &17.8&10.9 &  &      & 9 &20 &16 &  &             &9.0 &21 &10 &comp 1 \\
\hline
\end{tabular}
\\
$^a$ Average of four nights observations obtained on 2011 January 7, 2011 February 19, 2012
       January 16 and 2012 November 14. The $V_{\rm LSR}$ is an average of  $D_{\rm 2}$ and
      $ D_{\rm 1}$. \\  
\end{footnotesize}
\label{default}
\end{minipage}
\end{table*}

{\bf HD 76161:}   This star provides the  third example in which very strong Na D low velocity absorption seen in 1993 had very greatly weakened by 2011 - see
Figure 12: the 1993 saturated absorption  at about 9 km s$^{-1}$ has almost completely vanished by 2011. Weaker absorption at about $-9$ km s$^{-1}$ seems
unchanged over the 1993-2012 baseline. A weaker component at $+27$ km s$^{-1}$ in 1993 also weakened considerably by 2012.
  The 1993 Ca\,{\sc ii} K profile is almost identical to that of the Na D$_2$ 2011 profile.   The Ca\,{\sc ii} K profile reported by Wallerstein et al. (1980)
was very similar to that seen in 1993 by Cha \& Sembach (2000).
Table 6 summarizes the Gaussian fits.  High-velocity
features are absent from all 1993 and 2011 spectra.

\begin{figure*}
\includegraphics[width=6.5cm,height=6cm]{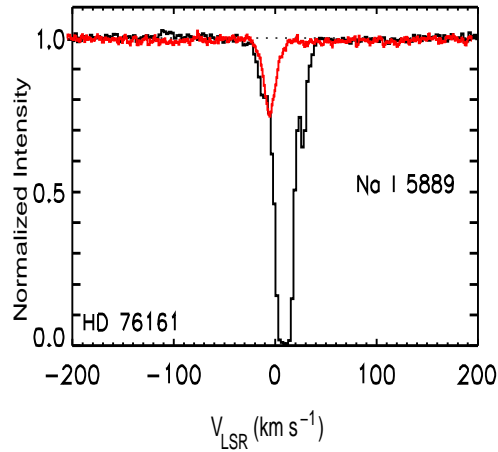}
\caption{The Na\,{\sc i} D$_2$ profiles of HD 76161 obtained on 2011 December 25
(red line) superposed on the profile obtained by Cha \& Sembach (2000)
 (black line) in 1993. }
\end{figure*}

\begin{table*}
\centering
\begin{minipage}{170mm}
\caption{\Large HD76161 Na\,{\sc i} and  Ca K absorption Lines  }
\begin{footnotesize}
\begin{tabular}{lcrrrcrrccrrrccr}
\hline
\multicolumn{1}{c}{}&\multicolumn{3}{c}{Cha \& Sembach}&\multicolumn{1}{c}{}&\multicolumn{1}{c}{}&
\multicolumn{3}{c}{Cha \& Sembach} &\multicolumn{1}{c}{} &\multicolumn{3}{c}{Present Observations$^a$}&\multicolumn{1}{c}{}\\
\cline{1-4} \cline{6-9} \cline{11-15}  \\
    &  & Ca\,{\sc ii} K &   &  & & &$D_{\rm 2}$&$ D_{\rm 1}$ & & & &$ D_{\rm 2}$& $D_{\rm 1}$ \\
\cline{1-4} \cline{6-9} \cline{11-14}  \\
  epoch&$V_{\rm LSR}$& Eq.w &  &  &epoch &$V_{\rm LSR}$&Eq.w&Eq.w&  &
epoch& $V_{\rm LSR}$& Eq.w &Eq.w& & \\
     & km s$^{-1}$ &(mA) &  & & &km s$^{-1}$ &(mA)& (mA)&  &   &km s$^{-1}$& (mA)&(mA)&&   \\
\hline
 1993 & -8  & 16  &   &  & 1993 & -6  &94 &32 &  &2011 &-8.9 &6 & 3  &comp 1 \\
      &  2  & 38  &   &  &      &     &   &   &  &     & 1.0 &60 &32 &comp 2  \\
      &     &     &   &  &      &  9  &395&365&  &     &     &   &   &       \\
      &     &     &   &  &      &     &   &   &  &     &12.6 &8 & 4  &comp 3 \\
      &     &     &   &  &      & 27  & 61&32 &  &     &     &  &    &    \\
\hline
\end{tabular}
\\
$^a$ :  Observations obtained on 2011 December 25.
       The $v_{\rm LSR}$ is an average of  $D_{\rm 2}$ and
      $ D_{\rm 1}$. \\
\end{footnotesize}
\label{default}
\end{minipage}
\end{table*}

{\bf HD 68243:} This star
               is the one example in our survey where Na\,{\sc i} D components are obviously stronger in 2011-2012
than in the spectrum obtained by Cha \& Sembach (Figure  13).
 Cha \& Sembach 's  observations  show an asymmetric Na D profile  which they modeled with
  components  at  LSR velocities of $ -8$ and $-2$ km s$^{-1}$. This asymmetry is also shown by their Ca K line profile.
  We  observed the star on
  on 2011 April  02 and 2012 February  14. Both observations show additional absorption at about $+8$ km s$^{-1}$  with a
  strengthening of the $-2$ km s$^{-1}$ component  (Table 7).  The assumption here is that changes in the D line profile reflect changes in the
  interstellar absorption along the line of sight to the star.  However, HD 68243 is a spectroscopic binary (Pourbaix et al.  2004;  Mason et al. 2009) and there is a  possibility that the changes are 
  the result of ejections of gas from one or both of the orbiting stars  but the implied low ejection velocity and narrow line width suggest this is a remote
  possibility.

\begin{figure*}
\vspace{0.3cm}
\includegraphics[width=7.5cm,height=7cm]{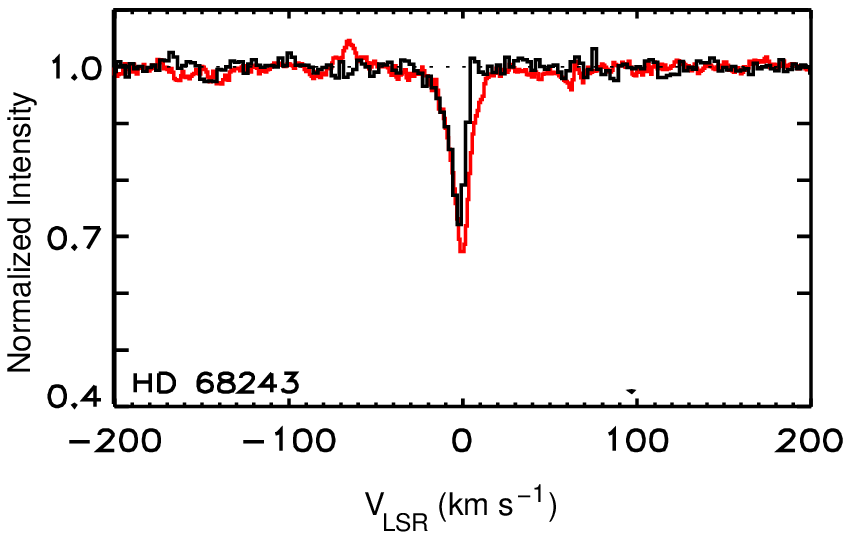}
\caption{The Na\,{\sc i} D$_2$ profiles of HD 68243. The profile obtained by
 Cha \& Sembach in 1994 (black line) is superposed on the profile obtained
 by us on 2012 February 14 (red line). Note the strengthening of the line in 2012
 due to  an extra absorption component in the red wing.  }
\end{figure*}
\begin{table*}
\centering
\begin{minipage}{170mm}
\caption{\Large HD68243 Na\,{\sc i} Absorption Lines  }
\begin{footnotesize}
\begin{tabular}{lcrrccrrrccr}
\hline
\multicolumn{1}{c}{}&
\multicolumn{3}{c}{Cha \& Sembach} &\multicolumn{1}{c}{} &\multicolumn{3}{c}{Present Observations$^*$}&\multicolumn{1}{c}{}\\
\cline{1-4} \cline{6-9}   \\
      & &$D_{\rm 2}$&$ D_{\rm 1}$ & & & &$ D_{\rm 2}$& $D_{\rm 1}$ \\
\cline{1-4} \cline{6-9}   \\
   epoch &$v_{\rm LSR}$&Eq.w&Eq.w&  &
epoch& $v_{\rm LSR}$ & Eq.w &Eq.w& & \\
      &km s$^{-1}$ &(mA)& (mA)&  &   &km s$^{-1}$& (mA)&(mA)&&   \\
\hline
       1994 &-8  &23 & 10 &  &2011-12&-11.5 &24 & 9.5 &comp 1 \\
            &-2  &33 & 25 &  &       &-2.1  &56 &32.5 &comp 2  \\
            &    &   &    &  &       & 7.6  &12 & 7   &comp 3 \\
\hline
\end{tabular}
\\
$^*$ : Average of two nights observations obtained on 2011 Apr 2, 2012 Feb 14, 2012.
       The $v_{\rm LSR}$ is an average of both $D_{\rm 2}$ and
      $ D_{\rm 1}$. \\
\end{footnotesize}
\label{default}
\end{minipage}
\end{table*}

\begin{figure*}
\vspace{0.3cm}
\rotatebox{90}{\hspace{1.2cm}Normalised Intensity}
\includegraphics[width=6.5cm,height=6cm]{shd72127a.eps}
\vspace{0.3cm}
\rotatebox{90}{\hspace{1.2cm}Normalised Intensity}
\includegraphics[width=6.5cm,height=6cm]{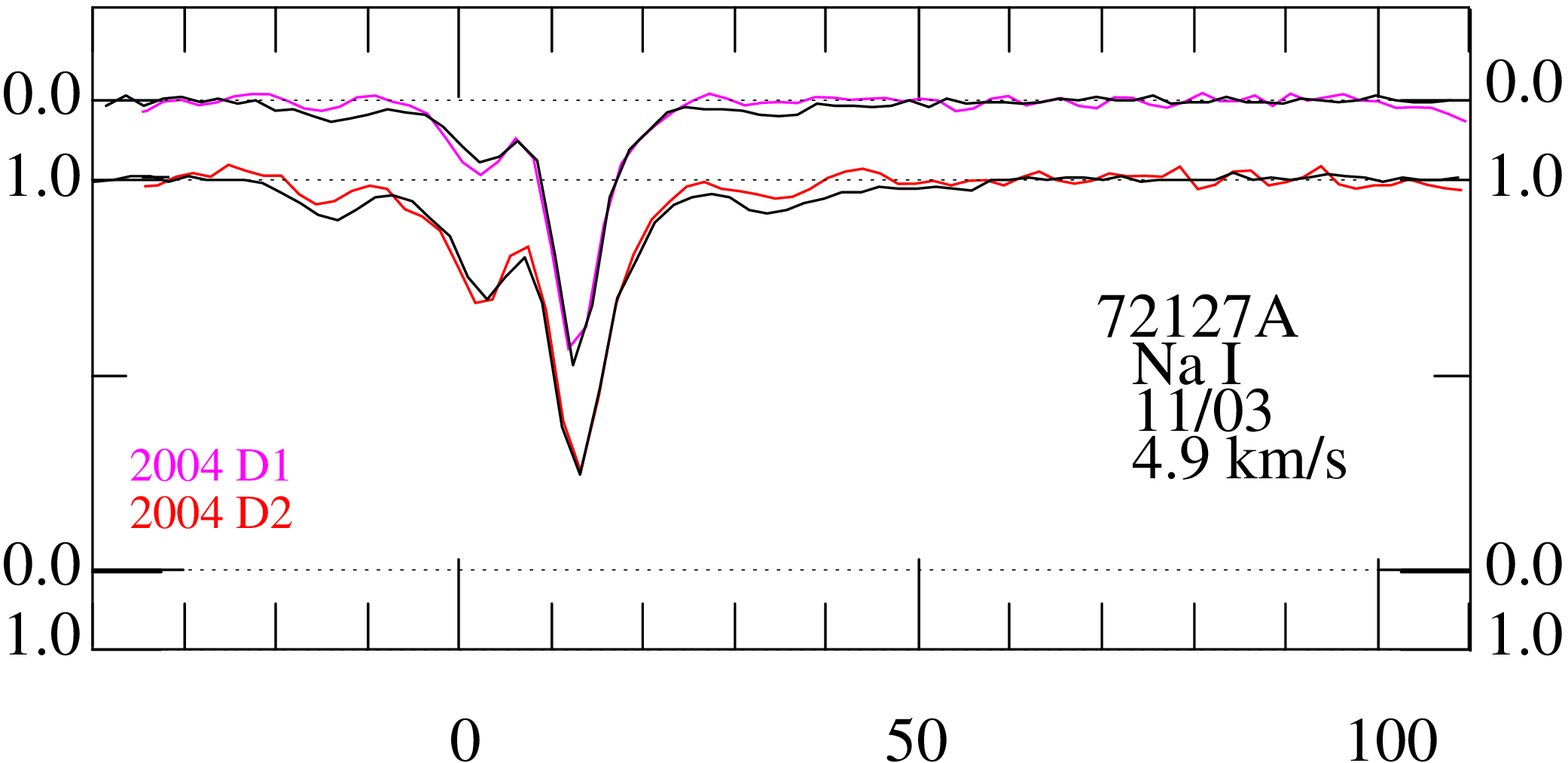}
\hspace{1cm} \hspace{8cm}$V_{\rm LSR}$(km s$^{-1}$) \\
\caption{ Left panel: The Na\,{\sc i} D$_2$ profile  of HD 72127A obtained in
 2007 (magenta) and in 2004 (red) superposed on the Na\,{\sc i} D$_2$  profile
 obtained by Cha \& Sembach in 1994 (black lines).  Note the appearance of the
 redward component at  22 km s$^{-1}$
in our spectrum  that was not prominent in
 1994 spectrum.  Right panel:  The Na\,{\sc i} D$_1$ and D$_2$ profiles (red) of
 HD72127A obtained by us in 2004 are
 coincided with  Na\,{\sc i} D$_2$,D$_1$  profiles  by Welty, Simon \& Hobbs (2008)
 observed  in 2003 November (black lines).}
\end{figure*}

{\bf  Other lines of sight:}   Cha \& Sembach noted two stars with  changes among the low velocity interstellar
components: HD 72127A and HD 73658 (see below).

HD 72127A and HD 72127B form a binary system with differences in their interstellar lines. HD72127A is also a spectroscopic binary by itself (Wallerstein,
Vanture \& Jenkins 1995). Variability  in HD 72127A was first reported
by Hobbs et al. (1991).  Cha \& Sembach drew attention to changes between 1994 and 1996 in the K
line profile but  their only Na D line profile was from 1994. Welty, Simon \& Hobbs  (2008) provide a comprehensive report on the interstellar lines from the ultraviolet to the optical in both stars.
Our observations confirm that the Na D lines are variable but the variations with respect to the 1994 profile (Cha \& Sembach 2000) and a 2003 UVES
spectrum (Welty et al. 2008) are not as dramatic as those reported earlier by Hobbs et al.  or Welty et al.  Figure 14  shows the Cha \& Sembach 1994 D$_2$ profile
along with VBT spectra from 2004 and 2007 (left panel) and the 2004 VBT and 2003 UVES spectra of D$_1$ and D$_2$ (right panel). There is a hint of variation in the
$-10$ km s$^{-1}$ component which is not surprising as this component was first shown to vary by Hobbs et al. (1991). The VBT spectra suggest a weak component
at about $+20$ km s$^{-1}$ has recently developed and at this velocity it falls within the complex of prominent K line components running from $-30$ to $+50$
km s$^{-1}$.

\subsection{Sight lines with high-velocity components}

A  feature of Cha \& Sembach's (2000) collection of Ca\,{\sc ii} K profiles was the appearance of high velocity features  to the red or blue of the
low velocity complex of  the main absorption lines.  Defining high velocity to be $|V_{\rm LSR}| \sim 100$ km s$^{-1}$,  14 of the 68 stars showed  one or more high velocity
K  line components.  Of this 14,  nine were observed twice in the 1993-1996 interval and five of these showed detectable variation in the high velocity
K line components.  ( One -- HD 73658 -- showed a large change at low velocity but no change in the weak high velocity component.)   Eleven of the 14
stars with  high velocity K line components were observed in the same interval as the Na D lines  with four showing variable high velocity absorption in also
the D lines.   The high velocity components  have somewhat larger equivalent widths in the K line than in the D lines. Such high-velocity features are quite
uncommon for sight lines through the diffuse interstellar medium and, therefore, their appearance around the Vela SNR suggests a causal association.
Here, we comment briefly on the  stars for which high velocity interstellar components have been detected.

{\bf HD 72067:}  High velocity components were not present in spectra of the K and D lines obtained by Cha \& Sembach in 1994.  Our spectra
from 2011 March 16 and December 24 show no high velocity components for the Na D lines.  The sole report of a high velocity component among the limited
observations of this star is by Hunter et al. (2006) who found a weak K line component with an equivalent width of 2.8$\pm$0.6 m\AA\ 
at a LSR velocity of 102 km s$^{-1}$ in a high S/N ratio VLT/UVES spectrum.
Hunter et al. illustrate the profiles of the very weak subordinate  Na\,{\sc i} 3303\AA\ lines but not the Na D lines but our
inspection of the archived spectrum  shows that the high velocity component in the K line is not present in the D lines.  The D lines have not changed their
low-velocity profiles over the 1994-2011 interval.


             The velocity resolution of the VLT observations is 3.75 km s$^{-1}$ not
 too different from that of Cha \& Sembach (2000)'s observations.
 The absence of high velocity component in Cha \& Sembach (2000)'s
 Ca\,{\sc ii} K profile is intriguing. The signal to noise limitation does
 not appear to
 be the  reason since Cha \& Sembach's observations are claimed to have S/N $\ge$ 100 (equivalent widths greater than 1  m\AA\ ).
 It is likely it was not present in 1994.

{\bf HD 72088:}  A 1994  spectrum of the K line showed a 20 m\AA\ line at $V_{\rm LSR}$ = +108 km s$^{-1}$ and no accompanying component in the
D lines.  Our  Na D profiles confirm the absence of this high velocity component  and the close similarity in the blended low-velocity components in the 1994 and 2011
spectra (Figure 4).

{\bf HD 72089:}  This star's Ca\,{\sc ii} K profile showed strong high-velocity components in 1993 and 1996 at  $+74$, $+87$, $+92$ and $+105$ km s$^{-1}$
with records of  this velocity gas extending back to before 1983 (Jenkins, Wallerstein \& Silk 1984;  Wallerstein \& Gilroy 1992).   In the Na D lines,
these components are much weaker with the $+76$ km s$^{-1}$ component having the same equivalent width in 1996 as in 1993 but the $+105$ km s$^{-1}$ having a much smaller equivalent width in 1996 than in 1993.  No earlier record of high velocity Na D appears to exist.  Our 2011 spectrum shows that the $+105$ km s$^{-1}$ 
component is undetectable (Figure 15) with the D$_2$ component having a equivalent width of less than 7 m\AA\ but  in 1996 23 m\AA\ was measured. The
 +75 km s$^{-1}$ component  with an equivalent width of  12-14 m\AA\ in 1993-1996  is not seen in 2011 (equivalent width of less than 7m\AA).     HD 72089 is a fine
example of short-term variability (1993-1996) in the interstellar K and D$_2$ lines with a longer term decline in the Na D lines.  New observations of the
K line would be of great interest.


\begin{figure*}
\vspace{0.3cm}
\includegraphics[width=7.5cm,height=7cm]{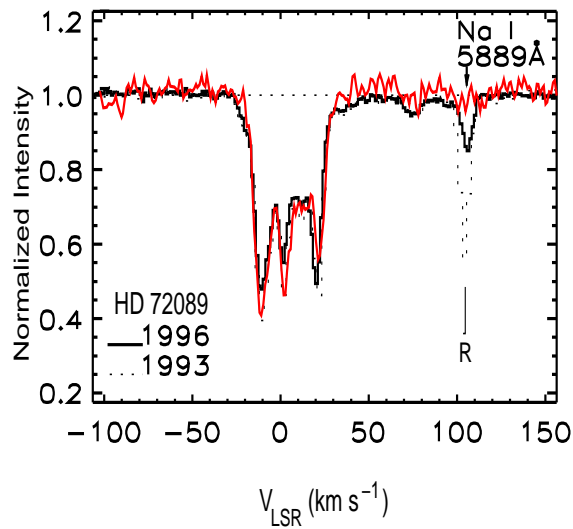}
\caption{Superposition of  the Na\,{\sc i} D$_2$ (red) profile of HD 72089 obtained on
 2011 March  20 is superposed on Na\,{\sc i} D$_2$ profile obtained in 1993
 (black dashed) and 1996 (black line) by Cha \& Sembach (2000). Note the weakening of
 and increased redshift of the high velocity component (marked R) from 1993 to 1996 and its absence in 2011.}
\end{figure*}

{\bf HD 72179:}  The 1996 Ca\,{\sc ii} K line profile showed high-velocity components  at $+61$, $+71$, $+78$,  $+84$ and $+127$ km s$^{-1}$.   Cha \& Sembach
did not obtain a matching Na D observation.  The K line components around $+80$ km s$^{-1}$ were unusually strong (Figure 2).
 Our 2011 Na D profile (Figure 2)  shows differences in the low-velocity and high-velocity components relative to the 1996 Ca\,{\sc ii} K line profile.  In particular, the  K line high-velocity component at $+84$ km s$^{-1}$ is  very weak in the D line and the
$+127$ km s$^{-1}$ line is not present in the D line.  High velocity absorption  around 72 km s$^{-1}$ in the D lines is exceptionally strong for such a component 
 suggesting that neutral  clouds do occur at high velocity. It 
 possibly might suggest  a shocked and recombined gas cloud.

{\bf HD72997:}  This star provided Cha \& Sembach with one of their seven examples of high-velocity variations in interstellar line profiles with a high-velocity
component strong at the K line and weaker in the D lines  and both exhibited  an approximately 2 km s$^{-1}$ shift to higher velocity over the interval 1991-1996 with little or no change in
equivalent width.  The 1991 K line profile was from Danks \& Sembach (1995).  Unfortunately, a VBT Na D spectrum is not available to constrain evolution of the D line.

{\bf HD 73658:}   This star qualifies as having variable interstellar line profiles. Changes have been seen in low velocity but not  high-velocity absorption components.
 The K line varied between 1993 and 1996 (Cha \& Sembach 2000, their Figure 7): the variable components were at
velocities of $-32$ and $-15$ km s$^{-1}$ and, in this respect, the variability might be more appropriately linked with the prominent changes
seen for  HD 63578, 68217 and 76161.   The Na D$_2$ profile in 2011  closely resembles that from 1993 (Figure 5).  In 1993 and 1996, the only
evidence for high-velocity interstellar gas was a weak  (5 m\AA) K line component at $-126$ km s$^{-1}$.  This did not then or in 2011 appear in the D line profiles.

{\bf HD 74194:}   This star was observed at the VBT in 2008 April. The star
showed several high-velocity blue-shifted components in the 1993 and 1996  K line spectra obtained by Cha \& Sembach.
Velocities and equivalent widths were considered unchanged over the three year interval and the star was not listed by them as having variable interstellar
components.   Their Na D observations did not show these high-velocity components. The VBT spectrum of the Na D$_2$ , although noisy, is a good match to
Cha \& Sembach spectrum (Figure 5) suggesting no variations.  

{\bf HD 74234:}  Cha \& Sembach (2000)  found high-velocity components in their 1996 K line observations but did not obtain D line observations.  Our D line
2011 observation and the 1996 K line observation are compared in Figure 2.  The D line shows high-velocity components at $+72$  km s$^{-1}$ which may be closely
related to the 1996 K line components at $+69$ km s$^{-1}$  (Table 2) suggesting, perhaps, mild
 acceleration of the cloud .

{\bf HD 74251:}  This line of sight in 1996  provided Cha \& Sembach  with weak K line components at $+78$ and $+86$ km s$^{-1}$.  Lack of a D line observation precluded 
comment on whether these components appeared also in the D lines. Our 2011 observation shows nothing at this (or other high velocities) in the
D lines (Figure 2). 

{\bf HD 74455:}  K line observations in 1994 and 1996 showed blended high-velocity components at $-166$, $-173$ and $-183$ km s$^{-1}$  and a 28 \% decrease
in equivalent width between 1994 and 1996 allowed Cha \& Sembach to label this as a line of sight with variable interstellar components. These blue-shifted components
 appear neither  in D line profiles from the same years nor in our profile from 2012 (Figure 5). 

\begin{figure*}
\includegraphics[width=8cm,height=8cm]{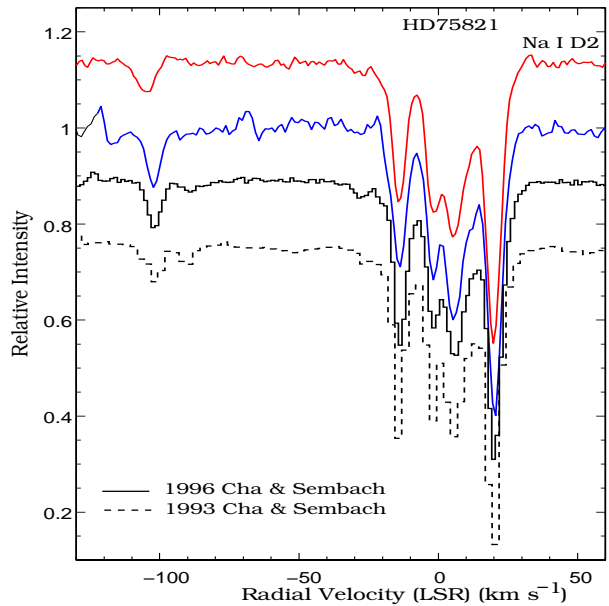}
\caption{The  Na\,{\sc i} D  profiles of HD 75821 obtained in
 1993, 1996,  2007 (blue curve)  and  2012 (red curve).
    }
\end{figure*}

\begin{figure*}
\vspace{0.3cm}
\rotatebox{90}{\hspace{1.2cm}Normalised Intensity}
\includegraphics[width=6cm,height=5.5cm]{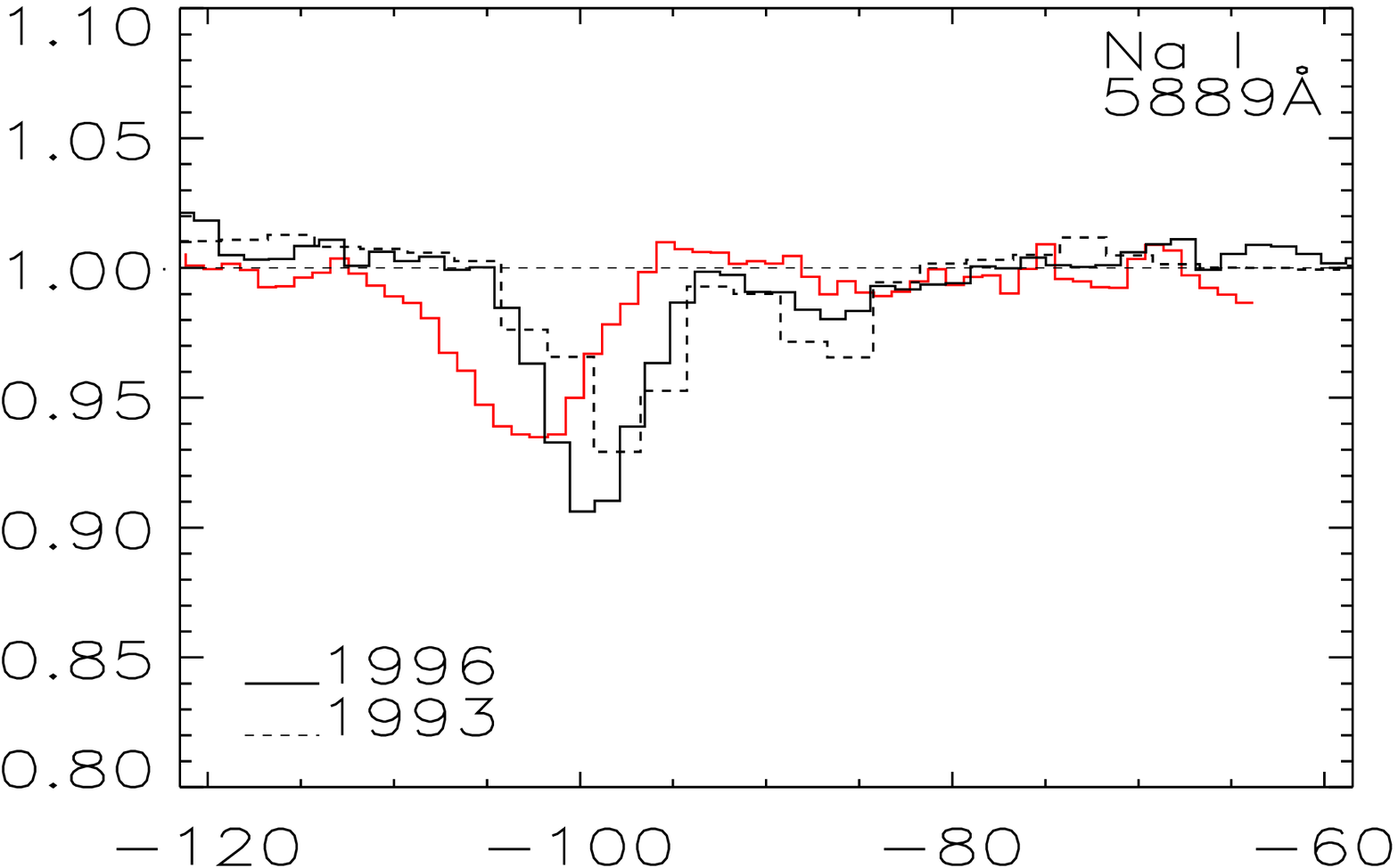}
\hspace{ 16cm}$V_{\rm LSR}$(km s$^{-1}$) \\
\vspace{0.1cm}
\caption{The $-$100 km s$^{-1}$ component of the   Na\,{\sc i} D  profile of HD 75821 obtained in
 1993, 1996 (Cha \& Sembach 2000),  and on 2012 (red).
  Note the
  increased blue shift of the high velocity component from 1993 to 2012.}
\end{figure*}

 {\bf  HD 74531:}  This line of sight is similar to that towards HD 74455.  A blue-shifted K line component  at $-141$ km s$^{-1}$ was seen not only in 1994
 and 1996 but  also earlier in 1973 (Wallerstein et al. 1980).  The equivalent width is reported as 34$\pm1$ m\AA\ for 1994 and 25$\pm1$ for 1996, a decreases
 of  26\% but apparently insufficient for Cha \& Sembach to deem it a variable component.   This blue-shifted component was not detectable in the D lines in
 1994 and it was similarly absent in our 2011 spectrum (Figure 5).

 {\bf HD 75129:}  The 1996 K line profile showed high-velocity components at $-67$ and $-98$ km s$^{-1}$ flanking stronger low-velocity absorption near
 0 km s$^{-1}$.  A Na D profile was not obtained by Cha \& Sembach. Our 2011 March D line profile shows no high-velocity components but this cannot
 be taken as an  evolutionary trend because high-velocity components generally appear weaker in the D  than the K line.
 
 {\bf HD 75309:}  This line of sight  has, perhaps, the most interesting  K line profile of those presented by Cha \& Sembach from their 1993 and 1996 observations.
 High-velocity components   were seen to the blue at $-120$ and $-74$ km s$^{-1}$ and to the red at $+81$, $+89$ and $+123$ km s$^{-1}$ . The blue-shifted
 components strengthened between 1993 and 1996  while the $+81$ and $+89$ km s$^{-1}$ components vanished between 1993 and 1996  and the $+123$ km s$^{-1}$ 
 component appeared unchanged between 1993 and 1996.  The 1993 D line profile shows none of these high-velocity components and their absence and the absence of
 all high-velocity absorption in the D lines is confirmed by our 2011 observation (Figure 6).   Of especial interest are the interstellar line profiles for HD 75309
 obtained by Pakhomov, Chugai \& Iyudin (2012) in 2008 March at a resolving power $R \simeq 88600$.  These observations confirm the presence in 2008 of the $-120$ and $+123$
 km s$^{-1}$ K line components.  Pakhomov et al. combine the 1993, 1996, and 2008 profiles to claim that the  blue- and red-shifted high-velocity components are 
 increasing their velocity separation at about 0.7 km s$^{-1}$ yr$^{-1}$ as though the centre of expansion were between their locations.  The D line  in 2008
 was devoid of high-velocity absorption. 
 
 {\bf HD 75821:}  This star provides the rare example of a line of sight in which high-velocity components are seen in both the K and the D lines.   HD 75821 was one of the  stars for which Cha \& Sembach  identified variable i high-velocity absorption  among a complex of K  and D lines at velocities  of  $-99$ and $-85$ km s$^{-1}$.
      These D line components are detected in VBT
 spectra from 2007,  2011 and 2012.
 The  trends seen from 1993 to 1996 continue;  the $-85$ km s$^{-1}$ line
 almost disappears by 2007 while $-99$ km s$^{-1}$ line gets stronger
 and is apparently accelerated by 2007.
 Figures 16 and 17 show the high-velocity component for the D$_2$ line from 1993, 1996,
 2007 and 2012 with Figure 17  showing that this component is being accelerated from $-98$ km s$^{-1}$ in 1993 to $102$ km s$^{-1}$ in 2012.
A   weak feature at $-28$ km s$^{-1}$ seems to be present
 in the D$_2$  line  from 1993 to 2012 and  to move to 
 more negative velocities with time.  Thus, not only high velocity features
 but even low velocity features are  accelerated.

 {\bf HD 76534:}  This line of sight provided a component at $-86$ km s$^{-1}$ in a 1994 K line observation with nothing detectable at this velocity in the D lines.  Our D
 line profile (Figure 6) confirms the absence of absorption at around $-86$ km s$^{-1}$.  However, there may be a weak component  about -150 km s$^{-1}$ in both
the  Cha \& Sembach's and VBT profiles.

In a simple  interpretation, high-velocity components are associated with the rim of the expanding SNR with blue-shifted lines originating from the nearside and
red-shifted lines from the far side.  Combining the SNR distance of 287 pc with the 7.3 degree angular size,  one obtains a SNR radius of about 18 pc.  Figure 18
shows the velocities of high and low velocity components seen in stars of different distances.   The striking feature  of Figure 18 is the absence of high-velocity
components in stars at  about the distance of the SNR, as indicated by the location of the pulsar,  and their first appearance  in stars at distances of about 600 pc, an observation
made earlier by Sushch et al. (2011).  The nearly 300 pc difference between the location of the pulsar and closest stars showing high velocity components
is an unexplained puzzle.

\begin{figure*}
\includegraphics[width=8cm,height=8cm]{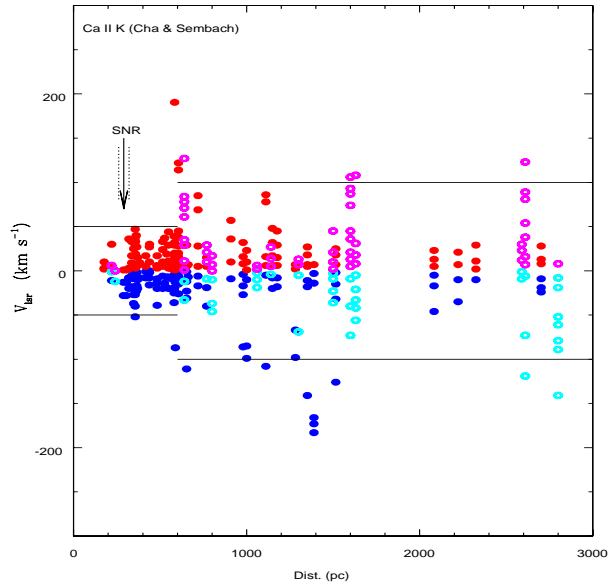}
\caption{$V_{LSR} $ of Ca\,{\sc ii} K components, positive(red) and negative
  (blue) plotted with respect to Hipparcos distances (solid symbols). When
 Hipparcos distances are not available spectroscopic distances are used 
 (denoted with circles -magenta and cyan for positive and negative 
 velocities, respectively). The horizontal lines denote $V_{LSR} $ of
 $\pm$50 and $\pm$100 km s$^{-1}$.  The location of the pulsar in the SNR is indicated by the arrow at 287 pc.}
\end{figure*}

\subsection{Other lines} 

The spectral region covered at the VBT includes other possible interstellar lines including 
the Li\,{\sc i} 6707\AA\ resonance doublet  and the K\,{\sc i} resonance lines at 7685 and 7698\AA.  

The lithium lines are of interest because energetic
particles in the SNR might produce Li  - both $^6$Li and $^7$Li - by spallation reactions.   Lithium isotopic ratios indicative of spallation have been reported for lines of sight near 
the SNR IC 348 and IC 343 (Knauth et al. 2000, 2003; Taylor et al. 2012). The resolving power and S/N ratio of the VBT spectra are inadequate for useful determinations of
the lithium isotopic ratio.  Visual inspection of the spectra does not indicate lines of sight with Li\,{\sc i} absorption of unusual strength.  In light of the fact that
regions within and near the SNR may have greater than average ionization rates for neutral lithium, analysis of the 6707\AA\ should be made relative to the
strength of a K\,{\sc i} line.

   Regarding  the K\,{\sc i} lines at 7665 and   7698 \AA,
   Pakhomov et al. (2012)  suggested that interstellar K\,{\sc i} lines  in the direction of the Vela SNR are not
  detected for stars with distances d $<$ 600 pc
  and then become sharply stronger for stars with distances d $>$ 600 pc.
   They interpret  this as due to an underpopulation  of dense interstellar  clouds
 until a distance of 600 pc.
  They go on to say that  this `hollow' in the distribution of K\,{\sc i} clouds 
 might be
  due to the dynamical evolution of Gum nebula. Apparently, the Gum supershell
  is known to be depopulated by neutral gas clouds between 350-570 pc 
  (Woermann, Gaylard \& Otrucek   2001).  This assertion is  contradicted by our observations, for example,
   the Hipparcos distance
  to HD 76838 is 336$\pm$ 59 pc yet  the sight line shows fairly strong K\,{\sc i}  lines
  (Figure 19).
     The distance to the star is about that of the Vela pulsar
 and  our observation suggests cold dense clouds are not uncommon at distances of less than 600 pc.
  We have detected the K\,{\sc i} 7698 \AA\ line in eight  stars out off 31 that
 are within a Hipparcos distance of 600 pc in our survey  and in two out of 21 below
 500 pc.
 We plan to discuss our
 observations of  the K\,{\sc i} and Li\,{\sc i}  lines and  diffuse interstellar
 bands (DIBs)  in a later paper . For the 
 present,  existence of cold dense clouds  in  the Vela SNR region is shown to be
  a reality.

\begin{figure*}
\includegraphics[width=8cm,height=8cm]{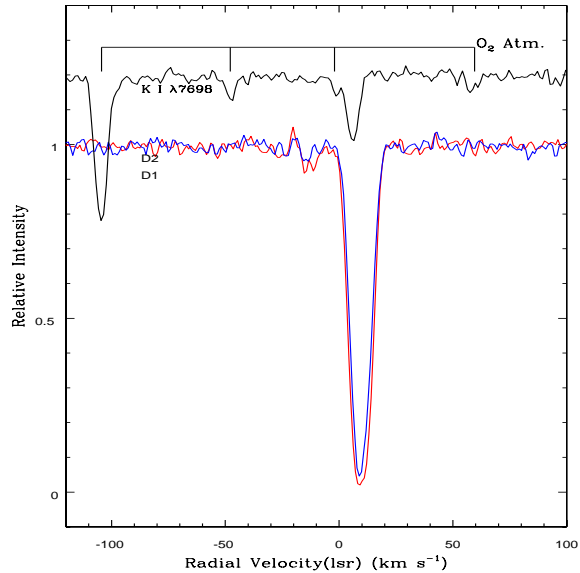}
\caption{ The  VBT Na\,{\sc i} D  profiles (red = D1, blue=D2) and K\,{\sc i} 7698 \AA\  interstellar  profiles of
  HD 76838 obtained on   2011 March 23.   The line across the top of the figure marks the locations of telluric O$_2$ absorption lines around the
  K\,{\sc i} line. }
\end{figure*}

\section{Concluding remarks}

Interstellar lines seen in spectra of stars lying
behind the Vela SNR  differ in two
principal respects from interstellar lines produced by the diffuse interstellar
gas along the typical of line of sight through the Milky Way. These two
differences, which are exhibited by the Na D lines whose evolution over nearly
two decades was the focus of this paper, suggest that the key to understanding
the two differences is to be found in interactions between the SNR remnant and
the gas now embedded in the remnant. Such gas may  be part of the remnant,  circumstellar gas
earlier ejected by the star or interstellar gas swept up by the expanding remnant.

\begin{figure*}
\includegraphics[width=10cm,height=8.5cm]{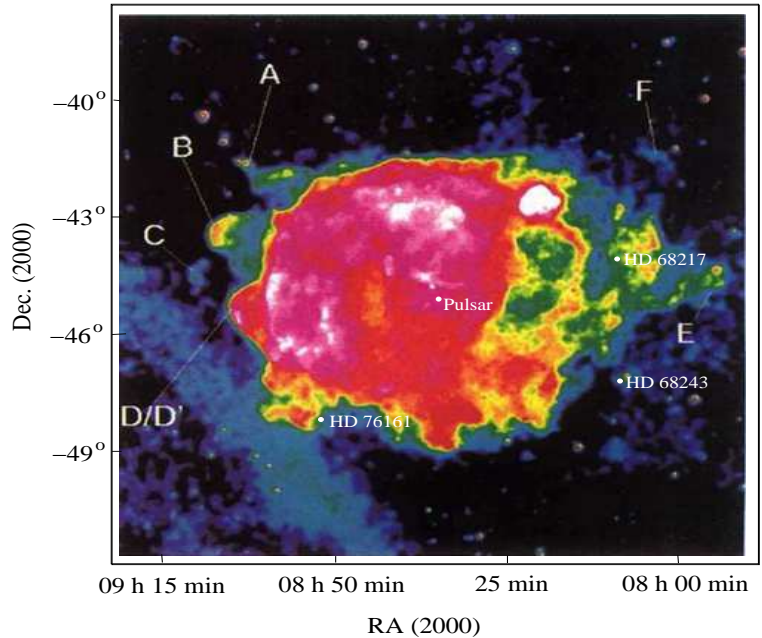}
\caption{ROSAT All-Sky survey image (0.1-2.4 Kev) of the Vels SNR
(Aschenbach et al. 1995). A-F are extended features outside the boundary
of the remnant (`bullets'). Light blue to white contrast represents
a contrast in surface brightness of a factor of 500. The location of
 three of the stars  that showed strong variable Na\,{\sc i} absorption in their ISM spectra. The blue stripe in the lower left hand corner
 is unrelated to SNR- apparently due to unremoved scattered solar X-rays.}
\end{figure*}

The first difference between Na D absorption along sight lines through the Vela SNR and typical sight lines in the Milky Way
concerns the presence of high-velocity components.  
Many of the observed lines of sight crossing the Vela
SNR show high-velocity absorption components.
 When multiple observations
are available, such components prove to be variable on a timescale of a
few years (Cha \& Sembach 2000). High-velocity absorption is particularly
well seen in ions such as Si\,{\sc iv} and C\,{\sc iv} through their
ultraviolet resonance lines
(Jenkins, Silk \& Wallerstein 1976; Jenkins et al. 1981;
Jenkins, Wallerstein \& Silk 1984). As Jenkins et al. (1984) remark ``The
velocity distribution of the triply ionized gas is probably representative
of the kinematics of the Vela remnant since gas at such a high ionization
is not commonly seen in the general interstellar medium." Unfortunately, a
general lack of repeat observations of high resolution ultraviolet spectra
means that reports of
the variability of the high velocity absorption are limited to
scarce repeat observations of the Ca\,{\sc ii} K line (Cha \& Sembach 2000)
and of the Na D line  but the latter almost never shows strong components at high velocity
(Cha \& Sembach 2000; this paper). The high velocity components are
considered to arise from the passage of the supernova's blast wave through
an inhomogeneous  medium- see Jenkins et al. (1984), Jenkins \& Wallerstein (1995) and
 Sushch et al. (2011).
Destruction of grains by the blast wave likely accounts
for the lower depletion of nonvolatile elements and, hence, the high ratio of 
 N (Ca\,{\sc ii}) to (Na\,{\sc i}).  However, it is
still unclear why high velocity components (in Ca\,{\sc ii} K) occur
  only in stars more distant than 600 pc when the SNR is at a distance
 of  290 $\pm30$ pc (Figure 18).

The second example of a distinct characteristic of interstellar
lines, specifically Na D lines, through and near the Vela SNR is that
four of the sightlines observed both with the VBT  and by Cha \& Sembach
show dramatic changes in the Na D profiles at low velocity.
Across the extensive literature on interstellar Na D lines there are
no reports of such large changes in Na D profiles from the quiescent
diffuse interstellar medium. The closest example may be the sightline to
the halo star HD 219188 (galactic latitude of $-50$ degrees)
which  showed
 a   factor of three
growth in Na D column density of the high-velocity gas from 1997 to 2000 but no detectable change in
several low velocity Na D components
(Welty \& Fitzpatrick 2001; Welty 2007). In addition to the few examples of sightlines
near the Vela SNR with large changes in Na D equivalent widths, some sightlines
show small changes in equivalent widths of low velocity components.
A clear example is HD 72127A (Hobbs et al. 1982; Welty et al. 2008).
There are a few examples of (small)
profile changes involving  the
general  interstellar medium - see, for example $\rho$ Leo (Lauroesch
\& Meyer 2003). Such changes are generally attributed to the line of sight to the star
 changing as star and interstellar cloud move and  different portions of an inhomogeneous cloud are sampled.
 Little attention has been given to
time dependent  changes of the physical conditions in the clouds.
For the four examples identified here of dramatic changes in the
Na D profiles, time dependent changes may be competing with geometric
changes.

In  three of the four 
cases  of large changes of equivalent width associated with the Vela SNR
 -- HD 63578, HD 68217 and HD 76161 -- the low velocity
blend observed in 1993-1996
was very much weaker by 2011-2012. In the fourth case -- HD 68243 -- the VBT
spectra  from 2011 and 2012
showed a strengthening relative to the 1994 profile obtained by
Cha \& Sembach.  Sight lines, as shown in Figure 20,  to three of these stars --
 HD 68217, HD 76161 and HD 68243 -- pass close to the edge of the ROSAT
0.1-2.4 keV image (Aschenbach et al. 1995).  HD 68217 may located behind a Vela SNR X-ray bullet, the
  extension E as
 shown in Figure 20. The Hipparcos distances of 383$\pm$30, 318$\pm$50,
 and 240 pc respectively of this trio suggests a
  physical connection  with the SNR (distance of 290 $\pm$30 pc).
  The strong Na D lines present in 1993-94 period imply large 
 column densities: N(Na\,{\sc i} ) of  about
  $\sim$1.6x10$^{13}$ cm$^{-2}$ for HD 76161 (Welsh et al. 2010) and a
 N(H\,{\sc i}) column density of $\sim$1.5x10$^{21}$ cm$^{-2}$, assuming
 the relation between N(Na\,{\sc i}) and N(H\,{\sc i}) for an average ISM cloud (Welty 2007).  For the size of a neutral cloud, as estimated by 
 Pakhomov et al. (2012) in the vicinity of HD 76161 in Vela of 0.85 pc would
 suggest a density of  N(H) of 580 cm$^{-3}$. Destruction of such a cloud
 by supernova blast wave as pictured in Vela SNR by Bocchino, Maggio \& Sciortino (1999)
 and Pakhomov et al. (2012) seem to require several thousands of years.
  The time scale of 17 to 18 years, as observed, for the disappearance of the
 clouds appears to be too short for  a supernova shock. Some other rapid
 cloud destruction  mechanisms seem to be  operative.
 However, one is tempted to argue
that physical conditions in these particular low velocity clouds were
affected by the SNR remnant: increased photoionization of Na by X rays or
removal of neutral Na atoms?

The sightline to
HD 63578, the fourth star
with a large change in Na D strength, does not pass near the SNR and its
 X ray emission. At the distance of 481 pc, HD 63578 is located behind the
wind swept bubble of $\gamma^2$ Vel. It is well known that high velocity
winds from Wolf-Rayet stars and, in particular, $\gamma^2$ Vel are clumpy
(Lepine, Eversberg \& Moffat  1999) and strong with wind velocities of 1500 km s$^{-1}$ (De Marco, Schmutz \& Crowther  2000). 
Perhaps, a clump was crossing the line of sight at the time of the VBT observations.

Continued pursuit of variations in the interstellar lines along sightlines
through and near the Vela SNR is to be encouraged. To expand the insights
into the principal variations described here it will be  helpful
to obtain high-resolution optical spectra over a broad wavelength range
such that Ca\,{\sc ii} K and Na D lines are both recorded along with a host
of weaker atomic and molecular lines. These spectra will serve to follow up
the principal discovery of this paper, namely large variations -- increases and
decreases --  in low velocity components of Na D lines and, in particular,
provide the first opportunity to map out long term changes in the Ca\,{\sc ii}
K line profiles relative to the 1993-1996 baseline established by Cha \&
Sembach (2000). More difficult to obtain but more insightful about the
high velocity components and  their variations will be high-resolution
ultraviolet spectra. In this case, the start of the  baseline would appear to be the
series of IUE spectra obtained in 1979-1981 and discussed by Jenkins
et al. (1984).

\section{Acknowledgements}

We thank the anonymous referee for a thorough and constructive report.  This research has made use of the SIMBAD database, operated
at CDS, Strasbourg, France. We would like to express our thanks to
 Davide de Martin for the permission to include the Vela SNR image
 shown in Figure 1. NKR would like to thank Melody and David
  Lambert, Vimala and Satya Mandapati for their hospitality during
  his stay in Austin, where part of this work was done. NKR would also
 like to thank Arun Mangalam for a discussion. We also would like to thank 
 staff of  the VBO at  Kavalur for their help with observations.

\end{document}